# HOW CAN WE MEASURE THE INFORMATION CREATED BY NATURAL SELECTION?

Joel R. Peck[a]
Department of Genetics
University of Cambridge
Cambridge, UK,

David Waxman[b]
Centre for Computational Systems Biology, ISTBI
Fudan University
Shanghai, PRC

(a) Email: joel.r.peck@gmail.com
(b) Email: davidwaxman@fudan.edu.cn

# Abstract

Natural selection can create information. In particular, because of the action of natural selection, we can often learn something about an environment by examining local organisms, and vice versa. For example, the characteristics of a cactus suggest that the local environment is relatively dry, and if a natural terrestrial environment is dry, then we will generally have an enhanced probability of finding drought-resistant plants (like cacti). Here, we propose a measure that can be used to quantify the information that is created by natural selection. We call the proposed quantity *reproductive information*, and we show that it has an intuitively satisfying relationship to standard quantitative definitions of information. Reproductive information is also approximately equal to a previously defined measure of biological adaptation. In addition, we explain how reproductive information can be measured using phenotypic characters, instead of genotypes. This could facilitate the measurement of reproductive information, and it could also allow for the quantification of the information that is created by natural selection on *groups* of organisms, instead of just selection on individuals. Thus, the concept of reproductive information has the potential to advance research on the "units of selection", the "major transitions in evolution", and the emergence of "superorganisms" via cooperation among group members.

One of the most fascinating aspects of life is the way that typical organisms seem to be complex mechanisms that have adapted in order to avoid destruction, and to reproduce[1–6]. It is generally accepted that this aspect of life is a result of the action of natural selection[2–4,6]. The adaptedness that results from natural selection creates correlations between the forms of organisms and their environments. Thus, for example, camels tend to occur in environments that are warm and dry. The correlations created by natural selection imply that the characteristics of organisms generally can provide information about their environments, and vice versa. Here, we describe a new measure that is intended to quantify the information created by natural selection. This information is, essentially, about what is required to survive and then reproduce in a particular environment. We therefore call the new measure *reproductive information*. We show that reproductive information has properties in common with standard measures of information[7–10], and that it is *mutual,* so that the reproductive information that an organism provides about its environment is approximately equal to the information that the environment provides about the organism[8,9]. These are characteristics what one would expect from a well-behaved measure of information. We note that the potential complexity of *groups* of organisms suggests that these groups may be able manifest substantially more reproductive information than is the case for individuals. Thus, the application of the concept of reproductive information to groups of organisms may provide powerful tools for analysis of the complex phenomena that enable the existence of contemporary life. These phenomena include mutualism, endosymbiosis, "altruism", and the "Major transitions in evolution"[11–24].

For all contemporary plants, animals, fungi, protists, and prokaryotes on Earth, DNA serves as the hereditary molecule[25]. The structure and the dynamics of DNA suggests that, within its genome, a typical organism can embody a very substantial amount of reproductive information. For example, specifying the nucleic-acid sequence of one human being would require about 700 megabytes of data, and this is a typical figure for mammals, fish, and flowering plants[26,27]. Furthermore, again taking humans as an example, it seems clear that at least 7% of the genetic sequence is subject to *purifying selection*, which means that small changes to these parts of the sequence will typically lead to a decrease in fitness[28]. Thus, a naïve interpretation suggests that the amount of reproductive information contained in the human genome is at least 7% of 700MB (i.e., 49MB). This is a substantial amount of information, as it is sufficient to encode the text from about 80 average-length English books. The impression of there being a very considerable amount of reproductive information embodied in typical organisms is reinforced by noting the number and complexity of the biochemical processes and interconnected parts that seem to be required for the functioning of contemporary organisms, particularly those that are multicellular. The fact that no known DNA-based organism has a genome with fewer than 150,000 base pairs also suggests that considerable information is required for the functioning of all contemporary DNA-based organisms[29].

The preceding "naïve" method for estimating reproductive information amounts to simply counting the sites under purifying selection, and then assigning two bits of reproductive information for each selected site. (It is two bits because there are four alternative nucleotides for DNA, and thus it takes (at most) two bits of data to specify one nucleotide[7,8]). This procedure is, in essence, very similar to the best-known method that has appeared in the scientific literature for measuring reproductive information (or something very much like it). This method was proposed by Adami and Cerf, who aimed to estimate the "*physical complexity*" of the genome[30–33]. These authors said that "the physical complexity measures the amount of information about the environment that is coded in the [genetic] sequence"[30]. In context, it is clear that Adami and Cerf consider that this "coding" is the result of natural selection, and thus their "physical complexity" seems to be more-or-less identical to the quantity that we call reproductive information in this work. A number of other investigators have made proposals that are, to a greater or lesser extent, along the same lines as what was suggested by Adami and Cerf[30,32,34–45]. Adami and Cerf's proposal, along with a number of other prominent ideas about how to measure the information created by natural selection, are discussed in Supplementary Note 1.

The method for measuring reproductive information just outlined, which consists of counting nucleotide sites that are subject to purifying selection, is relatively convenient. This is because it is easy to explain, and it can be used to produce estimates of reproductive information using existing databases[28,46–51]. However, this method is also unreliable. One of the reasons for this has to do with some results that emerged from studies of the *NK* model. The *NK* model was initially developed by Kauffman and his colleagues[52–54] in the late 1980's. We discuss the relevant results next.

# Epistasis and the measurement of reproductive information

In a typical implementation of the *NK* model[52–56], a haploid population starts by being *fixed* on a single randomly selected genotype. (Here, fixation means that all population members have the same genotype). The average relative fitness of the population then increases over time due to the occurrence of random mutations, followed by selection. Deleterious mutations become extinct, and beneficial mutations become fixed in the population. Evolution stops when the population becomes fixed on a *locally optimal genotype*, which is a genotype for which *any* single change of a nucleotide will cause a decrease in fitness. Thus, metaphorically speaking, a local optimum is the summit of a hill in the *"fitness landscape."*

The *NK* model is intended to facilitate the study the effects of *epistasis,* which occurs when the genotype at one nucleotide site within a genome alters the fitness effects of the genotypes at other nucleotide sites. As one might expect from basic physiological considerations, epistasis is ubiquitous in contemporary organisms[57–59].

Analysis of the *NK* model can be challenging. The situations that yield informative results most readily are the *no-epistasis case,* and the *maximal-epistasis case.* In the no-epistasis case there are no fitness-related interactions between nucleotide sites. As a consequence, a fitness-altering mutation at a given nucleotide site is either beneficial in all genetic backgrounds, or it is deleterious in all genetic backgrounds. In the maximal-epistasis case all nucleotide sites interact, so that changing the genotype at one site will alter the fitness effects of possible genetic changes at all other nucleotide sites.

In typical implementations of the *NK* model, every possible genotype has a unique fitness value, which is shared with no other genotype[53,54]. As a consequence, there is always a single genotype that has the highest fitness. This maximally fit genotype may be called the *globally optimal genotype.* The globally optimal genotype is, of course, also a locally optimal genotype.

Results from the *NK* model show that, in the no-epistasis case, the population always eventually becomes fixed on the globally optimal genotype[52–54]. With four possible nucleotides, a random initial haploid genotype can be expected to have a sub-optimal nucleotide present at about 75% of its nucleotide sites. This means that, in general, mutation and selection must change the genotype at (roughly) 75% of nucleotide sites before evolutionary change stops with the fixation of the optimal genotype. Thus, in some sense, when evolution stops, the genotype of a single randomly selected population member embodies a substantial amount of information. In particular, if the genome has a fixed length of $L$ nucleotide sites, then, after evolutionary changes stop happening, the genome of a single population member can tell us which of the $4^L$ possible haploid genomes is the most fit (assuming four possible nucleotides). Even for a bacterium with only 5,000,000 nucleotide sites, this is a lot of information. After all, if we use base 10, then $4^{5,000,000}$ has more than 3,000,000 digits. Thus, specifying any one particular genotype is likely to require a considerable amount of data.

However, the picture is very different in the maximal-epistasis case. Here, approximately $\ln(3L - 1)$ beneficial mutations are expected to become fixed before evolutionary change stops[54,60]. Thus, for a genome with 5,000,000 nucleotide sites, evolution is expected to cease after only (approximately) 17 beneficial mutations have fixed. After evolutionary changes have stopped, the population will be fixed on a locally optimal genotype. That locally optimal genotype will (almost certainly) be in the same neighbourhood of "genotype space" as the initial randomly selected genotype. This must be the case because, when evolution ceases, we can expect that only one nucleotide in (about) 300,000 will have changed from what it was in the initial randomly selected genotype.

These observations may lead one to conclude that, in the maximal-epistasis case, not much information accumulates during the course of evolution. A more formal information-theoretic analysis shows that this is the case, as we demonstrate in Supplementary Note 2. In fact, in the maximal-epistasis case, the genome of an individual drawn from the population after evolution ceases tells us little aside from giving us the approximate position in genotype space of the randomly selected initial genotype.

For our purposes, the crucial aspect to keep in mind about these results is that, under the *NK* model, when evolution ceases due to fixation on a locally optimal genotype, *any* change at *any* nucleotide site will result in a decrease in fitness. Thus, we have purifying selection occurring at *every* nucleotide site within the genome. This means, for example, that if we use the counting-sites-under-purifying-selection method, then, after evolutionary changes cease, we would assert that an organism with a genome with 5,000,000 nucleotide sites contains 10,000,000 bits of reproductive information. In the maximal-epistasis case, this is implausible, now that we know that, after evolution ceases, the vast majority of the genome has a *completely random* sequence that has *never* been altered by natural selection.

These considerations are elaborated in Supplementary Note 2, and this Note also shows that the conundrum cannot generally be resolved, even if one considers an infinite population in which simultaneous mutations at multiple nucleotide sites are allowed, so that all possible genotypes are created during every generation.

When considered from an information-theoretic perspective, the foregoing well-known results of the *NK* model suggest that, when epistasis is common, the method of counting nucleotide sites that are subject to purifying selection *cannot be relied upon*. A number of other considerations lead to this same conclusion, and three of these are presented in Supplementary Note 3. The present work is intended to point the way towards a more appropriate measure of reproductive information. We begin, in the next section, by providing a formal definition of reproductive information.

# A quantitative definition of reproductive information

In this section we will propose a formal and quantitative definition of reproductive information. Our aim is to define a quantity that can, at least in principle, be calculated, given access to the necessary data. We also seek a quantitative definition that corresponds, reasonably closely, to the informal statements made above, which say that reproductive information is "the information created by natural selection" and that reproductive information "is about what is required to survive and then reproduce in a particular environment."

The formal definition that we will propose here may be of particular interest because, as we shall see, it has a close quantitative relationship to key quantities from the mathematical theory of information[7–10]. It is also closely related to a previously published quantitative definition of adaptation[61]. Thus, the formal definition of reproductive information presented below has a certain 'external validity'.

We begin by providing an informal verbal definition as follows:

> **The reproductive information provided by the fitness landscape is the decrease in uncertainty about the characteristics of an organism that occurs when, in addition to the fitness landscape, we also take into account the theory of natural selection.**

In order to formalise this informal statement, we must define the term "theory." We must also say what we mean by a "decrease in uncertainty". In addition, we must explain what is meant by the "characteristics" of an organism. Finally, we must provide a precise definition of the "fitness landscape."

Let us start by explaining how we characterise individuals. We do this in terms of individual types (or *organismal types*). We can describe an individual's organismal type using continuous variables, discrete variables, or a combination of the two. Thus, an individual's organismal type may tell us about its genotype, or about its phenotype, or both. For present purposes we shall confine ourselves to the simple case in which organismal type is characterised by a single discrete variable with a finite number of possible values. This choice is fairly general because continuous variables can be 'discretised' by dividing possible measurements into groups. Furthermore, a discrete variable with an infinite number of categories can be turned into a discrete variable with a finite number of categories by simply designating a particular finite set of categories, and then classifying anything else as "other". We use $\Omega$ to represent the total number of categories (i.e., the number of possible organismal types). We number the organismal types as $1, 2, \ldots, \Omega$. The integer that is assigned to a particular organismal type is decided entirely at *random*. (Random assignment gives reproductive information certain desirable properties, as explained in Supplementary Note 4.) The integer assigned to a particular organismal type it called the *organismal-type index* for that type, and we will use $x$ to represent this quantity.

We can now formally define the fitness landscape. We use the vector $\vec{W}$ to represent the fitness landscape that can be used to characterise a particular location at a particular time. This vector has $\Omega$ elements, where the $i^{th}$ element, $W_i$, equals the absolute fitness of organismal type $i$. That is, $W_i$ is the expected reproductive success of type $i$ at a particular location, and at a particular time.

Next, let us say what we mean by a 'theory.' For our purposes, a theory is, essentially, a method for making predictions about observations of the real world. Theories contact the real world by means of these predictions. The veracity of a theory's predictions is a key factor in determining whether it is accepted, or not. With this in mind, we shall characterise theories solely in terms of their predictions.

The class of theories we consider in this work makes predictions in terms of *probabilities*. A particularly simple case is a theory that, for any given situation and any given organismal type, provides the probability that an organism found in that situation will have that particular organismal type. A theory of this sort provides, for any time and place, a total of $\Omega$ probabilities (one for each organismal type). These probabilities sum to unity.

Finally, we shall explain what is meant by a "decrease in uncertainty." Here, we follow the example of Shannon, who is the best-known figure in the development of the modern mathematical theory of information[7–9]. Shannon equated a decrease in uncertainty with the receipt of information, and we take the same view. The equivalence between a decrease in uncertainty and the receipt of information seems intuitively sensible. For example, our uncertainty about the weather outside is likely to decrease when we hear a thunderclap, and, in so doing, receive relevant information.

Next, let us consider how to quantify information. Fortunately, there is little need for innovation in this matter. This is because, due to the efforts of Shannon, along with various other researchers, a well-developed mathematical theory of information has existed since the mid-20$^{th}$ century[7–10,62,63]. This theory provides two different well-known quantitative definitions of information. One of these definitions was originally presented by Shannon in a seminal publication in 1948[7–9]. This type of information is often called *mutual information*. The concept of informational mutuality will be discussed below. However, here we will simply note that Shannon's notion of information is not unique in its "mutuality." We will, therefore, simply use the phrase *Shannon information* to refer to the type of information described by Shannon. Shannon information is calculated using the joint statistical distribution of two variables. At present, Shannon information is, by far, the most widely used quantitative definition of information[8].

In the 1960′s three researchers independently began to explore a different kind of information, which does not depend on statistical distributions. Instead, this second sort of information was intended to convey the amount of information that one *particular* object provides about another *particular* object[9,10,62,63]. This second type of information is sometimes called *algorithmic mutual information,* and it is most commonly associated with Kolmogorov. In this work, we will simply refer to this type of information as *Kolmogorov information.*

While there are technical differences between Shannon information and Kolmogorov information, both definitions are conceptually very similar. In particular, both definitions express information in terms of a decrease in the amount of data that must be provided to communicate a description of an observation[7,9,10]. Thus, for example, the amount of information that one requires to identify the location of a particular house decreases once one learns that the house is located in, say, Edinburgh. Further details regarding Shannon information and Kolmogorov information are provided below.

Let us now consider the question of how much data is required to specify the results of a particular observation. As noted above, this is the crucial question that lies at the heart of quantitative definitions of information. The answer, of course, depends on how we encode the various possible observations. However, because we are interested in "How much data is required?" we are naturally motivated to consider only efficient encoding schemes. These are coding schemes that use relatively short strings of symbols to encode data.

In line with information-theoretic tradition, we will consider only binary coding schemes[7,8]. Such coding schemes contain two symbols in their 'alphabets'. The results of information theory are more-or-less independent of the size of the alphabet used to encode data[7,8], and it has become traditional to utilise the smallest-possible useful alphabet, which is one that contains only two symbols. We will denote these two symbols by '0' and '1'.

On their own, two symbols are sufficient for a coding scheme only if there are $\Omega \leq 2$ possible organismal types. For larger values of $\Omega$ we must string zeros and ones together to form longer "code words". To ensure maximum efficiency, we will assign one code word to each organismal type. The assignment of particular code words to particular organismal types constitutes a *coding scheme*. For any value of $\Omega$, we consider only coding schemes that use the $\Omega$ shortest-possible code words. In particular, we will use the first $\Omega$ code words in a list of all possible binary code words, written in length-lexicographic order (which means that shorter sequences come first, and we order "alphabetically" among sequences with a particular length). For example, the first 14 binary code words in length-lexicographic order are shown Fig. 1.

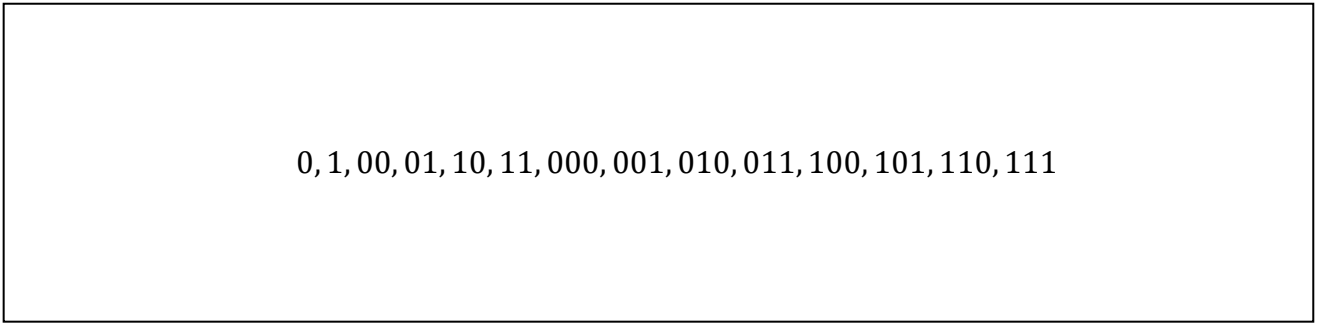

**Fig. 1 | The first 14 binary code words.** This figure shows the first 14 binary code words in length-lexicographical order. The code words are ordered from left to right. Shorter sequences come first, and the ordering is "alphabetical" among sequences with a particular length

We have now specified a definite way to produce the $\Omega$ different binary code words that we will use to encode the $\Omega$ different organismal types. The next question is, which binary code word do we assign to which organismal type? There are $\Omega!$ different ways to assign the $\Omega$ binary code words to the $\Omega$ different organismal types (corresponding to all possible permutations of the $\Omega$ organismal-type indices). Hence there are $\Omega!$ possible coding schemes. For example, if $\Omega = 14$, then there are $14! = 87,178,291,200$ different possible coding schemes.

It is important to keep in mind that *code words* are not the same as *organismal-type indices*. Organismal-type indices are integers that range from 1 to $\Omega$. They are labels that allow us to designate particular organismal types. A *coding scheme* for organismal types assigns a *unique binary code word* to each organismal-type index. That is to say, a coding scheme uses binary code words to label organismal-type indices.

For our purposes it is useful to divide the $\Omega!$ possible coding schemes into two groups. We will call one of these groups the *optimal coding schemes*. These are coding schemes in which the shortest sequences are associated with the fittest organismal types. A precise definition of optimality in this context is given in Supplementary Note 5. Note that whether a particular coding scheme is optimal for a particular time and place will depend on the fitness landscape that characterises that time and place. Furthermore, there will generally be multiple optimal coding schemes for any time and place. The only exception is when *every* organismal type has a unique fitness value that is shared with no other organismal type.

The theory of natural selection suggests that, all else being equal, the fittest organismal types (i.e., those with the highest expected reproductive success) are the ones that are most likely to be present in a population[2–4,6]. The optimal coding schemes are optimal in the sense that, if the theory of natural selection correctly predicts the probability of observing different organismal types, then an optimal coding scheme will tend to minimise the average length of the code words used to encode the organismal types of population members. (Note that, in this work, the word "average" always refers to the arithmetic mean.)

Consider a particular organismal type that has been observed at a particular location, and at a particular time. Note that different optimal coding schemes may assign different code words to this organismal type. Let $\bar{L}_s$ represent the *mean length* of the binary code words for this organismal type, if we use an optimal coding scheme. This mean length is obtained by averaging, with equal weight, over all of the coding schemes that, at that place and time, we can describe as optimal according to the theory of natural selection. Furthermore, let $\bar{L}_u$ represent the mean length of the binary code words for the same organismal type, when we average, with equal weight, over all $\Omega!$ coding schemes (this average will include both optimal and non-optimal coding schemes). Thus, $\bar{L}_u$ represents the expected code-word length when there is no preferred theory regarding the probabilities of the $\Omega$ organismal types.

Using this notation, we can now specify our formal quantitative definition of reproductive information. For a particular organism with organismal-type $x$, that is observed at a particular place and at a particular time, the reproductive information that the fitness landscape provides about an individual's organismal type is written as $I_R(\vec{W}:x)$, and is defined as:

$$I_R(\vec{W}:x) = \bar{L}_u - \bar{L}_s. \qquad (1)$$

The value of $I_R(\vec{W}:x)$ is the expected decrease in code-word length that is achieved if a particular organismal type ($x$) is encoded using a coding scheme that is optimal under the theory of natural selection, as compared to the expected code-word length for an arbitrarily chosen coding scheme. This decrease in code-word length provides a *quantitative measure* of the decrease in uncertainty about an organism's characteristics that occurs when we consider the theory of natural selection, along with the local requirements for survival and reproduction. Thus, $I_R(\vec{W}:x)$ is a quantitative definition of reproductive information that is firmly based in the concepts of information theory.

The range of possible values of reproductive information depends on $\Omega$, and on the fitness values that are associated with the various organismal types. The range of possible values is specified in Supplementary Note 6. Furthermore, Supplementary Note 7 discusses the generalisability of the methodology just described.

The definition of reproductive information that is embodied in equation (1) may be impractical in many circumstances. After all, it requires knowing the ordering of fitness values for (potentially) a very large number of organismal types. However, the definition seems to be in line with intuitive concepts related to information and natural selection. Also, as we shall see, it has satisfying and desirable mathematical characteristics. Thus, it may be able to serve as a "gold standard" for what, ideally, we would calculate when trying to quantify the information created by natural selection. We could then use this quantitative definition of reproductive information to guide the formulation of "proxy measures" that can allow us to approximate reproductive information while using measures that are more practical in field and laboratory situations.

In the next section we shall show that a known quantity that has been used to quantify adaptation can be used as a proxy measure for reproductive information. This known quantity is much more convenient to calculate than is reproductive information, and it is guaranteed to differ from reproductive information by no more than a few bits. However, using this quantity still requires some knowledge of the fitness landscape. Thus, in many cases, it would be helpful to have proxy measures that are even simpler to use. Mathematical investigations that compare these proxy measures to reproductive information should be able to reveal their shortcomings, and thus help us to specify conditions that must be met before they can be relied upon.

# Reproductive information is approximately equal to a measure of adaptedness

Equation (1) provides an explicit definition of the reproductive information that the fitness landscape provides about organismal type. However, equation (1) does not provide a convenient way to estimate reproductive information in various circumstances. Fortunately, it is possible to produce a simple expression that provides a very close approximation to $I_R(\vec{W}:x)$. To be specific, let us consider a particular organismal type, which we shall call the *focal type*. Say that, at a particular place and time, there are $Q_x$ organismal types that are *at least* as fit as the focal type. (Note that these $Q_x$ types include the focal type itself.) Thus, for example, if the focal type is the fittest possible type, and if no other types are as fit as the focal type, then $Q_x = 1$.) Using this definition of $Q_x$, we have the following expression, which is proved in Supplementary Note 8.

$$\left| I_R(\vec{W}:x) - \log_2\left(\frac{\Omega}{Q_x}\right) \right| < 4, \quad (2)$$

where $|y|$ represents the absolute value of $y$, and $\log_2(y)$ is the logarithm of $y$, when taken to base 2. Equation (2) tells us that the absolute difference between $I_R(\vec{W}:x)$ and $\log_2\left(\Omega/Q_x\right)$ can never be larger than four. In cases of biological interest, the amount of reproductive information will often be quite large. As such, the bound specified in equation (2) implies that, in many real-world situations, $\log\left(\Omega/Q_x\right)$ will be a highly accurate estimate of $I_R(\vec{W}:x)$ (at least in percentage terms).

Considering the result presented in equation (2), we note that, in a previous work, Peck and Waxman identified the quantity $\log\left(\Omega/Q_x\right)$ as a candidate for a quantitative measure of adaptation[61]. With this in mind, in the current work we shall use the phrase *'quantitative adaptedness'* to refer to the quantity $\log\left(\Omega/Q_x\right)$. Peck and Waxman showed that quantitative adaptedness has characteristics that correspond to the informal biological concept of adaptedness. Thus, it appears that there is a very close connection between reproductive information and adaptedness. This is, of course, exactly what one might expect. The approximate equality of reproductive information and quantitative adaptedness suggests that our definition of reproductive information has a close relationship to other key concepts in evolutionary biology.

Peck and Waxman were not the first to recognise the importance of quantitative adaptedness for systems that evolve by natural selection. In fact, $\log\left(\Omega/Q_x\right)$ is a special case of *functional information,* a concept that was first introduced by Szostak as "a quantitative means of comparing the functional abilities of different biopolymers"[64,65]. Furthermore, a similar (or identical) quantity was proposed by Adami, Ofria, and Collier as an alternative way to measure "physical complexity"[31]. The main method for measuring physical complexity that was suggested by Adami, Ofria, and Collier is, essentially, to count the number of nucleotide sites that are subject to purifying selection. In this, they take the same position as Adami and Cerf, which is discussed above[30]. Adami, Ofria, and Collier relegated description of a method using a quantity like quantitative adaptedness to an appendix, perhaps because counting sites under selection may seem to be more practical than carrying our measurements to estimate the value of $Q_x$. However, epistasis appears to be widespread in nature[57–59], and this makes the counting-sites-under-selection method unreliable. Furthermore, as we shall explain below, it may be much easier to measure $Q_x$ than one might imagine, at least if one uses phenotypes, rather than genotypes, to characterise organisms.

# Reproductive information is mutual

Above, we have defined $I_R(\vec{W}:x)$ to be the reproductive information that the fitness landscape provides about organismal type. Let us now consider whether knowing an individual's organismal type, along with the theory of natural selection, can tell us something about the fitness landscape. That is, we shall seek to calculate the reproductive information that the identity of an organismal type provides about the fitness landscape. We will use $I_R(x:\vec{W})$ to denote this kind of reproductive information.

To the best of our knowledge, no previous study of the information-theoretic aspects of natural selection has attempted to quantify the information that organismal types can provide about the fitness landscape. This may be because it is not immediately obvious how to classify fitness landscapes so that an information-theoretic analysis can be carried out. In Supplementary Note 9 we show how such a classification can be generated. The details of the classification procedure are somewhat involved, so here we will provide an outline of the procedure, and we will present a key result from the analysis that is detailed in Supplementary Note 8 and Supplementary Note 9.

The fitness-landscape-classification procedure is easiest to explain when each organismal type has a unique fitness value, and so, here, we will consider only this case. The general case, in which some organismal types may share a fitness value with other organismal types, is treated in Supplementary Note 8 and in Supplementary Note 9.

There are $\Omega!$ ways to order the $\Omega$ organismal types. We use each of these orderings to classify one set of possible fitness landscapes. For example, assume that $\Omega = 3$, and consider the ordering that puts organismal-type 2 in first place, and puts organismal-type 3 in second place, and puts organismal-type 1 in third place. This ordering would be the correct ordering to use to classify all fitness landscapes in which $\Omega = 3$, and organismal type 2 is the most fit, while organismal type 1 is the least fit.

We have seen that we can use the $\Omega!$ possible orderings of the organismal types to create $\Omega!$ possible classifications of fitness landscapes. That is, just as we have $\Omega$ possible organismal types, we also have $\Omega!$ possible *fitness-landscape types*. Let us use the first $\Omega!$ binary code words to encode these $\Omega!$ possible fitness-landscape types. (Here, "first" refers to the position in a length-lexicographical ordering of all possible binary code words.) There are $(\Omega!)!$ different ways to assign one of the first $\Omega!$ binary code words to each of the $\Omega!$ possible landscape types. Each of these $(\Omega!)!$ different ways constitutes a coding scheme that provides a binary code word for each of the $\Omega!$ different possible types of fitness landscape.

Consider a particular individual (the *focal individual*) that has been randomly selected from a population that lives in a particular habitat. Let us use the phrase *focal type* to refer to the organismal type of the focal individual. Let us use the phrase *focal landscape* to refer to the fitness landscape ($\vec{W}$) that characterises the environment in which the focal individual lives. If we have no theory about the nature of the focal landscape, then we should have no preference regarding which of the $(\Omega!)!$ coding schemes to use to encode the $\Omega!$ possible fitness-landscape types. Let $\bar{L}'_u$ represent the average length of the code word assigned to the focal landscape, if we average, with equal weight, over all $(\Omega!)!$ possible coding schemes. Thus, $\bar{L}'_u$ is the expected code-word length for the focal landscape, if we have no preferred theory regarding which fitness-landscape types are most likely to be found.

The theory of natural selection tells us that, all else being equal, we are more likely to find a relatively fit organismal type in a natural environment than we are to find a relatively unfit type[2–4,6]. Thus, if we know the focal type, and if we know *nothing else* about the environment in which focal individual was found, then, according to the theory of natural section, we should consider that fitness-landscape types in which the focal type is relatively fit are more likely to characterise the local environment than fitness-landscapes in which the focal type is relatively unfit. Thus, some coding schemes for the fitness landscapes are optimal, in that they encode fitness landscapes in which the focal type is particularly fit using relatively short binary sequences. All other coding schemes are not optimal. Let $\bar{L}'_s$ represent the average length of the binary code word that encodes the focal landscape, if we average, with equal weight, over all of the optimal coding schemes. (Precise definitions of optimal and non-optimal coding schemes for the current context are provided in Supplementary Note 9.)

We now have the ingredients needed to define the reproductive information about the fitness landscape that is provided by the observation of an organismal type ($I_R(x:\vec{W})$) as follows:

$$I_R(x:\vec{W}) = \bar{L}'_u - \bar{L}'_s. \qquad (3)$$

Thus, $I_R(x:\vec{W})$ is equal to the expected decrease in the length of the code word that we use to encode the fitness landscape that characterises a particular environment, if we take into account the theory of natural selection, along with the identity of an organismal type that has been observed in that environment. This decrease is in comparison to the expected code-word length that obtains if we have no preferred theory regarding which fitness landscapes are most likely.

A direct evaluation of $I_R(x:\vec{W})$ by calculating $\bar{L}'_u$ and $\bar{L}'_s$ might prove challenging, in part because of the immense number of fitness-landscape categories that can arise, even when the magnitude of $\Omega$ is relatively modest. With this in mind, it is worth asking whether there is a convenient way to estimate the value of $I_R(x:\vec{W})$. We show in Supplementary Note 8 that such an approximation can be produced. In particular, we prove the following relationship:

$$\left| I_R(x:\vec{W}) - \log_2\left(\frac{\Omega}{Q_x}\right) \right| < 4. \qquad (4)$$

Here, as before, $Q_x$ refers to the number of organismal types that are at least as fit as the focal type. Note that, in Supplementary Note 8, we show that equation (4) is correct even when some organismal types have exactly the same fitness values.

Equation (4) tells us that, like $I_R(\vec{W}:x)$, the value of $I_R(x:\vec{W})$ is approximately equal to $\log\left(\frac{\Omega}{Q_x}\right)$, which is a quantity that has been proposed as a measure of the adaptedness of the focal type[61]. Along with equation (2), equation (4) leads to the following relationship:

$$\left| I_R(x:\vec{W}) - I_R(\vec{W}:x) \right| < 8. \qquad (5)$$

Note that we expect that the reproductive information embodied in the organismal type of an individual living in a contemporary natural population will typically be thousands of bits (or more). Thus, typically, the maximum possible difference between $I_R(\vec{W}:x)$ and $I_R(x:\vec{W})$ will be, comparatively speaking, quite small (on the scale of one percent or less).

Equation (5) shows us that reproductive information is approximately mutual. In other words, for any fitness landscape and for all organismal types, $I_R(\vec{W}:x)$ will always be approximately equal to $I_R(x:\vec{W})$. Thus, for each of the $\Omega$ different organismal types, the amount of reproductive information that the fitness landscape provides about the organismal type is approximately equal to the amount of reproductive information that the organismal type provides about the fitness landscape. This mutuality accords with intuitive notions of information. For example, if reading the weather report provides information about the weather, then observing the weather should provide information about what we are likely to read in a weather report. In addition, mutuality is a characteristic of standard measures of information. In particular, Shannon has been proven to be mutual, and Kolmogorov information has been proven to be approximately mutual[7–10].

At first sight, the mutuality of reproductive information may seem somewhat surprising. After all, if $\Omega$ is substantial in magnitude, then the number of possible fitness-landscape types will be many orders of magnitude greater than $\Omega$, the number of possible organismal types. Thus, we can expect that $\bar{L}'_u$ and $\bar{L}'_s$ will typically be many orders of magnitude greater than $\bar{L}_u$ and $\bar{L}_s$. Nevertheless, equation (5) tell us that the difference between $\bar{L}'_u$ and $\bar{L}'_s$ will always be very close in value to the difference between $\bar{L}_u$ and $\bar{L}_s$.

# Reproductive information is closely related to both Shannon information and Kolmogorov information

Shannon information and Kolmogorov information are both ways to characterise the relationship between two variables in information-theoretic terms. However, unlike reproductive information, neither of these "classical" approaches is intrinsically associated with the theory of natural selection. Nevertheless, both Shannon information and Kolmogorov information have a close relationship to reproductive information. This is explored in some detail in Supplementary Note 4 and in Supplementary Note 10. Here, we will simply provide a brief and somewhat informal description of some of the relevant relationships.

It is clear that neither Shannon information nor Kolmogorov information can, in general, serve as a method for assessing the degree to which the theory of natural selection provides information about organismal types (or about fitness landscapes). A key reason for this is that Shannon information and Kolmogorov information measure the degree to which one variable is associated with another, but they do not necessarily depend on the *nature* of that association. Thus, if we use Shannon information or Kolmogorov information to calculate the amount of information that fitness landscapes provide about organismal types, then, if relatively fit types tend to occur in every environment, we may find that both Shannon information and Kolmogorov information tend to take on relatively large values. However, both Shannon information and Kolmogorov information can take on similar large values if, in every environment that we examine, we tend to find organismal types that are among the *least* fit. In contrast, reproductive information only takes on near-maximal values (on average) when we tend to find the fittest types in every environment. If, instead, we tend to find the least-fit types, then average reproductive information will tend to take on values that are close to zero.

Further details regarding the relationship between Shannon information and reproductive information are provided in Supplementary Note 10. These details include the specification of cases in which Shannon information and average reproductive information are guaranteed to be equal (to a close approximation).

Reproductive information is more similar to Kolmogorov information than it is to Shannon information. This is because both Kolmogorov information and reproductive information focus on the amount of information that one *particular* object provides about another *particular* object. In contrast, Shannon information focuses on a joint frequency distribution of two types of object (e.g., fitness landscapes and organismal types), and it tells us how much information the nature of one type of object can provide, *on average,* about the nature of the other type of object. Precise definitions of Kolmogorov information and Shannon information are provided in Supplementary Notes 4 and 10, respectively.

Kolmogorov information, unlike Shannon information, has been mostly used for theoretical endeavours, and not for real-world science-and-engineering purposes[8,9]. This is because there are substantial limitations on the practicality of Kolmogorov information. One problem is that there can never be a general method by which Kolmogorov information can be calculated. This is due mathematical results that, ultimately, are related to Gödel's famous incompleteness theorems[9,66]. Another problem is that, in calculating Kolmogorov information, one must make some arbitrary decisions about the "description language" that one uses to describe objects[8,9]. These decisions can have a large effect on the value of Kolmogorov information that is obtained. However, a key theorem shows that Kolmogorov information becomes (more or less) independent of the decisions about description language in the limit as the amount of Kolmogorov information becomes large[8–10,67].

It should be noted that, although reproductive information is similar to Kolmogorov information, it does not suffer from the problem of incalculability that hinders the practical applicability of Kolmogorov information, and it also does not involve arbitrary decisions about description languages. Thus, reproductive information is practical in a sense that Kolmogorov information is not. The reasons for these differences between Kolmogorov information and reproductive information are discussed in Supplementary Note 11.

As noted above, a high value for the Kolmogorov information between a fitness landscape and an organismal type does not imply that reproductive information will also take on a high value. However, what about the opposite? Does a large amount of reproductive information imply that the Kolmogorov information between the fitness landscape and an organismal type must also be large? We address this point in Supplementary Note 4. We find that precise results are not possible unless reproductive information is sufficiently large. This is normal for comparisons with Kolmogorov information, and it is a consequence of the arbitrary decisions about description language that occur when Kolmogorov information is calculated[8,9].

What happens when reproductive information is large? To understand what a circumstance like this implies for Kolmogorov information, it is useful to consider a situation in which there are very many organismal types for which reproductive information is very large (i.e., for which reproductive information is in the limit where it goes to infinity). In this case, for each of these organismal types, reproductive information is almost guaranteed to be approximately equal to the amount of Kolmogorov information that the fitness landscape provides about the organismal type. Here, "approximately equal" means that, in comparison to the numerical magnitude of both Kolmogorov information and reproductive information, the difference between these two quantities is infinitesimal, in percentage terms. Thus, the ratio of these two quantities is almost guaranteed to be extremely close to unity. (Note that, in this context, the phrase "almost guaranteed" means that the near-equality of reproductive information and Kolmogorov information must hold for all-but-an-infinitesimal fraction of these organismal types.)

Given the incalculability of Kolmogorov information, it may seem surprising that a reasonably precise result (such as the one just outlined) can be obtained. If fact, in the situation just described, for the organismal types for which reproductive information is very large, there is generally a possibility that Kolmogorov information is much larger than reproductive information (i.e., much larger than $I_R\left(\overrightarrow{W}:x\right)$). Furthermore, due to the incalculability of Kolmogorov information, it is typically impossible to test, in a definitive manner, whether this large difference in value exists for a given organismal type. Nevertheless, one *can* show that there must be, at most, only an infinitesimal proportion of the organismal types in question for which near equality of Kolmogorov information and reproductive information does not hold. An example that involves one of these very rare organismal types is provided in Supplementary Note 12.

# **Discussion**

The title of this work asks how one can measure the information that is created by natural selection. We have proposed a specific answer to this question. This answer is in the form of a quantity that we call reproductive information. Having studied the properties of reproductive information, it seems appropriate to now ask whether reproductive information is an adequate answer to the question posed by the title.

First, let us ask whether reproductive information actually measures information. Within modern information theory, there are two well-developed and generally accepted definitions of information[8]. These are Shannon information, and Kolmogorov information[7,10,62,63]. Both of these definitions quantify the information that one object (say object *A*) provides about another (say object *B*) as the reduction in the length of a complete description (or designation) of object *B* that is possible if one has access to object *A*. The reduction is achieved by using a code that is expected to be efficient, given the state of object *A*. (For Shannon information the reduction is measured on average, while for Kolmogorov information the reduction is measured for a particular pair of objects.)

Reproductive information measures information using the same basic procedure that is specified by Shannon information and Kolmogorov information. In particular, reproductive information measures information in terms of the reduction in the length of the description (or designation) of an organism that is possible if one knows the state of the fitness landscape (a list containing the fitness of each possible type of organism). The reduction is achieved by using the fitness landscape to encode the description of the organism with a code that is expected to be efficient in light of the theory of natural selection (and the state of the fitness landscape). Thus, given the similarity of the procedures used by Shannon information, Kolmogorov information, and reproductive information, it seems fair to say that reproductive information is a real measure of information, at least insofar as Shannon information and Kolmogorov information are real measures of information.

Next, we note that the two most-widely accepted definitions of information (Shannon information and Kolmogorov information) have been proven to be mutual[7–10]. This means that that information that object *A* provides about object *B* is approximately the same as the information that object *B* provides about object *A*. (Again, Shannon information measures this information using averages, while Kolmogorov information measures it for particular pairs of objects.) Mutuality is a natural feature of information. For example, very rough water in the middle of the ocean suggests that the wind in that area is blowing hard, and very strong winds suggest that any nearby large bodies of water will have a rough surface. It seems plausible that *any* reasonable measure of information will be mutual in this sense. With this in mind, it is worth noting that reproductive information is mutual, as we have demonstrated. This further supports the idea that reproductive information is a legitimate measure of information. This idea is also supported by the observation that, when reproductive information is sufficiently large, it is almost guaranteed to be approximately equal to the Kolmogorov information that the fitness landscape provides about organismal type, or vice versa (See Supplementary Note 4).

Given that reproductive information is a legitimate measure of information, it is worth asking whether reproductive information specifically measures the information created by natural selection. We note that, when average reproductive information exists at an appreciable level, there is the tendency for fitter organismal types to be more common. It is, indeed, this tendency that is measured by the average value of reproductive information. Clearly, the existence of such a tendency does not absolutely prove that natural selection has been at play. For example, in theory, the unbiased process of random genetic drift can create *any* sort of correlation between environments and organismal types. However, the large amounts of reproductive information that are apparent in even the simplest contemporary organisms cannot be explained by any plausible sequence of chance events, unguided by natural selection. Thus, it seems legitimate to use reproductive information to measure the information created by natural selection. This is because there is no other known natural phenomenon that can plausibly account for highly adapted organisms, and thus for high values of reproductive information[2,4,6,68].

Collectively, it seems reasonable to claim that these observations constitute a prima-facie case that reproductive information is an adequate measure of the information that is created by natural selection.

We will conclude by considering three matters that relate to the practical application of reproductive information, and to its possible implications for the future of evolutionary research.

# Measuring the reproductive information associated with a small part of the genome

In this work we have considered the calculation of the reproductive information associated with entire genomes. However, genomes consist of very long genetic sequences, and thus the number of possible alternative genomes is generally huge. Our method for measuring reproductive information (when applied to genotypes) requires some knowledge regarding the fitness of all of these alternative genomes, and this may often be impractical. One possible way to address this problem would be to focus on short sequences within the genome. This can greatly decrease the number of alternative sequences that must be considered, and thus it can make the calculation of the reproductive information associated with those sequences into a much more practical endeavour.

Consider a genetic sequence that occurs at a particular location within a particular genome. We will call this sequence the *focal sequence.* The location within the genome of the focal sequence will be called the *focal locus.*

In principle, we can produce a fitness landscape that specifies the fitness associated with each possible focal sequence that could occur at the focal locus. As usual, this fitness landscape can be expected to depend on local environmental conditions. However, due to epistasis, the fitness landscape may also depend upon the genetic background. (Here, "genetic background" refers to all of the genome, except for the focal locus.) Strictly speaking, the reproductive information associated with the focal sequence can only be calculated with respect to a particular set of environmental conditions *and* a particular genetic background.

While calculating the reproductive information associated with various possible genetic sequences at the focal locus may be a practical endeavour, it is important to keep in mind that these calculations are only valid for a particular genetic background. Furthermore, in the absence of a detailed knowledge of epistatic effects within the genome, such calculations cannot be "scaled up" so as to estimate the reproductive information associated with the entire genome. Nevertheless, such calculations may be of considerable interest, particular when calculated with respect to genetic backgrounds that are very common in a particular population.

# Measuring reproductive information using phenotypic characteristics

Previous attempts to measure the information that is created by natural selection have generally focussed on genotypes[30–34,37–39,43–45,69,70]. This is understandable, given the similarity between genetic sequences and the abstract symbolic strings that are the main focus of elementary information theory. However, if one wishes to measure reproductive information, then there are some difficulties that are associated with the nature of genotypes. One of these, as mentioned above, is the immense number of possible genotypes, even if one is working with a highly constrained situation. For example, even if we only consider the set of genotypes possible by altering just 10 nucleotide sites in a haploid organism, calculating reproductive information means that we must evaluate the level of fitness associated with more than one million genotypes.

One approach that might be utilised to deal with the immense number of possible genotypes is to attempt to evaluate the fitness associated with each genotype by means of calculations, rather than experiments. After all, the genetic constitution of an organism is a physical system that causes its effects via physical interactions. However, despite recent advances, the question of how genes lead to the functional shapes of particular proteins is not fully understood[71,72]. Even if this issue of "protein-folding" were to be completely solved, there would still be huge challenges such as understanding how various combinations of proteins lead to the phenotypes upon which natural selection acts[73–76]. This problem is made more difficult by the fact that, in general, we can expect that any given genotype will be associated with a *distribution* of different phenotypes. To calculate the fitness associated with a particular genotype we would, in principle, have to predict the probability of each possible phenotype for that genotype, and then evaluate the fitness of each of these phenotypes (or, at least, evaluate fitness for all of those phenotypes that have a non-negligible probability of arising).

It seems possible that these problems might be ameliorated if one characterises organismal types using phenotypes, rather than genotypes. This immediately eliminates the problem of trying to assign a particular distribution of phenotypes to every viable genotype. Instead, one would work directly with the phenotypes.

While characterising organismal types using phenotypes solves some problems, it also creates new challenging issues. For example, many phenotypic characters are most naturally measured using *continuous* scales. These include mass, length, growth rate, etc. These continuously distributed phenotypic characters can take any one of an *infinite* number of possible values. At first, this seems like a situation that is even worse than the enormous (but finite) number of possible genotypes that can occur for genomes of a certain size. However, we generally expect continuous characters to exhibit continuity with respect to their causal relationships. In particular, though there may be exceptions, we generally expect that a very small change in a particular continuously distributed phenotypic character will have, at most, a very small effect on fitness. This continuity of causal effects allows us to interpolate. Thus, if we are satisfied with a close estimate, then we need not consider an infinite number of phenotypes when we are dealing with a continuously distributed phenotypic character. Rather, we can make measurements for a finite (and, hopefully, manageable) number of phenotypic values, and rely on continuity to allow us to approximate the effects for intermediate values for which no measurements were made.

Working with phenotypes also may have enormous advantages if we attempt to estimate the fitness effects of various organismal types using physics-based calculations. When we work directly with phenotypes, there is no need to consider the protein-folding problem or the influence of various proteins on phenotypes. Rather, one may be able to use the results of physics-based calculations to estimate the fitness effects of various phenotypes. For example, one might estimate the effect of plant height on the dispersal of seeds or pollen. Similarly, one might estimate the effects of the length of an eyeball on visual acuity, and thus, on prey-capture efficiency. Furthermore, if calculation the reproductive information associated with an entire organismal phenotype proves to be too challenging, then it should also be possible to calculate the reproductive information associated with various alternative forms of some particular part of an organism. However, just as with the effects of epistasis among genetic loci, it is important to recognise that the reproductive information associated with a particular form of a particular part of an organism can, in general, be expected to depend on the form of the rest of the organisms. This is because of the fitness-altering effects of interactions between different body parts.

Despite the advantages just mentioned, it may appear that there are substantial technical problems standing in the way of estimating reproductive information using continuously distributed phenotypic variables. In particular, the details of the way in which we measure a phenotypic trait can affect our estimate of reproductive information. For example, we might get a different answer if we measure height, as compared to a situation in which we choose to measure the square root of height. This is so even though we can easily calculate height from the square root of height, and vice versa. However, recent results in a study carried out by Peck and Waxman[77] appear to provide a route to a satisfactory solution to this conundrum. The nature of the problem and the proposed solution are explained in Supplementary Note 13.

# Reproductive information
# and the emergence of "superorganisms"

It now seems certain that much of the adaptive complexity that is apparent in contemporary organisms is the result of evolutionary events in which groups of independent organisms become united, so that, to some extent, they can collectively be viewed as a "higher-level" organism[11,19,78–80]. One well-known example of this sort of process is the phenomenon of *endosymbiosis,* in which one organism comes to live within the body of another, thus producing benefits for one or both of the organisms. This process led to the evolution of the eukaryotes, which include all contemporary animals, plants, fungi, and protists[2,12,13]. Another well-known process that can lead to "higher-level" organisms is the evolution of *eusocial* animal colonies, in which many individuals do not directly reproduce, but instead are specialised in performing various tasks that increase the fitness of a relatively small number of reproductive individuals. The non-reproductive "workers" in eusocial colonies have often been compared to the somatic cells within multicellular organisms. They are specialists that are crucial for the functioning of the colony, but they do not, themselves, reproduce[2,14–16].

From the perspective of the current work, an intriguing aspect of any group of organisms is that, in comparison to a single organism, they typically have considerably more potential to manifest reproductive information. Thus, for example, while a single haploid individual with a genome consisting of $L$ nucleotides might have any of $4^L$ different genomic sequences, for a group of $n$ distinct individuals, each with a genome of $L$ nucleotides, there are $4^{nL}$ different possible genetic constitutions. Thus, the maximum-possible amount of reproductive information is at least $n$ times larger for such a group than what is the case for a single individual. In addition, groups or organisms may have important "emergent" characteristics that describe the *relationship* between individuals, and that cannot be reduced to a collection of individual measurements. A simple example of such a characteristic is the average distance between randomly selected pairs of group members.

The foregoing considerations notwithstanding, a group of organisms does not automatically constitute a *superorganism* (that is, a "higher-level" organism that contains multiple individuals). For example, the organisms that form a group may have no biologically meaningful interaction other than resource competition. Another possibility is that they have actively antagonistic interactions, with some members of the group preying on others. This leads to the question of whether there might be a general and quantitative way to decide whether specific groups of organisms can reasonably be considered to be superorganisms. It would also be good to have a measure that would allow us to express, in quantitative terms, the extent to which a group of organisms has progressed towards a full integration such that the characteristics of each individual group member are determined by the interests of the group as a whole.

We propose that the concept of reproductive information could serve as a quantitative measure that can detect the emergence and development of superorganisms. The basic idea is quite simple. First, one would specify a set of characteristics that can be used to characterise groups of organisms. These characteristics could be based on the genotypic constitution of the group. However, it might be much more convenient to categorise groups in terms of the phenotypes of the group members. In addition, consideration of the "extended phenotype" of the group is also likely to worthwhile. The extended phenotype includes various effects on the natural world that are generated by the group. Beaver dams and the nests of colonial birds are examples of aspects of the extended phenotypes of groups[81–83]. Note that, in research of this sort, it may be very useful to apply the recent information-theoretic results mentioned above[77], which have to do with continuously distributed phenotypic traits (See Supplementary Note 13).

It may be worth noting that reproductive information could be used to study the evolution of groups that include members of multiple species. By using reproductive information in this context, it should be possible to address the question of whether there are entire ecosystems that can legitimately be considered to be superorganisms that have evolved to be highly integrated and cooperative[17,18,20,84,85].

In order to use the concept of reproductive information to study the emergence of superorganisms, it will be necessary to assign a single fitness value to each of a number of possible configurations of a group. Fitness is a measure of reproductive success. To define the fitness of a group of organisms, we must have a procedure by which, for each group in the present, we can identify one or more groups in the past that are the "parents" of the present group. Of course, there are very many different ways to do this. However, in general, these different ways of assigning parental groups to present groups will not be equivalent. In light of Darwin's theory of natural selection, it seems clear that, all else being equal, we should choose a method for assigning parenthood to groups that maximises "group-level heritability"[3,22,85]. That is, we should choose a method that maximises the resemblance between parental groups and the groups that we designate as being their offspring. The best way to do this will depend on the details of the organisms involved, and, for any particular situation, it seems likely that considerable research may be necessary before an optimal (or nearly optimal) method for assigning parenthood can be developed.

An alternative method for assigning parental groups to offspring groups would be to consider a variety of possible methods of assignment, and then choose the method that maximises average group-level reproductive information. It seems likely that this method would lead to conclusions that are similar to those that would be derived from the group-level-heritability method just mentioned. However, in any given circumstance, one of these two methods is likely to prove to be more practical than the other.

The creation of practical and useful methods for revealing reproductive information at the group level is unlikely to be an easy project. However, such an endeavour may ultimately prove to have been well worth the effort. This is because the emergence of group-level adaptations seems to have played a central role in some key moments during the development of life on earth. These moments include the origin of the eukaryotes and the emergence of eusociality, as mentioned above. They also include the emergence of contemporary genetic systems and the evolution of complex multicellular animals, plants, and fungi[88–92]. In addition, group-level adaptations are important for understanding topics that have occupied a considerable amount of the attention of evolutionary biologists during recent decades. These topics include the evolution of phenomena such as cooperation, symbiosis, mutualism, and "altruism"[11,14–16,22,80,86].

In this study we have described a measure, reproductive information, which, we believe, can provide a satisfactory way to quantify the information that is created by natural selection. If reproductive information can effectively serve its purpose, then we will have a new and useful tool for understanding adaptation. The development of such an understanding seems like a worthy endeavour because adaptation can be considered to be the key distinguishing feature of life[61,87–91]. In this final section we have noted that adaptation, which is usually conceived of in the context of individual organisms, is a phenomenon that can also be manifest in the characteristics of *groups* of organisms. This includes groups in which members come from very different species that may differ from each other in size and lifespan by orders of magnitude. It seems plausible that the concept of reproductive information could help to illuminate processes by which such groups can evolve adaptations that favour the reproduction of the group as a whole. As such, the assessment of reproductive information may turn out to be crucial for understanding some of the most intriguing and conceptually challenging phenomena that are currently under study within the field of evolutionary biology.

## Author contributions
Both authors contributed equally to all aspects of this paper, namely conceptualisation, formal analysis, investigation, methodology and writing of the original draft and supplementary information.

## Supplementary Information
This work is accompanied by Supplementary Information. This consists of fully referenced notes that provide background material on the main text. Some notes contain additional comments, while others contain mathematical proofs of results mentioned in the main text.

# SUPPLEMENTARY INFORMATION

## HOW CAN WE MEASURE THE INFORMATION CREATED BY NATURAL SELECTION?

Joel R. Peck[1] and David Waxman[2]

[1]Department of Genetics,
University of Cambridge, Cambridge, UK

[2]Centre for Computational Systems Biology, ISTBI,
Fudan University, Shanghai, PRC

This document is divided into notes that provide background material on the main text. Some notes contain additional comments, while others contain mathematical proofs of results mentioned in the main text.

# Contents

# Supplementary Note 1: Previous proposals for measuring the information that is created by natural selection

In this Note we shall consider some previously published studies that suggest methods for measuring the information that is created by natural selection. It is best to read this note after reading the main text, at least up until the start of the Discussion. This is because this Note makes reference to results and concepts that are presented in the main text.

The idea that the forms of organisms provide information about the nature of their environment is intimately intertwined with the recognition of biological adaptation[1–4]. As such, this idea has arisen many times in the work of various thinkers and natural historians. The idea certainly predates Darwin[5–8]. However, attempts to *quantify* the information that is manifest in the relationship between organisms and their environment could not begin until the 20th century, when the foundations of the formal theory of information were laid[9–14]. An early example of this work is a paper by Kimura, which appeared in 1961, only 13 years after the publication of Shannon's seminal 1948 paper, in which he introduced the quantity that, in the main text, we call Shannon information[9,15].

After publication of Kimura's 1961 paper, various other researchers proposed methods to quantify the information that is created by natural selection[1,16–27]. Among these proposals, one stands out from among the others. This is the work of Adami, and his colleagues[28–33]. This body of work appears to be more developed, more nuanced, and more grounded in conventional information-theoretic ideas than the other well-known previously published proposals. It also appears to be more frequently cited than the work surrounding other proposals. The published research of Adami and his colleagues was particularly helpful as we decided on how to define reproductive information.

Here, we will describe and critique the work that Adami and his colleagues carried out on the information-theoretic effects of natural selection. We will then provide a brief discussion of several studies on this same topic that have been carried out by other authors (including Kimura). Further analysis related to these previously published proposals is provided in Supplementary Note 14.

The studies that we describe here were chosen because they are relatively well known, and because their authors expressed an interest in measuring the information that is created by natural selection (or something very similar). However, this *does not* mean that the intentions of these authors were identical to ours. Reproductive information is intended to measure the information created by natural selection in a way that conforms with the two main quantitative concepts of information that are in use today (i.e., Shannon information and Kolmogorov information[34–36]). Here, we will ask whether the work of various researchers is adequate with regard to our *objectives*. We will conclude that, in this regard, the work of these authors has substantial shortcomings. However, this *does not* mean that the work of these authors is lacking with regard to *their own* objectives. We certainly do not wish to disparage the work of these researchers. However, their words suggest that they intended to explore questions that are, at the very least, in the same arena as those addressed by reproductive information. We will, therefore, ask whether their formulations are helpful with regard to measuring reproductive information. However, we do this without pretending that the answer to this question addresses the adequacy the work of these authors with respect to *their own* objectives.

## "Physical complexity" and the research of Adami and colleagues

In a study published in 2000, Adami and Cerf described a concept that they called physical complexity[28]. The authors say that "physical complexity measures the amount of information about the environment that is coded in the [genetic] sequence." From context, it seems clear that, in Adami and Cerf's view, the information in the genetic sequence is largely generated by the process of natural selection. Indeed, elsewhere, Adami tells us that "natural selection increases physical complexity"[30]. Given these observations, it is apparent that physical complexity is intended to be a quantity that is very similar to reproductive information. However, there are both technical and conceptual differences between physical complexity and reproductive information, as we shall see presently.

Adami and Cerf develop their concept of physical complexity in three stages. First, they identify physical complexity with a modified version of the Kolmogorov information that the environment provides about an individual's genotypic sequence. In the process of describing their modified version of Kolmogorov information, they hypothesise the existence of a "universal" computer that does not have access to the "rules of mathematics"[28,35]. Unfortunately, this concept is not explained in detail by Adami and Cerf. In information theory, a universal computer is, in effect, a set of instructions about how to process inputs and produce outputs[35]. It seems hard to understand how such a set of instructions could be implemented without reference to mathematical concepts such as number and the notion of ordering (i.e., direction). Without such concepts, how, for example, could the computer implement an instruction to advance a data-storage device forward by, say, five "cells", and then read the instruction found by doing so?

Another difficulty is that Adami and Cerf provide no system for classifying environments. As such, it is impossible to tell whether their physical complexity is "mutual", as has been shown to be the case for Kolmogorov information, Shannon information, and reproductive information[34,35].

Despite any possible problems with their method for calculating physical complexity, Adami and Cerf's initial approach (i.e., using a modified version of Kolmogorov information) is intriguing. This is because it is very different from the much-more popular method of simply counting genomic sites that are under purifying selection (which is described in the main text)[18–24,28–33,37]. As demonstrated in the main text, the counting-sites method can be problematic. However, instead of just counting sites under purifying selection, Adami and Cerf's initial approach involves *explicitly* focusing on the *information* that the environment can provide about a particular genotype.

Note that, as it is initially defined, the concept of physical complexity differs from the concept of reproductive information because physical complexity involves the information embodied in an individual's characteristics that is about the environment *in general*[28,29]. Thus, physical complexity may reflect aspects of the environment that do not affect fitness. For example, the information measured by physical complexity may reflect the patterns of mutation or gene conversion that prevail in a particular environment. On the other hand, the reproductive information embodied in an organism's characteristics quantifies *only* information that is related to the details of how natural selection is operating in the organism's environment.

After defining physical complexity, Adami and Cerf say that using their modified version of Kolmogorov information to calculate physical complexity is "not practical." While their reasons for this assertion are not entirely clear, it seems likely that they have to do with the fact that there is no general procedure that can be guaranteed to calculate a quantity known as *algorithmic complexity*[34,35]. In brief, the algorithmic complexity of a sequence of symbols is defined with respect to some chosen hypothetical computer, and it is equal to the length of the shortest program that, when provided as input to that computer, leads to the output of the sequence of symbols[12,34,35]. The calculation of algorithmic complexity is an essential step in the calculation of Kolmogorov information[12,35].

Despite the impracticality of Kolmogorov information, Adami and Cerf say that they *can* calculate the *average* physical complexity for a *collection* of genomes, at least under certain circumstances. This leads to the second stage of their analysis.

Adami and Cerf's method for calculating average physical complexity is based on their assertion that the average algorithmic complexity for a collection of symbolic strings is approximately equal to the entropy of that collection of symbolic strings (see their equation (6)). They say that this assertion holds so long as the strings are sufficiently long, and we have "near-optimal coding"[28].

Entropy measures the extent to which a probability distribution is dispersed over a set of possible outcomes. For example, if the symbolic strings in question represent genomes, and if there are $\Omega$ different possible genomic sequences, then the genotypic entropy ($H$) in some particular population is given by

$$H = \sum_{g=1}^{\Omega} p(g) \log_2 \left( \frac{1}{p(g)} \right) \tag{S1.1}$$

where $p(g)$ represents the proportion of population members that have genotype $g$.

Unfortunately, Adami and Cerf's claim about the approximate equality of average algorithmic complexity and entropy is not generally true. Regardless of string length and "near-optimal coding", approximate equality is guaranteed *only* for certain types of probability distributions[35,36]. Evolution by natural selection tends to generate highly peaked genotypic distributions, such that, even if population size is extremely large, only a very small fraction of possible genomes have a non-negligible frequency[38,39]. It has been shown that these sorts of highly peaked distributions are *exactly* the type for which there is *absolutely* no guarantee of an approximate equality between entropy and the average algorithmic complexity[35,36]. This lack of equality for peaked distributions is, in fact, obvious. Consider a distribution in which all but one possible genomic sequence has a frequency of zero (so that only one genomic sequence is present in the population). Clearly, the entropy associated with this distribution is equal to zero (see equation (S1.1)). However, the one existing type of genome can have any sequence, and it can be arbitrarily long. This implies that the algorithmic complexity of the sequence can be equal to any positive integer, no matter how large[34,35]. This unequivocally demonstrates that, contrary to the assertion of Adami and Cerf[28], there is no mathematical law that forces the average algorithmic complexity of a population of sequences to be equal to its entropy[35,36].

Based (in part) on the foregoing faulty foundations, Adami and Cerf's presentation implies that, in a sufficiently large population, the average physical complexity of the population's genomes ($\bar{C}$) can be estimated by the expression

$$\bar{C} \approx L - H \tag{S1.2}$$

where $L$ is the number of loci (or nucleotide sites) in the organism's genome, and $H$ is the entropy of the frequency distribution of the population's genomes. However, the formulation of this expression appears to involve an implicit assumption that only two variants (i.e., alleles or nucleotides) are possible at each locus. Alternatively, it may be that formulation of equation (S1.2) assumes that logarithms are taken to base $U$, where $U$ is the number of alleles (or nucleotides) per locus. If, instead, we use logarithms to base two, then the expression becomes

$$\bar{C} \approx L \log_2(U) - H. \tag{S1.3}$$

Unfortunately, as equations (S1.2) and (S1.3) only provide estimates for the *average* physical complexity, they do not allow us to calculate the physical complexity associated with a particular genomic sequence. In addition, we can expect that genotypic entropy will typically not change by much immediately after an environmental change, and thus the estimate of average physical complexity can be completely inaccurate in a population in which genotype frequencies are not at equilibrium. (Adami and Cerf's text indicates that that were aware of this difficulty[28].)

However, perhaps the most troubling problem with using equation (S1.2) or equation (S1.3) to estimate average physical complexity is that, as explained above, if epistasis is common, then a fitness landscape can contain very many local optima. Populations can become stuck in the vicinity of these local optima for very long periods of time. In fact, a sexual population can become *permanently* stuck (and thus fully equilibrated) in the immediate vicinity of a local optimum even if mutations are occurring, and the population is effectively infinite in size, so that all possible genotypes are always present in the population (see Supplementary Note 15). As noted above, when epistasis is common, the genome of an organism from a population that is stuck in the vicinity of a local optimum may convey very little information about the selective environment. However, according to equations (S1.2) and (S1.3), a population that is nearly fixed on a local optimum (which will have near-zero entropy ($H$) will always have nearly maximal average physical complexity, regardless of whether it is nearly fixed on the fittest possible genotype, or it is nearly fixed on the least-fit of a huge number of local optima. This is so even though, according to Adami, Ofria, and Collier, the genomic entropy ($H$) that appears in equations (S1.2) and (S1.3) "takes into account all epistatic correlations between sites"[29]. Thus, according to Adami, Ofria, and Collier, in a sufficiently large population, equations (S1.2) and (S1.3) should provide a good estimate of average physical complexity, regardless of whether the fitness landscape is highly epistatic, or not. These observations provide additional reasons to doubt the effectiveness of equations (S1.2) and (S1.3) as useful tools for estimating average physical complexity.

Adami and Cerf motivate the third stage of their analysis by saying that equation (S1.2) (or the equivalent) provides a reasonable estimate of average physical complexity only if population sizes are at least on the order of the number of possible genotypes. This requirement on population size is typically impossible to fulfil in realistic circumstances, and so Adami and Cerf provide an alternative that they recommend using in real populations. This alternative involves using equation (S1.2) or (S1.3), but substituting in the following estimate for $H$:

$$H \approx \sum_{i=1}^{L} H_i \qquad (S1.4)$$

where $H_i$ is the entropy of the distribution of alleles (or nucleotides) at the $i^{th}$ locus. According to Adami and Cerf, this is only an estimate because, unlike equation (S1.2), it does not take into account the effects of "correlations within a sequence"[28].

Unfortunately, the approximation provided by equation (S1.4) does not ameliorate the problems with Adami and Cerf's approach to calculating the information created by natural selection. In particular, regardless of whether one uses this approximation or not, Adami and Cerf's formulation always assigns a high degree of physical complexity to a population with a large genome, so long as that genome is fixed (or nearly fixed) on a local optimum. As we have seen, this is not the sort of result that would be expected from a well-behaved measure of the information created by natural selection. In addition, Adami and Cerf's suggested measure (i.e., equation (S1.4)) does not produce a quantity that allows us to determine the physical complexity of individual genomes. Thus, even if it was not otherwise problematic, its utility would be quite limited.

As mentioned in the main text, in an Appendix of their paper, Adami, Ofria, and Collier consider measuring physical complexity by using a quantity that is similar (or identical) to quantitative adaptedness[1,29]. (Recall that the quantitative adaptedness of a particular organismal type, $x$, is given by $\log_2(\Omega/Q_x)$, where $\Omega$ is the total number of possible organismal types, and $Q_x$ is the number of organismal types that are at least as fit as organismal type $x$.) In contrast to the other previously published approaches discussed above (and in the main text), the use of quantitative adaptedness escapes all of the shortcomings that we have just discussed. In particular, quantitative adaptedness changes immediately and appropriately when the environment changes. It also applies to individual organismal types, and not just to populations. Furthermore, in a highly epistatic landscape, quantitative adaptedness tends to associate a relatively low number of bits to typical local optima, and it only awards a high number of bits to organismal types that are truly exceptional in terms of being associated with a superior level of fitness.

Given the advantages of quantitative adaptedness, it may seem surprising that Adami, Ofria, and Collier did not feature quantitative adaptedness as a way of measuring physical complexity. One possible reason for their decision has to do with practicality. The main method recommended by Adami, Ofria, and Collier involves using equation (S1.2) or equation (S1.3), along with equation (S1.4), and this requires nothing more than knowing the size of the genome, along with knowing the distribution of entropy values among the organism's various loci. On the other hand, using quantitative adaptedness means estimating what fraction of all possible organismal types are at least as fit as some focal organismal type. Given the immense number of possible genotypes for a typical DNA-based organism, it might be very challenging to produce such an estimate with a reasonable level of precision.

The foregoing notwithstanding, the situation regarding the practicality of quantitative adaptedness appears to be very different if, instead of focussing on genotypes, we consider phenotypes. This topic is discussed in the main text.

Next, we will delve into the literature a bit more in order to provide a more comprehensive overview of published research that is relevant to the measurement of the information that is created by natural selection.

## Additional approaches to measuring the information created by natural selection

We will now briefly discuss three other well-known attempts to develop methods to measure the information created by natural selection. These three attempts were carried out (in chronological order) by Kimura[15], Frank[26], and by Barton and his colleagues[18,19]. Kimura said that he wished to measure "*genetic information*." While Kimura did not explicitly define genetic information, he did say that: "New genetic information was accumulated in the process of adaptive evolution, determined by natural selection acting on random mutations"[15]. Thus, it is clear that the information that Kimura intended to measure is information that is created by natural selection. Similarly, Frank says that his measure calculates "the information accumulated by natural selection"[26]. Finally, Barton and his colleagues tell us that they are addressing the question of "how much information can selection accumulate and maintain in the genome?"[19].

All three of the aforementioned studies share certain characteristics that are different from the characteristics of reproductive information. With this in mind, we will henceforth refer to their authors, collectively, as *KFB*.

One example of the differences between reproductive information and the work of KFB is that, while average reproductive information depends on the relationship between the distribution of organismal types and the distribution of fitness landscapes, the studies of KFB all calculate their measures of "information" by using mathematical formulae to compare two distributions of organismal types. For Kimura and Frank, one of these distributions is observed before an episode of selection, and the other is observed after an episode of selection. For Barton and his colleagues, the two distributions are averages of organismal-type distributions that occur when natural selection is present, as compared to when there is no natural selection[15,19,26].

Unlike reproductive information, the studies of KFB do not quantify information by using the traditional method of calculating the difference in the length of codes when information is present, as opposed to when it is absent[15,19,26,34,35]. Instead, each of these studies chooses a different quantitative method to measure information. However, traditional information theory serves to measure the amount of information that one phenomenon (e.g., environmental characteristics) provides about another (e.g., organismal types). As KFB's methods only consider data arising from one phenomenon (i.e., organismal-type distributions), KFB are unable to use traditional methods, and so they invent new methods. Unfortunately, it seems that each of the quantities used by KFB to calculate information can produce results that are difficult to reconcile with the classical quantitative theory of information (see Supplementary Note 14). In particular, basic information-theoretic considerations suggest that, if $\Omega$ different genomic sequences are possible, then it should be possible to accumulate no more than $\log_2(\Omega)$ bits of information[34,35]. Regardless, it can be shown that the methods used by KFB can allow for the apparent accumulation of *any finite amount* of information, regardless of the value of $\Omega$.

Another distinction between reproductive information and the studies of KFB is that reproductive information refers to the information in a *particular* organismal type that is about a *particular* fitness landscape (or vice versa). It is not clear whether this can be done for the three aforementioned studies, as the calculations involve entire frequency distributions, and not the relationship of a single organismal type to the fitness landscape[15,19,26]. In addition, the work of KFB resembles some of the work of Adami and colleagues in that it does not deal with the phenomenon of epistasis in a way that seems to be in line with the basic principles of information theory (see the discussion of epistasis in the main text). In another similarity to the work of Adami and colleagues, KFB do not provide any means to test whether their measures of information are "mutual". That is, they do not allow us to assess whether the information that the form of an organism provides about the environment is approximately the same as the amount of information that the environment provides about the form of the organism. As mentioned above, this sort of mutuality is a characteristic of Shannon information, Kolmogorov information, and reproductive information[34,35]. Finally, like some of the work of Adami and his colleagues, it appears that the studies of Kimura[15] and those of Barton and his colleagues[18,19] do not provide measures that respond, immediately and appropriately, to environmental changes that affect the fitness landscape. This is because genotypic distributions generally take some considerable time to respond fully to environmental changes[39]. On the other hand, Frank[26] discusses environmental change in terms of a "change in coordinates." However, Frank does not discuss and develop this concept in detail.

*Conclusion*

The main suggestion of Adami and his colleagues with respect to measuring the information that is created by natural selection has to do with measuring organismal-type distributions. It does not involve considering the relationship between organismal types and environmental conditions. The same is true for the work of KFB. With this in mind, it is not surprising that these approaches apparently have very limited utility, at least when it comes to measuring reproductive information (or something similar). Reproductive information is the information embodied in the genotypes and phenotypes of organisms that is about what is required to survive and reproduce in a particular environment. It is hard to see how this can be adequately assessed if one looks only at data that is about the organisms, and thus uses a measure that does not explicitly consider environmental conditions.

Attempting to measure reproductive information (or something similar) without measuring environmental conditions is like watching weather reports in an air-conditioned and soundproof room that has no windows. The reports might contain a lot of information about the weather outside, but they could also be completely made up, and thus contain *no information at all* about the actual weather. The only way to know for sure is to make independent measurements of the weather, and then compare these measurements to the weather reports in question. In the same way, it is plausible that any truly reliable measure of reproductive information would have to consider organismal types, environmental conditions, and the relationship between these two phenomena.

# Supplementary Note 2: An information-theoretic analysis of the NK model

Is it reasonable to use the number of nucleotide sites that are subject to purifying selection as a measure of reproductive information? In this Note we will demonstrate that the answer to this question depends strongly on the structure of the *fitness landscape*. (In this context, the fitness landscape may be envisioned as a list that specifies the fitness of each genotype in some set of possible genotypes. Fitness, in turn, measures expected reproductive success.)

In nature, it seems likely that many nucleotide sites are neutral, in that variation at these sites has no effect on fitness[38,40,41]. Nevertheless, in the interests of clarity and simplicity, in the present section we will consider organisms in which *all* nucleotide sites are subject to selection. The resulting analysis can be easily generalised to situations in which some nucleotide sites are neutral.

To clarify matters, let us consider a simple genetic model, which is a particular case of the *NK model*, as developed by Kauffman and colleagues[42,43]. We will draw upon results derived from this model to explain problems associated with the calculation of reproductive information by counting nucleotide sites under purifying selection.

We confine ourselves to consideration of fitness landscapes for which every possible genotype has a different fitness. This is a typical assumption in studies of the NK model[42,43]. In particular, we will consider a haploid species for which there are four possible nucleotides at each of $L$ different nucleotide sites.

Because all possible genotypes differ from each other in fitness, there must be one genotype that has a higher fitness than all other genotypes. We will call this the *globally optimal genotype*. Clearly, any change to any one of the nucleotides that constitutes the globally optimal genotype will produce a decrease in fitness. Thus, for the globally optimal genotype, all $L$ nucleotide sites are subject to purifying selection. Generally, any genotype for which the change of any *single* nucleotide causes a decrease in fitness is called a *locally optimal genotype*. The globally optimal genotype is thus also a locally optimal genotype. A difference between the globally optimal genotype and other locally optimal genotypes is that, for these latter local optima, it is possible to improve fitness by altering *multiple* nucleotides. By doing this one could, for example, create the globally optimal genotype.

Analyses of the NK model typically assume (either explicitly or implicitly) that the population starts by being fixed on one particular genotype (which is chosen at random)[42–46]. In this context, 'fixed' means that every population member has the same genotype. Individual mutations then sequentially occur at random, and any mutation that enhances fitness becomes fixed in the entire population. Mutations that decrease fitness are lost from the population. Evolution stops when a locally optimal genotype becomes fixed in the population because, at that point, any single additional mutation will cause a decrease fitness. Thus, when evolution stops, all $L$ nucleotide sites are subject to purifying selection. Some analyses also consider the possibility of multiple mutations, but these rare events typically do not alter the overall nature of the results, unless one considers biologically unrealistic cases where, for example, substantial fractions of the genome often mutate at once[42,43]. For now, we will follow common tradition and ignore multiple mutations while also assuming that the population is initially fixed on a randomly selected genotype. We will use the phrase *ultimate genotype* to refer the locally optimal genotype that eventually becomes fixed in the population. After the ultimate genotype is fixed, no further evolution occurs.

Of course, in nature, we cannot expect an organism to survive if its genome is simply a random set of monomers that have been unaffected by natural selection, as is the case for the initial population mentioned above. However, this is a biological reality that is generally ignored in analyses of the NK model, in the interests of gaining insight into fundamental evolutionary processes. In any case, the assumption may hold true if, instead of considering an entire genome, we confine ourselves to a small part of a genome, such as a single gene.

The NK model was designed to allow researchers to explore the consequences of "landscape ruggedness", where a rugged landscape is just one for which there are many locally optimal genotypes[42,43,45]. Usually, these local optima are not input explicitly. Rather, they arise naturally from the way that genotypes at different nucleotide sites interact to affect fitness. This sort of genetic interaction is called *epistasis*. When the nucleotide sites do not interact at all (which we call the *no-epistasis case*) there is only one local optimum (which is the global optimum). In this case every possible mutation is either beneficial or deleterious, and these effects on fitness do not depend on the state of the rest of the genome. In contrast, when every nucleotide site interacts with every other nucleotide site (which we call the *maximal-epistasis case*) the number of local optima is relatively large, and it is known that the probability that a randomly selected genotype is a local optimum is $\frac{1}{3L+1}$, where $L$ is the number of nucleotide sites in the genome[42,43]. To construct a maximally epistatic fitness landscape, one can independently assign fitness values to all genotypes, using a continuous distribution of fitness values (for example, a uniform distribution).

Analysis of the NK model can be challenging, but it has been possible to obtain reasonably comprehensive analytical results for both the no-epistasis case, and for the maximal-epistasis case. We will focus on these two cases. Analysis and computer simulations suggest that results from cases with intermediate levels of genetic interaction tend to lie between these two extremes, as one might expect[42,45,46].

The central point that we wish to make in the remainder of this section is that, unless one has reasonably detailed knowledge of the fitness landscape (i.e., knowledge about the level of epistasis), counting nucleotide sites that are subject to purifying selection may not reveal much about the reproductive information embodied in a genotype. We will use an informal discussion to demonstrate the veracity of this point.

For the sake of concreteness, let us arbitrarily assume that the genome consists of 5,000,0000 base pairs, which is a typical genome size for a bacterium[47]. Initially, we assume that the population is fixed on a particular randomly chosen genome. Thus, we expect that, because there are four monomers, there will, by chance, be approximately 25% of all nucleotide sites that are initially fixed for the same monomer that the globally optimal genotype has at that site. In the no-epistasis case, we are then guaranteed that, eventually, the monomers at the other (approximately) 75% of all nucleotide sites will change as a result of mutation and natural selection, and the population will become fixed upon the globally optimal genotype. At that point, after natural selection has "rewritten" the constitution of approximately 3,750,000 nucleotide sites, evolution will stop. The same final outcome would result no matter where the population begins, even if, initially, not a single nucleotide site was fixed on the monomer present at that nucleotide site in the globally optimal genotype. Thus, in the no-epistasis case, the state of the population after evolution has stopped depends 100% upon the fitness landscape, and it does not depend on the initial starting genotype of the population. In other words, in the no-epistasis case, the ultimate genotype is always the globally optimal genotype.

Clearly, in the no-epistasis case, after evolution has stopped, examining the genome of any population member provides a very considerable amount of information about the fitness landscape. In particular, it gives us a definite answer to the question: What is the fittest possible genome, given the nature of the fitness landscape? Note that this question is about the *entire* fitness landscape, and it can be answered by observing the genome of a single individual, even if the population started at a 'position' that is very far from the globally optimal genotype in 'sequence space.'

Let us now turn to the maximal-epistasis case. Here, the character of evolution is very different. Instead of a single local optimum, in the maximal-epistasis case there are an enormous number of local optima, so long as the number of nucleotide sites ($L$) is substantial. In particular, there are $4^L$ possible genotypes, and the probability of a randomly selected genotype being a local optimum is $\frac{1}{3L+1}$, and so the number of local optima is approximately equal to $\frac{4^L}{3L+1}$. For $L = 5,000,000$, this means that there are more than $10^{3,000,000}$ local optima.

As stated above, evolution is presumed to proceed until the population becomes fixed on a locally optimal genotype, and then it stops. However, in the maximal-epistasis case, evolution stops after only a relatively small number of beneficial mutations have accumulated. In fact, for substantial values of $L$, evolution is likely to proceed until approximately $\ln(3L-1)$ beneficial mutations have become fixed in the population[42,48]. When $L = 5,000,000$, this means that evolution is expected to stop after approximately 17 beneficial mutations have been fixed in the population. This is in very stark contrast to the no-epistasis case, where, typically, evolution does not stop until millions of beneficial mutations have been fixed. In the maximal-epistasis case an evolutionary run of more than 17 beneficial mutations is certainly possible. However, it is nearly impossible that it would take more than, say, 50 beneficial mutations before a local optimum is achieved, and evolution stops. Indeed, even for a species with a genome of three billion nucleotide sites that is initially fixed on a randomly selected genotype, evolution is expected to stop (under maximal epistasis) with the achievement of a local optimum after only about 23 beneficial mutations have become fixed.

Because, in the maximal-epistasis case, only a handful of beneficial mutations are fixed before evolution stops, the sequence of the ultimate genotype must be quite close to the random initial sequence that we have assumed was fixed in the population when evolution began. Indeed, after evolution stops, fewer than one nucleotide in 100,000 is likely to have changed from the nucleotide present in the initial random genotype (presuming $L = 5,000,000$). Thus, in the maximal-epistasis case, the sequence of the ultimate genotype cannot possibly tell us anything about any part of the fitness landscape except for a very tiny region that happens to be close (in a mutational sense) to the random initial position of the population. Indeed, the sequence of the random initial genotype determines almost everything about the sequence of the ultimate genotype. After evolution stops, only a very a few nucleotide sites have had their final state determined by natural selection. Thus, it seems that the sequence of the ultimate genotype tells us almost nothing, except for the fact that the observed genotype corresponds to one of the vast number of local optima that exist in the fitness landscape.

The foregoing observations suggest that much more reproductive information tends to accumulate over the course of evolutionary history when fitness landscapes are relatively "smooth" (e.g., the no-epistasis case) as opposed to when they are rugged (e.g., the maximal-epistasis case). We can make this insight more quantitative and gain understanding about what is going on from an information-theoretic perspective if we follow the example of Watkins[24], and consider how we might use the relationship between the fitness landscape and the nature of evolved genotypes as a communications system in which information is sent and received. In information theory, this sort of analysis is the "acid test" for measuring the information-theoretic aspect of the relationship between two phenomena (in this case, between the fitness landscape and the sequence of the ultimate genotype)[9,24,34].

Continuing with the NK model, as described above, we shall simplify matters by restricting attention to communication systems in which the person receiving a message can always decide (given an appropriate "code book") *exactly* which message was sent by the sender. To accomplish this, we assume that the sender has $M$ different possible messages, and the sender prepares a code book which associates each of these messages with one or more of the $4^L$ possible genotypes. The sender shares this codebook with the receiver. At some later time, the sender decides to send one of the $M$ possible messages. To do this, the sender chooses an appropriate fitness landscape, puts the population into its initial random state, and allows the population to evolve until evolution stops because it has become fixed on a locally optimal genotype - the ultimate genotype. A population member is then sent to the receiver who, after determining the genotype of the organism received, uses to the code book to reconstruct the original message with 100% accuracy[24].

If the sender uses a non-epistatic landscape, then to send a particular message, the sender simply picks a fitness landscape for which, in the code book, the globally optimal genotype is associated with the intended message. The sender is then guaranteed that the communication will be successful because the population will definitely converge to the global optimum. There are $4^L$ different possible genotypes, and by an appropriate choice of a non-epistatic landscape, each of these could be the globally optimal genotype. It follows that for a non-epistatic landscape, the maximum possible number of different messages that could be sent is $M = 4^L$.

If there are multiple local optima, then the sender will not, in general, know the identity of the ultimate genotype in advance. This is because, by assumption, the population begins fixed on a randomly chosen genotype. As we have seen, when there are multiple optima, the identity of this initial genotype can determine the identity of the ultimate genotype. All that the sender knows for sure is that the ultimate genotype will be a locally optimal genotype. Thus, the only way to ensure flawless communication is to pick a fitness landscape that associates *every* locally optimal genotype with the message that is to be sent. Thus, the code book must be redundant, with multiple genotypes associated with the same message.

If, for example, the sender is only allowed to use fitness landscapes that have two local optima, then flawless communication requires that each possible message in the codebook is associated with a minimum of two genotypes. (A minor caveat about this is presented in Supplementary Note 16.) If each possible message is associated with precisely two genotypes, then the total number of possible messages that can be flawlessly sent is half of what it was in the no-epistasis case, where each message was associated with only one genotype. That is to say, if the sender must use landscapes with two local optima, then the maximum possible value for the number of possible messages, $M$, is reduced to $M = \frac{4^L}{2}$.

For a flawless-communications system of the sort that we are discussing here, we can measure the information-transmission capacity of the system by simply taking the logarithm of the number of possible messages[9,34]. Thus, for example, in the no-epistasis case we can send, on average, about $\log_2(4^L) = 2L$ bits of information per message (where $\log_2(a)$ represents the logarithm of $a$ to base 2). On the other hand, if the sender can only use fitness landscapes that contain two local optima, then the information-transmission capacity of the system is reduced to $\log_2(\frac{4^L}{2})=(2L-1)$ bits of information.

In general, if the number of different locally optimal genotypes for each fitness landscape is denoted by $V$, then the maximum number of different messages that can be flawlessly sent by a single use of the 'fitness-landscape-and-ultimate-genotype' messaging system is approximately $M = \frac{4^L}{V}$. For the maximal-epistasis case, each genotype has a probability of being a local optimum that is equal[42,43] to $\frac{1}{3L+1}$. Thus, the total number of local optima in the landscape is approximately $V = \frac{4^L}{3L+1}$, and so $M$ is reduced to a huge extent in comparison to the no-epistasis case. In particular, we have $M \approx 3L + 1$. For our example of $L = 5,000,000$, this means that our estimate for information-transmission capacity is $\log_2(3L+1) = \log_2(15,000,001) \approx 24$ bits. This is dramatically different from the no-epistasis case for which, when $L = 5,000,000$, information-transmission capacity is equal to $10,000,000$ bits (because $\log_2(4^L) = 2L = 10,000,000$).

The information-transmission capacity of the system just described can be used to estimate the reproductive information that is associated with the ultimate genotype. This is because, if the ultimate genotype is associated with a high degree of reproductive information, then, by the definition of reproductive information, this means that there is a close relationship between structure of the fitness landscape, and the identity of the ultimate genotype. Furthermore, the more closely the structure of the fitness landscape controls the identity of the ultimate genotype, the more effectively the system can be used to transmit information.

In the preceding example we see that both informal observations and semi-formal quantitative considerations suggest that the amount of reproductive information can differ by many orders of magnitude when we compare the no-epistasis case with the maximal-epistasis case. However, the critical fact to note is that, in both of these cases, the ultimate genotype is a locally optimal genotype, and thus, once genetic evolution stops, every single nucleotide site will be found to be subject to purifying selection. This means that the naive method (which is based on simply counting the number of nucleotide sites that are subject to purifying selection) will give *exactly the same estimate* for the amount of reproductive information in both cases (i.e., $2L = 10,000,000$ bits). In light of the preceeding considerations, this estimate is reasonable in the no-epistasis case, but it can be completely wrong in the maximal-epistasis case. Epistasis seems to be ubiquitous, and thus rugged fitness landscapes should be common[49–51]. Collectively, these observations demonstrate that simply counting nucleotide sites that are subject to purifying selection is a method that cannot be relied upon to give reasonable estimates for reproductive information.

Up to this point we have followed traditional practice among analyses of the NK model by ignoring the possibility that multiple mutations simultaneously occur in a single newborn individual. This is reasonable if the population size is small enough, and if we exclude consideration of very long periods of time. However, for populations that are sufficiently large, multiple mutations cannot be ignored, and, indeed, in an infinite asexual population we are likely to find that only the globally optimal genotype is stable. However, it is worth noting that, if reproduction is sexual, then a population can remain nearly fixed on *any* local optimum for an indefinite period of time, even if the population is effectively infinite in size, so that all possible genotypes are present at all times, and all possible multiple-mutation events occur every generation. This fact is demonstrated in Supplementary Note 15. The observation that local optima can be stable in sexual populations of any size suggests that the information-limiting effects of epistasis should not be ignored, even when considering truly enormous sexual populations.

The impact of epistasis is only one factor that compromises the usefulness of estimating reproductive information by counting genetic sites that are subject to purifying selection. There are various other difficulties with this approach, and three prominent examples of these difficulties are presented in Supplementary Note 3.

# Supplementary Note 3: Additional reasons to question the utility of counting sites under purifying selection in order to estimate reproductive information

The impact of epistasis is only one factor that compromises the usefulness of estimating reproductive information by counting genetic sites that are subject to purifying selection. Another consideration is that purifying selection is typically assessed by examining genotypic distributions within and/or between species. Indeed, use of these distributions is the method recommended by Adami and his colleagues[28–30]. A problem with this is that, while environmental conditions can change quickly, genotypic distributions may require many generations to respond after an environmental change has altered the fitness landscape. Just as a change the course of a river will typically decrease the amount of information that a map provides about waterways, so it is that changes in the fitness landscape should typically decrease the amount of information that common genotypes provide about that fitness landscape. If, for example, there is a very large environmental change in a particular location, then, immediately after the change, the genotype of a local organism may tell us very little about which types of organisms are most fit in the new environment that now prevails in the organism's location. However, a sudden environmental change may have no immediate effect on genotypic distributions. Instead, the genotypic changes may happen gradually, over time. In this case, immediately after the environmental change, using genotypic distributions may be a very inaccurate way to measure the number of genetic sites that are under purifying selection. As such, this method may initially fail to record the decreases in reproductive information that results from the environmental change.

Another problem is that purifying selection is not the only type of natural selection. The method for measuring reproductive information (or physical complexity) that was recommended by Adami and Cerf involves, in essence, making the measurement by counting the number of nucleotide sites with relatively low genetic variation[28–30]. As a result, Adami and Cerf's suggested method can actually show an apparent decrease in reproductive information (or "physical complexity") during episodes of adaptive evolution. This is because adaptive evolution often involves a transitional period during which genetic variation becomes high at some genetic sites. As an example, a case in which average reproductive information apparently decreases during adaptive evolution is apparent in Figure 4 of a study by Adami et. al in 2000[29]. Clearly, a well-behaved measure of average reproductive information should *always* increase during episodes of adaptive evolution, and so we have another reason to treat the counting-selected-sites method with circumspection.

One additional problem with using genotypic distributions in attempts to calculate the information created by natural selection has to do with the efficiency of coding. Imagine two complex and well-adapted species that are phenotypically identical. Assume further that one of these species uses an inefficient genetic code, so that it uses twice as many nucleotide sites to encode phenotypes as the other species. Let us assume that every nucleotide site that is responsible for producing the phenotypes are fixed (or nearly fixed) in both species. For simplicity, let us also assume that all other nucleotide sites have a uniform distribution of the possible nucleotides. Finally, assume that the genome of the species with the inefficient genetic code is twice the size as the genome of the other species. In this situation, using the counting-selected-sites method, we would find that genomes of species with the inefficient genetic code apparently embodies about twice the amount of reproductive information (or physical complexity) as the genomes of the species with the efficient code. As selection acts on phenotypes, and these two hypothetical species are assumed to be phenotypically identical, this conclusion seems to defy common sense.

# Supplementary Note 4: Reproductive information and Kolmogorov information

In this Note we consider the relationship between reproductive information and Kolmogorov information. We start by providing a formal definition of Kolmogorov information. We will then provide a proof of a result mentioned in the main text. Informally speaking, this result shows that, when the reproductive information associated with a particular organismal type is sufficiently large, we are almost guaranteed that the reproductive information provided by the fitness landscape about that organismal type will be approximately equal to the Kolmogorov information that the fitness landscape provides about that organismal type.

This Note utilises certain results that are presented in the main text. With that in mind, we suggest that readers do not read this Note until they have read the main text at least until the end of the section that is entitled: "Reproductive information is approximately equal to a measure of adaptedness."

We use $K\left(\vec{W} : x\right)$ to refer the Kolmogorov information that $\vec{W}$ provides about $x$. Here, $\vec{W}$ refers to the fitness landscape at a particular time and place, and $x$ is the organismal-type index of an individual that is present at that time and place. The mathematical definition of $K\left(\vec{W} : x\right)$ is as follows[35]:

$$K\left(\vec{W} : x\right) = K(x) - K\left(x | \vec{W}\right). \tag{S4.1}$$

In this equation, $K(x)$ is the unconditional algorithmic complexity of $x$. That is, $K(x)$ is the length, in binary digits, of the shortest binary program that, when provided as input to a particular hypothetical computer, will output the value of $x$. We assume that the hypothetical computer is reasonably capable and efficient. In particular, we assume that it is "additively optimal and universal". (See Li and Vitányi's textbook[35] for definitions of "additively optimal" and "universal" in this context.) In equation (S4.1), $K\left(x | \vec{W}\right)$ refers to the conditional algorithmic complexity of $x$ (conditional on access to $\vec{W}$). This is the length, in binary digits, of the shortest binary program that, when provided to the additively optimal and universal computer mentioned above, will output the value of $x$. The difference between $K(x)$ and $K\left(x | \vec{W}\right)$ is that we assume that, for the calculation of $K(x)$, the computer has access to no external data (except for the computer program of length $K(x)$ that outputs the value of $x$). However, for the calculation of $K\left(x | \vec{W}\right)$, we assume that the computer has access to $\vec{W}$ (e.g., $\vec{W}$ may be recorded on a tape-memory device that is connected to the computer)[35,36].

If $K\left(x | \vec{W}\right) \ll K(x)$ then we conclude that the fitness landscape "contains" a lot of information about the value of $x$, and thus, in this case, $K\left(\vec{W} : x\right)$ will be large. The justification for saying that $\vec{W}$ contains a lot of information about $x$ is that, in this case, the number of bits required to specify $x$ is greatly reduced when the computer has access to $\vec{W}$, as compared to when the computer does not have access to $\vec{W}$.

Before providing a proof of the main result for this Note, we will provide a formal statement of the result. In preparation for this formal statement, we note that, in the limit as $\Omega \to \infty$, the maximum-possible amount of reproductive information that the fitness landscape can possibly provide about an organismal type is approximately equal to $\log_2(\Omega)$. This is because the reproductive information about organismal type that is contained in the fitness landscape (i.e., $I_R\left(\overrightarrow{W} : x\right)$) is defined by $I_R\left(\overrightarrow{W} : x\right) = \overline{L}_u - \overline{L}_s$. Furthermore, as shown in Supplementary Note 6, the quantity $\overline{L}_u$ cannot differ from $\log_2(\Omega)$ by more than 3 bits, and the smallest-possible value of $\overline{L}_s$ is unity.

With the foregoing observations in mind, let us consider a particular environment in which a particular fitness landscape is manifest. let us assume that there exists at least one organismal type that has reproductive information that is a non-negligible fraction of $\log_2(\Omega)$, while still not being very close to $\log_2(\Omega)$. To be more precise, we assume that there is at least one organismal type which, given the fitness landscape, has reproductive information ($I_R\left(\overrightarrow{W} : x\right)$) with a value that satisfies

$$\varepsilon_1 \log_2(\Omega) < I_R\left(\overrightarrow{W} : x\right) < (1 - \varepsilon_2) \log_2(\Omega) \tag{S4.2}$$

where $\varepsilon_1$ and $\varepsilon_2$ are arbitrarily chosen small and positive real numbers. (That is, $0 < \varepsilon_1, \varepsilon_2 \ll 1$).

Let us define a *qualifying organismal type* to be an organismal type for which $I_R\left(\overrightarrow{W} : x_i\right) > \varepsilon_1 \log_2(\Omega)$. Furthermore, let $P$ represent the proportion of qualifying organismal types for which the reproductive information ($I_R\left(\overrightarrow{W} : x\right)$) differs substantially (in proportional terms) from the Kolmogorov information in the fitness landscape that is about the organismal type (i.e., $K\left(\overrightarrow{W} : x\right)$). In particular, we consider that $I_R\left(\overrightarrow{W} : x\right)$ differs substantially from $K\left(\overrightarrow{W} : x\right)$ (in proportional terms) if

$$\frac{K\left(\overrightarrow{W} : x\right)}{I_R\left(\overrightarrow{W} : x\right)} > 1 + \varepsilon_3 \tag{S4.3}$$

or if

$$\frac{K\left(\overrightarrow{W} : x\right)}{I_R\left(\overrightarrow{W} : x\right)} < 1 - \varepsilon_3 \tag{S4.4}$$

where $\varepsilon_3$ is another arbitrarily chosen small and positive real number.

With this preparation, we can now state the main result for this section. In particular:

$$\lim_{\Omega \to \infty} P = 0. \tag{S4.5}$$

We shall now demonstrate that equation (S4.5) is correct. To do this, we begin by proving that the proportion of qualifying types that satisfy inequality (S4.3) must go to zero as $\Omega \to \infty$. The strategy of this proof to define a set of qualifying types that contains *all* qualifying types that satisfy inequality (S4.3). We will then show that the proportion of qualifying types that are part of this set must go to zero as $\Omega \to \infty$. We then complete the proof by carrying out a similar procedure with respect to inequality (S4.4).

In the main text we presented the result

$$\left| I_R\left(\overrightarrow{W} : x\right) - \log_2\left(\frac{\Omega}{Q_x}\right) \right| < 4, \tag{S4.6}$$

where $Q_x$ is the number of organismal types that are at least as fit as organismal type $x$. Using inequality (S4.6) along with equation S4.1, we can show that, as $\Omega \to \infty$, inequality (S4.3) will be satisfied if and only if

$$\frac{K(x) - K\left(x|\vec{W}\right)}{\log_2(\Omega) - \log_2(Q_x)} > 1 + \varepsilon_3. \tag{S4.7}$$

Note that, when substituting from inequality (S4.6) to create the denominator on the left side of inequality (S4.7) it is permissible to ignore the term that is no greater in magnitude than four because inequality (S4.3) is calculated for a qualifying type, which means that $\varepsilon_1 \log_2(\Omega) < I_R\left(\vec{W} : x\right)$. This implies that, for qualifying types, $I_R\left(\vec{W} : x\right)$ becomes large as $\Omega$ becomes large. In light of inequality (S4.6) we know that $\log_2(\Omega) - \log_2(Q_x)$ also becomes large as $\Omega$ becomes large. These observations imply that, in the limit as $\Omega \to \infty$, the term in question becomes negligible in comparison to $\log_2(\Omega) - \log_2(Q_x)$, and thus it can be ignored.

We can rewrite inequality (S4.7) as

$$\frac{[K(x) - \log_2(\Omega)] + \left[\log_2(Q_x) - K\left(x|\vec{W}\right)\right]}{\log_2(\Omega) - \log_2(Q_x)} > \varepsilon_3. \tag{S4.8}$$

We know that $x$ is an integer that satisfies $1 \leq x \leq \Omega$, and $K(x)$ is equal to the smallest-possible length of a program that outputs $x$ for some arbitrarily selected "additively optimal and universal" computer. A program that outputs $x$ can simply be a bare-bones program that contains a print command followed by the value of $x$. The minimum-possible length of such a program is no greater than $O(1) + \log_2(x)$ (because $\log_2(x)$ is approximately equal to the number of digits in $x$, when it is written as a binary number). Thus, $K(x)$ cannot be larger than $O(1) + \log_2(\Omega)$. This implies that, in the limit as $\Omega \to \infty$, the maximum value of $[K(x) - \log_2(\Omega)]$ is $O(1)$.

Next, let us turn our attention to the denominator on the left side of equation (S4.8). For qualifying organismal types we know (by the definition of these types) that that $I_R\left(\vec{W} : x\right) > \varepsilon_1 \log_2(\Omega)$. Thus, from equation (S4.6) we have, for qualifying types, $\log_2(\Omega) - \log_2(Q_x) > \varepsilon_1 \log_2(\Omega) + O(1)$. In the limit as $\Omega \to \infty$ this implies that $\log_2(\Omega) - \log_2(Q_x) > \varepsilon_1 \log_2(\Omega)$. Combining the foregoing considerations we find that, in the limit as $\Omega \to \infty$, for inequality (S4.8) (and thus inequality (S4.3)) to be satisfied for a qualifying organismal type it is necessary (but not sufficient) to have, for that type,

$$\frac{\log_2(Q_x) - K\left(x|\vec{W}\right)}{\log_2\left(\Omega^{\varepsilon_1}\right)} > \varepsilon_3. \tag{S4.9}$$

In the limit as $\Omega \to \infty$, inequality (S4.9) must be satisfied if inequality (S4.3) is to be satisfied. However, satisfaction of inequality (S4.9) does not guarantee satisfaction of inequality (S4.3).

Let $Q^*$ represent the largest value of $Q_x$ for any qualifying organismal type. This implies (from the definition of $Q_x$) that $Q^*$ is also the total number of qualifying organismal types. With this in mind it is clear that, In the limit as $\Omega \to \infty$, for any value of $x$ that is associated with a qualifying organismal type and that allows satisfaction of inequality (S4.9) (and thus, potentially, inequality (S4.3)), it must

be the case that
$$\frac{\log_2(Q^*) - K\left(x|\vec{W}\right)}{\log_2(\Omega^{\varepsilon_1})} > \varepsilon_3. \tag{S4.10}$$

Let us define $d_x$ as $d_x = \log_2(Q^*) - K\left(x|\vec{W}\right)$. Using this we can rewrite inequality (S4.10) as

$$\frac{d_x}{\log_2(\Omega^{\varepsilon_1})} > \varepsilon_3. \tag{S4.11}$$

From inequality (S4.11) we can see that, for any value of $d_x$ to allow for satisfaction of inequality (S4.3), we must have $d_x > \log_2(\Omega^{\varepsilon_1\varepsilon_3})$. With this in mind, let $d^* = \log_2(\Omega^{\varepsilon_1\varepsilon_3})$. Thus, all values of $d_x$ that allow satisfaction of inequality (S4.3) must be larger than $d^*$.

To a close approximation, the total number of binary sequences that are, at most, $L$ digits long is equal to $2^{L+1}$. From the definition of $d_x$ and $d^*$ we know that, for all qualifying organismal types that satisfy inequality (S4.3), we must have $K\left(x|\vec{W}\right) < \log_2(Q^*) - d^*$.

From the preceding we can see that the total number of binary sequences that are less than or equal to $\log_2(Q^*) - d^*$ digits long is approximately equal to $2\left(2^{\log_2(Q^*)-d^*}\right)$. For a qualifying organismal type, any value of $K\left(x|\vec{W}\right)$ that, in the limit as $\Omega \to \infty$, allows for satisfaction of inequality (S4.3) must be the length of one of these sequences. Thus, in the limit as $\Omega \to \infty$, the total number of qualifying organismal types that are associated with a value of $K\left(x|\vec{W}\right)$ that allows for satisfaction of inequality (S4.3) must be less than or equal to (approximately) $2\left(2^{\log_2(Q^*)-d^*}\right)$. More simply stated, the total number of such qualifying organismal types must be less than or equal to (approximately) $\frac{2Q^*}{2^{d^*}}$.

The total number of qualifying organismal types is equal to $Q^*$. Thus, by dividing $\frac{2Q^*}{2^{d^*}}$ by $Q^*$ we obtain an approximate upper limit on the proportion of qualifying organismal types that satisfy inequality (S4.3). This approximate upper limit is equal to

$$\frac{1}{2^{d^*-1}}. \tag{S4.12}$$

We know that $d^* = \log_2(\Omega^{\varepsilon_1\varepsilon_3})$, and thus, as $\Omega \to \infty$, the proportion of qualifying organismal types that satisfy inequality (S4.3) must go to zero.

Let us now calculate an upper bound for the proportion of qualifying types that satisfy inequality (S4.4).

Using similar logic to that which led to inequality (S4.8) we can show that, in the limit as $\Omega \to \infty$, inequality (S4.4) will be satisfied if and only if

$$\frac{[\log_2(\Omega) - K(x)] + \left[K\left(x|\vec{W}\right) - \log_2(Q_x)\right]}{\log_2(\Omega) - \log_2(Q_x)} > \varepsilon_3. \tag{S4.13}$$

Here, $K\left(x|\vec{W}\right)$ represents the algorithmic complexity of $x$ for a computer that has access to the fitness landscape, $\vec{W}$. The value of $K\left(x|\vec{W}\right)$ cannot be greater than $\log_2(Q_x) + O(1)$. This is because $K\left(x|\vec{W}\right)$ is the length of the shortest-possible program (for some qualifying computer) that outputs $x$. Furthermore, we can always (in principle) write a computer program of length $\log_2(Q_x) + O(1)$ (or less) that will output $x$, if the computer for which we are writing the program has access to the fitness landscape ($\vec{W}$). This can be done by using the fitness landscape to order all of the $\Omega$ possible values of $x$ from most fit to least fit, putting any organismal types that are equal in fitness in the length-lexicographic order of their organismal-type indices, so that, after the ordering, there are no two types that differ in fitness for which the fitter one comes later in the ordering. The rank (i.e., the position) of a particular organismal type (say type $x$) in this ordering will always be equal to or less than $Q_x$. This follows from the definition of $Q_x$ (it is the number of types that are at least as fit as organismal type $x$). The program can thus simply instruct the computer to print out the organismal-type index that is associated with the rank of type $x$ in the ordering of organismal-types. A binary representation of this rank will be no longer than (approximately) the logarithm of the rank (where the logarithm is taken to base 2). Furthermore, the logarithm of the rank cannot be larger than $\log_2(Q_x)$. This justifies the claim that we can always write a computer program of length $\log_2(Q_x) + O(1)$ (or less) that will output $x$. The $O(1)$ terms represent the length of the part of the computer code that includes the print command, but that does not include the value of the rank. The length of this part of the code does not vary with $Q_x$. The length of this part of the code also does not vary with $\Omega$, as $\Omega$ can be derived from the fitness landscape, to which the computer is assumed to have access.

Because the value $K\left(x|\vec{W}\right)$ cannot be greater than $\log_2(Q_x) + O(1)$, the term $K\left(x|\vec{W}\right) - \log_2(Q_x)$ in inequality (S4.13) cannot be greater than $O(1)$. Furthermore, as shown above, we know that the denominator on the left side of inequality (S4.13) cannot be less than $\log_2(\Omega^{\varepsilon_1})$. This means that the denominator grows without limit as $\Omega \to \infty$, and so any term in the numerator that is $O(1)$ (such as, in this case, $K\left(x|\vec{W}\right) - \log_2(Q_x)$) can be ignored, as it is negligible in the limit as $\Omega \to \infty$. Combining the foregoing considerations we find that, in the limit as $\Omega \to \infty$, for satisfaction of inequality (S4.4) by a qualifying organismal type we must have, for that organismal type,

$$\frac{\log_2(\Omega) - K(x)}{\log_2(\Omega^{\varepsilon_1})} > \varepsilon_3. \tag{S4.14}$$

In the limit as $\Omega \to \infty$, inequality (S4.14) must be satisfied if inequality (S4.4) is to be satisfied. However, satisfaction of inequality (S4.14) does not guarantee satisfaction of inequality (S4.4).

Let us define $\widetilde{d}_x$ as $\widetilde{d}_x = \log_2(\Omega) - K(x)$. Using this we can rewrite inequality (S4.14) as

$$\frac{\widetilde{d}_x}{\log_2(\Omega^{\varepsilon_1})} > \varepsilon_3. \tag{S4.15}$$

Clearly, for any value of $\widetilde{d}_x$ to allow for satisfaction of inequality (S4.15), we must have $\widetilde{d}_x > \log_2(\Omega^{\varepsilon_1 \varepsilon_3})$. Recall that $d^* = \log_2(\Omega^{\varepsilon_1 \varepsilon_3})$. Thus, all values of $\widetilde{d}_x$ that allow satisfaction of inequality (S4.4) must be larger than $d^*$.

Note that, from inequality (S4.15) and from the definition of $\widetilde{d}_x$ and $d^*$ we know that, for qualifying organismal types that, in the limit as $\Omega \to \infty$, satisfy inequality (S4.4), we have $K(x) < \log_2(\Omega) - d^*$. In light this, we know that the proportion of qualifying organismal types that satisfy inequality (S4.4) can be no larger than the proportion of qualifying organismal types for which $K(x) < \log_2(\Omega) - d^*$. With this in mind, it is important to note that, in the limit as $\Omega \to \infty$, the proportion of qualifying organismal types for which $K(x) < \log_2(\Omega) - d^*$ must be *exactly* equal to the proportion of organismal types for which $K(x) < \log_2(\Omega) - d^*$ among the total collection of *all* $\Omega$ organismal types. The reason for this is that, while the determination of whether an organismal type is a qualifying organismal type depends on the fitness of the organismal type in question, the value of $K(x)$ for a given organismal type depends *only* on the value of the organismal index, $x$. (This is due to the definition of $K(x)$, which assumes a computer that has access to no external data other than the program that has length $K(x)$). Furthermore, in the main text we assumed that the values of $x$ are assigned to organismal types entirely at random. Thus, the probability that, for a given organismal type, $K(x) < \log_2(\Omega) - d^*$, is independent of the fitness of that type, and therefore it is independent of whether or not that type is a qualifying organismal type. Furthermore, the number of qualifying organismal types ($Q^*$) becomes very large in the limit as $\Omega \to \infty$. This is because we have assumed that there is at least one organismal type for which $I_R\left(\vec{W} : x\right) < (1 - \varepsilon_2)\log_2(\Omega)$. In the limit as $\Omega \to \infty$, this implies that, for this type, $\log_2(\Omega) - \log_2(Q_x) < (1 - \varepsilon_2)\log_2(\Omega)$, and thus $Q_x > \Omega^{\varepsilon_2}$. From the definition of $Q_x$ we know that there are $Q_x$ types that are at least as fit as the type for which $x$ is the organismal index. Furthermore, all of these are qualifying types (we know this from the definition of qualifying types). As $Q_x > \Omega^{\varepsilon_2}$, this implies that the number of qualifying types becomes very large as $\Omega \to \infty$. This observation allows for application of the law of large numbers to determine the proportion of qualifying organismal types for which $K(x) < \log_2(\Omega) - d^*$. In particular, application of the law of large numbers, along with the foregoing observations, demonstrates that, in the limit as $\Omega \to \infty$, the proportion of qualifying organismal types for which $K(x) < \log_2(\Omega) - d^*$ must be equal to the proportion of organismal types for which $K(x) < \log_2(\Omega) - d^*$ among the *total* collection of *all* $\Omega$ organismal types.

The foregoing considerations demonstrate that, to show that the proportion of qualifying organismal types that satisfy inequality (S4.4) goes to zero as $\Omega \to \infty$, we need only show that, among *all* organismal types, the proportion for which $K(x) < \log_2(\Omega) - d^*$ goes to zero as $\Omega \to \infty$.

The total number of binary sequences that are less than or equal to $\log_2(\Omega) - d^*$ digits long is approximately equal to $2^{\log_2(\Omega) - d^* + 1}$. Each value of $K(x)$ for which $K(x) < \log_2(\Omega) - d^*$ must be equal to the length of one of these sequences. Thus, the total number of organismal types that are associated with a value of $K(x)$ for which $K(x) < \log_2(\Omega) - d^*$ must be less than or equal to (approximately) $2^{\log_2(\Omega) - d^* + 1}$. More simply stated, the total number of such organismal types must be less than or equal to (approximately) $\frac{2\Omega}{2^{d^*}}$.

The total number of organismal types is equal to $\Omega$. Thus, by dividing $\frac{2\Omega}{2^{d^*}}$ by $\Omega$ we obtain an approximate upper limit on the proportion of all organismal for which $K(x) \leq \log_2(\Omega) - d^*$. This approximate upper limit is equal to

$$\frac{1}{2^{d^* - 1}}. \tag{S4.16}$$

We know that $d^* = \log_2(\Omega^{\varepsilon_1 \varepsilon_3})$, and thus, as $\Omega \to \infty$, the proportion of all organismal types for which $K(x) < \log_2(\Omega) - d^*$ must go to zero. As noted above, this implies that the proportion of qualifying organismal types that satisfy inequality (S4.4) goes to zero as $\Omega \to \infty$. We have already seen that the proportion of qualifying organismal types for which inequality (S4.3) is satisfied goes zero as $\Omega \to \infty$, and thus we have now proven that equation (S4.5) is correct, as per our original objective.

# Supplementary Note 5: Formal definition of an optimal code plus an alternative definition of an optimal code

In the main text we discuss *optimal coding schemes* in the context of the definition of reproductive information. As explained in the main text, a *coding scheme* for organismal types is a way of assigning one of the first $\Omega$ binary code words to each of the $\Omega$ organismal types. An optimal coding scheme is a coding scheme that does this in a way such that the fittest organismal types will tend to be associated with the shortest binary code words.

How, specifically, do we decide whether a particular coding scheme for organismal types is optimal in the context of reproductive information? To answer this question, we imagine ordering the organismal types so that they are in the same order as the length-lexicographical ordering of their associated binary code words (where these associations are specified by the coding scheme in question). Thus, for example, if $\Omega = 14$, then we would order the 14 organismal types so that the organismal types associated with the binary code words 0, 1, 00, and 01 come first, second, third, and fourth, respectively (see Fig. 1 in the main text). The last organismal type (in $14^{th}$ place) would be associated with the binary code word 111. After creating this ordering, we then determine whether the associated coding scheme is optimal by looking to see if there is any pair of organismal types that differ in fitness, and for which the organismal type with the lower fitness comes first in the ordering. If this is the case for *any pair* of organismal types, then the coding scheme is non-optimal. Otherwise, the coding scheme is optimal. This automatically satisfies the requirement that, in optimal coding schemes, fitter types will generally be associated with shorter code words.

Our definition of an optimal coding scheme depends, for every pair of organismal types that differ in fitness, on whether or not the fitter type comes first in a particular ordering of the organismal types (as just described). A possible alternative definition (which we *did not* adopt) would decide the question of optimality by comparing the *length* of the code words associated with each pair of organismal types that differ in fitness. For this alternative method, a coding scheme would be optimal if there is no pair of organismal types that differ in fitness for which the fitter one is associated with a longer code word. In fact, the method we adopted (as described above) and this alternative method are, effectively, very similar. In particular, they both result in the optimal coding schemes being ones in which the fittest organismal types are assigned the shortest code words. However, the method adopted in this work allows for a particular ordering of the organismal types to be unambiguously labelled as optimal or non-optimal, *even if one changes between "alphabets" that use different numbers of symbols to encode organismal types.* This independence from the size of the alphabet is the rationale behind choosing the definition of optimal coding schemes that we use in this work.

## Supplementary Note 6: The range of possible values of reproductive information

In this Supplementary Note we give restrictions on the ranges that several quantities of interest can take. The restrictions follow from special cases of results presented in Supplementary Note 8.

1. In Supplementary Note 4, on page 20, we state that $\overline{L}_u$ cannot differ from $\log_2(\Omega)$ by more than 3 bits. This follows from equation (S8.13) when inequalities (S8.6) and (S8.7) are used with $c = 1$.

2. In Supplementary Note 4, on page 20, we state that $\overline{L}_s$ has a smallest possible value of unity. This follows from inequality (S8.15) with $Q_x = 1$.

3. From inequality (S8.17) we have that the reproductive information, $I_R(\overrightarrow{W} : x)$ will always lie within the range $\log_2(\Omega/Q_x) - 4 < I_R(\overrightarrow{W} : x) < \log_2(\Omega/Q_x) + 4$. However, as stated at the end of Supplementary Note 8, it should be possible to establish tighter bounds for $I_R(\overrightarrow{W} : x)$, should this be required.

## Supplementary Note 7: The generalisability of the methodology used to define reproductive information

The methodology described in the section entitled "A quantitative definition of reproductive information" can be generalised. In particular, in any case where any theory makes predictions about the relative probability of a finite number of possible events, the methodology can be used to measure the amount of information about the events that is provided by the theory.

## Supplementary Note 8: Approximate properties of reproductive information (including mutuality)

In this note we determine bounds on reproductive information that allow us to show that:
(i) reproductive information is approximately equal to quantitative adaptedness and
(ii) reproductive information is approximately mutual.

We begin with the two forms of reproductive information considered in the main text:

1. the reproductive information that the landscape ($\vec{W}$) provides about organismal type ($x$), written as $I_R(\vec{W} : x)$, and

2. the reproductive information that the focal type ($x$) provides about the fitness landscape ($\vec{W}$), written as $I_R(x : \vec{W})$.

We shall determine relations that these two forms of reproductive information obey. We proceed by first establishing some preliminary results for binary code words, then determining relevant inequalities, and then applying the inequalities to the two forms of reproductive information.

**Preliminaries**

Using an index $b$, which takes values in the positive integers, i.e., $1, 2, 3, ...$, the associated binary code words, in *length-lexicographic order*[a], are

| index, $b$ | binary code word |
|:---:|:---:|
| 1 | 0 |
| 2 | 1 |
| 3 | 00 |
| 4 | 01 |
| 5 | 10 |
| 6 | 11 |
| 7 | 000 |
| 8 | 001 |
| 9 | 010 |
| ⋮ | ⋮ |

Henceforth, when we refer to the $b$'th binary code word, or binary code word $b$, we mean the binary code word with index $b$, as contained in the above table of length-lexicographically ordered binary code words.

Length-lexicographic order has the property that binary code word $b$ has a length that is less than or equal to the length of binary code word $b + 1$, for any positive integer, $b$.

We note that there are

$2^1$ binary code words of length 1 (namely 0 and 1)

$2^2$ binary code words of length 2 (namely 00, 01, 10 and 11)

$2^3$ binary code words of length 3 (namely 000, 001, ..., 111)

---

[a]The length of a binary code word in the number of binary digits it contain. Under length-lexicographic ordering, strings are first sorted according to length, and then strings of a given length are sorted lexicographically.

$$\vdots$$

$2^p$ binary code words of length $p$

We denote the length of binary code word $b$ by $\lambda(b)$. We have that

$$\lambda(b) = \text{ length of binary code word } b$$

$$= \lfloor \log_2 (b+1) \rfloor \tag{S8.1}$$

where $\log_2(b)$ denotes the logarithm of $b$ to base 2 and $\lfloor b \rfloor$ denotes the largest integer that is less than or equal to $b$. To illustrate the working of $\lambda(b)$, the 14th and 15th binary code words are 111 and 0000, respectively, and $\lambda(14) = \lfloor \log_2(15) \rfloor = 3$ while $\lambda(15) = \lfloor \log_2(16) \rfloor = 4$.

For the purposes of this work, we require the length of binary code word $b$, namely $\lambda(b)$, and also the *mean length* of the first $b$ binary code words, which we write as $\mu(b)$. We have

$$\mu(b) = \text{mean length of first } b \text{ binary code words}$$

$$= \frac{1}{b} \sum_{a=1}^{b} \lambda(a). \tag{S8.2}$$

The sum in the above equation is given by

$$\sum_{a=1}^{b} \lambda(a) = 2 + (2+b)\lambda(b) - 2^{\lambda(b)+1} \tag{S8.3}$$

as may be verified[b], and we thus obtain

$$\mu(b) = \lambda(b) - \Delta(b) \tag{S8.4}$$

where

$$\Delta(b) = 2\frac{2^{\lambda(b)} - 1 - \lambda(b)}{b}. \tag{S8.5}$$

**Inequalities**

Anticipating matters, we shall shortly encounter the quantities $\mu(b) - \lambda(c)$ and $\mu(b) - \mu(c)$ where $b$ and $c$ are positive integers with $b \geq c$. We shall make use of the inequalities

$$\mu(b) - \lambda(c) > \log_2(b/c) - 4 \tag{S8.6}$$

and

$$\mu(b) - \mu(c) < \log_2(b/c) + 4 \tag{S8.7}$$

which we now prove.

---

[b]With $S(b) = 2 + (2+b)\lambda(b) - 2^{\lambda(b)+1}$, we have that $S(b+1) - S(b) = \lambda(b+1) + T$ with $T = (2+b)[\lambda(b+1) - \lambda(b)] - 2\left(2^{\lambda(b+1)} - 2^{\lambda(b)}\right)$. There are two cases: (i) $\lambda(b+1) = \lambda(b)$ and (ii) $\lambda(b+1) = \lambda(b) + 1$. In case (i) we directly have $T = 0$. Case (ii) occurs when $b+2 = 2^k$ for $k = 2, 3, 4, ...$ and also leads to $T = 0$. Thus $S(b+1) - S(b) = \lambda(b+1)$. Also, we have $S(1) = 1 \equiv \lambda(1)$ thus $S(b) = \sum_{a=1}^{b} \lambda(a)$.

**Proofs of the inequalities**

First we generally have

$$0 \leq \Delta(b) < 2 \tag{S8.8}$$

as can be straightforwardly shown[c].

Next, for $b \geq 1$ we have

(i) $\lfloor \log_2(b+1) \rfloor \leq \log_2(b+1) \leq \log_2(b) + 1$,

(ii) $\lfloor \log_2(b+1) \rfloor > \log_2(b+1) - 1 > \log_2(b) - 1$.

Combined, (i) and (ii) lead to $\log_2(b) - 1 < \lfloor \log_2(b+1) \rfloor < \log_2(b) + 1$ which can be written as

$$\log_2(b) - 1 < \lambda(b) \leq \log_2(b) + 1. \tag{S8.9}$$

We now consider inequality (S8.6) for $\mu(b) - \lambda(c)$. Using equation (S8.4) we have

$$\mu(b) - \lambda(c) = \lambda(b) - \lambda(c) - \Delta(b) \tag{S8.10}$$

To find a lower bound on $\mu(b) - \lambda(c)$ we use the lower bounds of each of the three terms on the right hand side of equation (S8.10). That is, we add the lower bounds of $\lambda(b)$, $-\lambda(c)$ and $-\Delta(b)$, that follow from equations (S8.9) and (S8.8), namely $\log_2(b) - 1$, $-(\log_2(c) + 1)$ and $-2$, respectively. This yields $\mu(b) - \lambda(c) > \log_2(b/c) - 4$ which is the result in equation (S8.6).

Consider now inequality (S8.7) for $\mu(b) - \mu(c)$. Using equation (S8.4) yields

$$\mu(b) - \mu(c) = \lambda(b) - \lambda(c) - \Delta(b) + \Delta(c). \tag{S8.11}$$

To establish an upper bound on $\mu(b) - \mu(c)$ we use the upper bounds of the four terms on the right hand side of equation (S8.11) that follow from equations (S8.8) and (S8.9). We obtain $\mu(b) - \mu(c) < \log_2(b/c) + 4$ which is the result in equation (S8.7).

Note that to establish the results in equations (S8.6) and (S8.7), we adopted the procedure of using the lower bounds of the *separate* terms on the right hand side of equation (S8.10), and the corresponding upper bounds of the terms in equation (S8.11). This procedure is simple and produces results that are adequate for our purposes, but it is not optimal. Tighter bounds would follow if, in the expression $\lambda(b) - \Delta(b)$ that occurs in equations (S8.10) and (S8.11), we did not use the separately calculated bounds of $\lambda(b)$ and $-\Delta(b)$ but instead used the bound of the whole expression, $\lambda(b) - \Delta(b)$. Indeed there is numerical evidence (not presented) that suggests that tighter bounds exist. Analytically determining tighter bounds will be left for future work, should the need arise.

---

[c]With $f(x) = 2^x - 1 - x$ we have $f(1) = 0$ and $f'(1) = 2\ln(2) - 1 > 0$. Thus for $x \geq 1$ we have $f(x) \geq 0$. Also $2^{\lambda(b)} - 1 - \lambda(b) < 2^{\lambda(b)} - 1 \leq 2^{\log_2(b+1)} - 1$ i.e., $2^{\lambda(b)} - 1 - \lambda(b) < b$. The above two results lead to $0 \leq \Delta(b) < 2$.

## The reproductive information $I_R(\vec{W} : x)$

We shall now put bounds on the value reproductive information that the fitness landscape provides about organismal type, as given by

$$I_R(\vec{W} : x) = \bar{L}_u - \bar{L}_s. \tag{S8.12}$$

The quantity $\bar{L}_u$ in this equation is the mean binary code word length when all $\Omega$ binary code words are used, i.e.,

$$\bar{L}_u = \mu(\Omega). \tag{S8.13}$$

By contrast, the quantity $\bar{L}_s$ depends on the form of the fitness landscape, as encapsulated in $\vec{W}$, and the specific organismal type, $x$. We shall consider forms of $\vec{W}$ that lead to $\bar{L}_s$ taking its smallest and largest possible values.

Proceeding, we take organismal type $x$ to correspond to there being $Q_x$ organismal types (including type $x$) whose fitness equals or exceeds the fitness associated with organismal-type $x$.

If all fitness values in $\vec{W}$ are different, then in this case $\bar{L}_s$ has its largest possible value (because with some fitness values coinciding, $\bar{L}_s$ will generally obtain contributions from different length code words and will not be maximal). The value of $\bar{L}_s$, when all fitness values are different, is simply the *length* of the single binary code word corresponding to organismal-type index $x$. The corresponding code word length is $\lambda(Q_x)$. Thus we generally have

$$\bar{L}_s \leq \lambda(Q_x). \tag{S8.14}$$

By contrast, if the binary code words, with indices 1, 2, ..., $Q_x$, all have the same fitness value associated with them, and other fitness values are smaller than these, then $\bar{L}_s$ will take its smallest possible value, corresponding to the *mean* of the first $Q_x$ binary code words. This is $\mu(Q_x)$ and we generally have

$$\bar{L}_s \geq \mu(Q_x). \tag{S8.15}$$

The inequalities for $\bar{L}_s$ in equations (S8.14) and (S8.15) can jointly be written as $\mu(Q_x) \leq \bar{L}_s \leq \lambda(Q_x)$. From this we arrive at $I_R(\vec{W} : x)$ lying in the range

$$\mu(\Omega) - \lambda(Q_x) \leq I_R(\vec{W} : x) \leq \mu(\Omega) - \mu(Q_x). \tag{S8.16}$$

From equations (S8.6) and (S8.7) we obtain $\log_2(\Omega/Q_x) - 4 < \mu(\Omega) - \lambda(Q_x)$ and $\mu(\Omega) - \mu(Q_x) < \log_2(\Omega/Q_x) + 4$ thus equation (S8.16) yields $\log_2(\Omega/Q_x) - 4 < I_R(\vec{W} : x) < \log_2(\Omega/Q_x) + 4$ which can be written as

$$\left| I_R(\vec{W} : x) - \log_2(\Omega/Q_x) \right| < 4. \tag{S8.17}$$

# The reproductive information $I_R(x : \vec{W})$

We shall now put bounds on the value reproductive information that the focal type[d] provides about the fitness landscape, as given by

$$I_R(x : \vec{W}) = \bar{L}'_u - \bar{L}'_s. \tag{S8.18}$$

In what follows, we shall repeatedly meet the quantity $(\Omega - 1)!$ Accordingly, we shall designate this quantity by a symbol, and we set

$$\Lambda = (\Omega - 1)! \tag{S8.19}$$

Proceeding, the quantity $\bar{L}'_u$ in equation (S8.18) is the average binary code word length of the first $\Omega! \equiv \Omega \times \Lambda$ binary code words (see Supplementary Note 9). That is,

$$\bar{L}'_u = \mu(\Omega\Lambda). \tag{S8.20}$$

The quantity $\bar{L}'_s$ depends on the focal type. As previously, we use $Q_x$ to denote the number of organismal types whose fitness equals or exceeds that of the focal type, i.e., of organismal type $x$.

A method for classifying fitness landscapes is specified in Supplementary Note 9, and this Note also explains how to identify an optimal coding scheme for a particular fitness landscape. Under an optimal coding scheme, the ordering index that characterises the focal-individual's environment must be assigned one of the first $Q_x \times (\Omega - 1)! \equiv Q_x \times \Lambda$ binary code words. (See Supplementary Note 9 for the definition of ordering indices.) The smallest possible value of $\bar{L}'_s$ occurs when all $Q_x$ organismal types, whose fitness equals or exceeds that of the focal type, have values of their fitness that exactly equals that of the focal type. In this case the ordering index that characterises the focal-individual's environment is equally likely to be assigned any of the first $Q_x \times \Lambda$ binary code words. In this case $\bar{L}'_s$ will therefore equal the *mean length* of these binary code words. We thus have

$$\bar{L}'_s \geq \mu\left(Q_x \Lambda\right). \tag{S8.21}$$

On the other hand, the largest code word that can be associated with the focal-individual's environment is the one that occurs at position $\Omega \times \Lambda$ in the list of binary code words. That is

$$\bar{L}'_s \leq \lambda\left(Q_x \Lambda\right). \tag{S8.22}$$

The inequalities for $\bar{L}'_s$ in equations (S8.21) and (S8.22) can jointly be written as $\mu\left(Q_x\Lambda\right) \leq \bar{L}'_s \leq \lambda\left(Q_x\Lambda\right)$. From this we arrive at $I_R(x : \vec{W})$ lying in the range

$$\mu(\Omega\Lambda) - \lambda(Q_x\Lambda) \leq I_R(x : \vec{W}) \leq \mu(\Omega\Lambda) - \mu(Q_x\Lambda). \tag{S8.23}$$

From equations (S8.6) and (S8.7) we obtain $\log_2(\Omega/Q_x) - 4 < \mu(\Omega\Lambda) - \lambda(Q_x\Lambda)$ and $\mu(\Omega\Lambda) - \mu(Q_x\Lambda) < \log_2(\Omega/Q_x) + 4$ thus equation (S8.23) yields $\log_2(\Omega/Q_x) - 4 < I_R(x : \vec{W}) < \log_2(\Omega/Q_x) + 4$ which can be written as

$$\left| I_R(x : \vec{W}) - \log_2(\Omega/Q_x) \right| < 4. \tag{S8.24}$$

---

[d]The *focal type* is the organismal type of an individual that has been sampled (or observed) in a particular environment.

**Overall conclusions**

We have established that both forms of reproductive information ($I_R(\vec{W}:x)$ and $I_R(x:\vec{W})$) differ from the quantitative adaptedness, $\log_2(\Omega/Q_x)$, by less than 4 bits. This means that $I_R(\vec{W}:x)$ and $I_R(x:\vec{W})$ differ from each other by less than 8 bits. With additional work it should be possible to establish more refined bounds that show that $I_R(\vec{W}:x)$ and $I_R(x:\vec{W})$ differ by less than 8 bits.

# Supplementary Note 9: A method for classifying fitness landscapes plus related observations

In this Supplementary Note we specify a methodology for classifying fitness landscapes. We also make a number of quantitative observations that emerge from this methodology.

To begin, we note that the fitness landscape, $\vec{W}$, is a vector of real numbers that are equal to the fitnesses of the $\Omega$ organismal types. We observe that the predictions of natural selection depend on the *relative* fitness of various types of organisms[6,38–40]. With this in mind, we shall characterise fitness landscapes in terms of which organismal types are the most fit (i.e., lie at the first rank in fitness), which types lie at the second rank in fitness, which types lie at the third rank, and so forth.

To describe a scheme for classifying fitness landscapes, we consider all possible ways to order the $\Omega$ organismal-type indices. As noted above, there are a total of $\Omega!$ such orderings, each of which can be represented by an ordered list of organismal-type indices. These ordered lists can be arranged in lexicographic order. The position of a particular list in this lexicographic arrangement of orderings will be denoted by the value of $e$, which we call the *ordering index*, and which can take the values $1, 2, \ldots, \Omega!$. As we shall see, the ordering indices can be used to describe the conditions within environments. An example of the assignment of ordering indices can be seen in Table 1.

Lexicographic arrangement of the possible orderings of organismal types when $\Omega = 4$.

| ordering index, $e$ | | | | | | | | | | | | | | | | | | | | | | | |
|---|---|---|---|---|---|---|---|---|---|---|---|---|---|---|---|---|---|---|---|---|---|---|---|
| 1 | 2 | 3 | 4 | 5 | 6 | 7 | 8 | 9 | 10 | 11 | 12 | 13 | 14 | 15 | 16 | 17 | 18 | 19 | 20 | 21 | 22 | 23 | 24 |
| **order of types:** | | | | | | | | | | | | | | | | | | | | | | | |
| 1 | 1 | 1 | 1 | 1 | 1 | 2 | 2 | 2 | 2 | 2 | 2 | 3 | 3 | 3 | 3 | 3 | 3 | 4 | 4 | 4 | 4 | 4 | 4 |
| 2 | 2 | 3 | 3 | 4 | 4 | 1 | 1 | 3 | 3 | 4 | 4 | 1 | 1 | 2 | 2 | 4 | 4 | 1 | 1 | 2 | 2 | 3 | 3 |
| 3 | 4 | 2 | 4 | 2 | 3 | 3 | 4 | 1 | 4 | 1 | 3 | 2 | 4 | 1 | 4 | 1 | 2 | 2 | 3 | 1 | 3 | 1 | 2 |
| 4 | 3 | 4 | 2 | 3 | 2 | 4 | 3 | 4 | 1 | 3 | 1 | 4 | 2 | 4 | 1 | 2 | 1 | 3 | 2 | 3 | 1 | 2 | 1 |

Table 1: This table covers the case of $\Omega = 4$ organismal types. The table shows all $4! = 24$ possible orderings of organismal types, written in lexicographic order. In the table, the ordering index, $e$, associated with each ordering of organismal types is also shown.

We can use the ordering indices to classify any fitness landscape, given the fitness of each organismal type (i.e., given all components of $\overrightarrow{W}$). To see how this can be done, let us begin by assuming that, in a given environment, each of the $\Omega$ organismal types has a unique fitness. (That is, no two organismal types have the same fitness.) In this situation there is exactly one way to order the organismal types from most fit to least fit. This particular ordering is just one of the $\Omega!$ possible orderings of organismal types. We classify the fitness landscape by simply assigning to it the ordering index (the value of $e$) that is associated with this particular ordering of organismal types. In this case the ordering index provides information about the fitness landscape because it indicates, for any two organismal types, which is fitter, and which is less fit.

Next, consider a fitness landscape for which multiple organismal types have exactly the same fitness. In this case, to classify the fitness landscape we arrange the organismal types so that in no case do we have two organismal types with different fitnesses where the one with the lower fitness comes first in the ordering. That is, the ordering goes from most fit to least fit, and there may be some organismal types that have the same fitness as another organismal type that is next to them in the ordering. Of course, as there are multiple types that are equal in fitness, there must be multiple orderings that can emerge from this procedure. For example, if only two organismal types are equal in fitness, then there are two orderings that can emerge. We regard any ordering that can emerge from this procedure to be a correct ordering to characterise the fitness landscape in question. Thus, we can use the ordering index for any of these correct orderings as a descriptor for that fitness landscape (i.e., as a value of $e$ that correctly applies to that fitness landscape). Of course, if multiple organismal types share the same fitness value for some fitness landscape, then there will be multiple ordering indices that can be correctly used to classify that fitness landscape. Indeed, for a fitness landscape in which *all* of the organismal types have exactly the same fitness, all of the $\Omega!$ ordering indices are applicable.

Even when there are multiple types that have the same fitness, an ordering index that can be correctly associated with a fitness landscape provides some information about the nature of that fitness landscape. In particular, we know that the organismal types that come first in the ordering associated with the ordering index will always be at least as fit as those that come later in the ordering.

Ordering indices are integers that can be used to provide information about a fitness landscape. Let us now consider how we might encode the ordering indices using an efficient digital code. An ordering index, $e$, can take the values $1, 2, \ldots, \Omega!$. We shall use the first $\Omega!$ binary codewords (in length-lexicographic order) to encode these ordering indices. Using this choice of binary codewords is maximally efficient, compared other choices of binary code words, since it tends to minimise the length of the code words.

There are $(\Omega!)!$ different ways to assign the first $\Omega!$ binary code words to the $\Omega!$ ordering indices, such that, when the assignment has been done, each code word is uniquely assigned to one ordering index. Each of these $(\Omega!)!$ assignments constitutes a different encoding of the ordering indices. If we ignore the theory of natural selection (and all other theories), then we have no hypothesis about which fitness landscapes are most likely, and thus there is no reason to prefer any of these encodings. As such, in this theory-free context, any of the first $\Omega!$ binary code words can be used to encode any of the ordering indices. Thus, in the absence of a theory, the average length of all the code words that could (under the system described here) be assigned to any particular ordering index is simply the mean of the length of the first $\Omega!$ binary code words. We will use $\bar{L}'_u$ to refer to this average length. Thus, $\bar{L}'_u$ can be considered to be the expected code-word length for an ordering index when we do not take the theory of natural selection (or any other theory) into account.

Let us assume that we have observed the organismal type of an individual that was selected at random from a population living in some natural environment. We will call this individual the *focal individual*, and we will call its organismal type the *focal type*. The identity of the focal type, along with the theory of natural selection, suggests that certain ways of encoding the ordering indices should be preferred because, if the theory correctly predicts the association between fitness landscapes and organismal types, then the preferred encodings will tend to lead to shorter code words being used to describe the fitness landscape that characterises the environment in which the focal individual was found.

In order to find an encoding of the ordering indices that is optimal in light of the theory of natural selection, we must decide which binary code words to assign to which of the ordering indices. Recall that the first of the binary code words (in length-lexicographic order) are the shortest. Thus, since the theory of natural selection tells us that randomly selected organisms will tend to have relatively fit organismal types, we should use the first binary code words to encode ordering indices associated with orderings in which the focal type comes first. There are $(\Omega - 1)!$ orderings of organismal types where the focal type come first, and thus, for an optimal encoding of ordering indices, the ordering indices associated with these $(\Omega - 1)!$ orderings are assigned to the first $(\Omega - 1)!$ binary code words. Of course, there are multiple ways in which these assignments can be made.

Logic similar to the foregoing suggests that, in an optimal encoding of the ordering indices, the binary code word that is in position $(\Omega - 1)! + 1$ in a length-lexicographic list of binary code words should be assigned to an ordering index for an ordering of the organismal types in which the focal type comes second in the ordering. There are $(\Omega - 1)!$ of these ordering indices, and so these $(\Omega - 1)!$ ordering indices should be assigned to the binary codewords that come between position $(\Omega - 1)! + 1$ and position $2(\Omega - 1)!$ in a length-lexicographic list of binary code words. The next $(\Omega - 1)!$ binary code words should be assigned to ordering indices that are associated with orderings in which the focal type comes third, and so on.

It is straightforward to show that, of the $(\Omega!)!$ possible ways to assign the first $\Omega!$ binary codes to ordering indices, a total of $[(\Omega - 1)!]^\Omega$ of them are optimal, given the theory of natural selection and the identity of the focal type. All other coding schemes are non-optimal. As an example, when $\Omega = 4$ there are 4! = 24 possible orderings of the organismal types, and thus there are 24 ordering indices. There are 24! ≈ 6.2 × 10$^{23}$ ways to assign one of the first 24 binary code words to each of these 24 ordering indices. Of these (approximately) 6.2 × 10$^{23}$ ways of assigning code words to ordering indices, a total of $(3!)^4 = 1296$ of them are optimal in light of the identity of the focal type, and the theory of natural selection. Table 2 shows one of the 1296 optimal assignments of binary code words to ordering indices, under the assumption that the focal individual has an organismal-type index that is equal to three.

An optimal assignment of the first 24 binary code words to ordering indices when $\Omega = 4$

| | binary code word | | | | | | | | | | | | | | | | | | | | | | | |
|---|---|---|---|---|---|---|---|---|---|---|---|---|---|---|---|---|---|---|---|---|---|---|---|---|
| | 0 | 1 | 00 | 01 | 10 | 11 | 000 | 001 | 010 | 011 | 100 | 101 | 110 | 111 | 0000 | 0001 | 0010 | 0011 | 0100 | 0101 | 0110 | 0111 | 1000 | 1001 |
| OI | 18 | 13 | 15 | 17 | 14 | 16 | 4 | 23 | 9 | 3 | 24 | 10 | 12 | 20 | 1 | 22 | 6 | 7 | 8 | 2 | 11 | 21 | 5 | 19 |
| | 3 | 3 | 3 | 3 | 3 | 3 | 1 | 4 | 2 | 1 | 4 | 2 | 2 | 4 | 1 | 4 | 1 | 2 | 2 | 1 | 2 | 4 | 1 | 4 |
| | 4 | 1 | 2 | 4 | 1 | 2 | 3 | 3 | 3 | 3 | 3 | 3 | 4 | 1 | 2 | 2 | 4 | 1 | 1 | 2 | 4 | 2 | 4 | 1 |
| OT | 2 | 2 | 1 | 1 | 4 | 4 | 4 | 1 | 1 | 2 | 2 | 4 | 3 | 3 | 3 | 3 | 3 | 3 | 4 | 4 | 1 | 1 | 2 | 2 |
| | 1 | 4 | 4 | 2 | 2 | 1 | 2 | 2 | 4 | 4 | 1 | 1 | 1 | 2 | 4 | 1 | 2 | 4 | 3 | 3 | 3 | 3 | 3 | 3 |

Table 2: For this table, we assume that the organismal-type index of the focal individual is three. The table shows one of the 1296 possible optimal assignments of the first 24 binary code words to the *ordering indices* (OI). For convenience, the orderings of the *organismal types* (OT) that are associated with each of the ordering indices are also shown.

# Supplementary Note 10: Reproductive information and Shannon information

The best-known quantitative expression of the concept of information is the quantity we call Shannon information. The key publication that defined Shannon information and explored its potential was published by Shannon in 1948[9]. Shannon information has proved to have great utility in practical situations, and it has played a key role in various scientific and engineering endeavours[34,35,52,53]. However, despite its power and utility, Shannon information has features that make it impractical as a measure of the information that is created by natural selection. This will be explained in the present section. We will also explore the quantitative relationship between reproductive information and Shannon information.

Shannon information uses probability distributions to measure the extent to which the state of one variable tells us something about the state of another variable. To see how this works in a biological context, let us return to the case of a population of organisms that have $\Omega$ different possible organismal types. For each organism we will assume we know their organismal type index, $x$. This is an integer that satisfies $1 \leq x \leq \Omega$. We will also assume that we know $e$, the type of the environment in which that organism lives. As explained in Supplementary Note 9, $e$ is an integer that satisfies $1 \leq e \leq \Omega!$. As also explained in Supplementary Note 9, when multiple organismal types have the same value of fitness in a particular environment, there will be multiple values of $e$ that can correctly be used to label that environment.

For the purposes of this Supplementary Note, we will sometimes treat $x$ and $e$ as *realisations* of random variables $X$ and $E$, respectively.

For the first part of this Supplementary Note we simplify matters by assuming that every organismal type has a *unique fitness value*. This assumption applies throughout the present Supplementary Note, except in the last section, within which we will explicitly relax this assumption. The assumption that each organismal type has a unique fitness value simplifies some of the analysis we present in this Supplementary Note. However, in general, we *do not* make this assumption.

Note that our current assumptions imply that each individual is characterised by a single value of $x$ (which indicates the individual's organismal type) and a single value of $e$ (which indicates the individual's environment, as characterised by ordering of fitnesses within the environment).

Let us assume that our population lives in a complex habitat, where different individuals may experience different environments, as characterised by different fitness landscapes. Let $p_{E,X}(e,x)$ represent the probability that a randomly picked member of the population has organismal type $x$, and lives in an environment of type $e$. Furthermore, let $p_E(e)$ be the probability that a randomly picked member of population lives in an environment of type $e$. Thus, $p_E(e) = \sum_{x=1}^{\Omega} p_{E,X}(e,x)$. Finally, let $p_X(x)$ be the probability of a randomly picked member of the population has organismal type $x$. Thus, $p_X(x) = \sum_{e=1}^{\Omega!} p_{E,X}(e,x)$. Using these definitions, we can now define a quantity which, in the world of Shannon information, has been used to characterise the extent to which $e$ provides information about $x$, and vice versa. This quantity is sometimes called *pointwise mutual information*, and we use $f(e,x)$ to represent it[54,55]. We have

$$f(e,x) = \log_2 \left( \frac{p_{E,X}(e,x)}{p_E(e) p_X(x)} \right). \tag{S10.1}$$

There are a number of mathematically equivalent ways to define Shannon information. One of these is in terms of $f(e,x)$. In particular, Shannon information, which we denote by $I_S(\vec{W}; X)$, is simply the mean value of $f(e,x)$. That is

$$I_S(\vec{W}; X) = \sum_{e=1}^{\Omega!} \sum_{x=1}^{\Omega} p_{E,X}(e,x) f(e,x). \tag{S10.2}$$

## Can we use pointwise mutual information to measure the information created by natural selection?

Shannon information itself cannot serve to measure reproductive information because it is not measured on individual organisms and their environments. Thus, it cannot tell us how much information the form of a *particular* organism embodies about a *particular* environment. However, Shannon information is the mean of pointwise mutual information, which relates more directly to particular organisms and particular environments. Thus, we might hope that pointwise mutual information could serve as a measure of the information created by natural selection. However, as we shall now see, there are several difficulties associated with this approach.

The first problem with using $f(e,x)$ to measure the information created by natural selection is that $f(e,x)$ does not have a natural and precise interpretation in terms of the length of code words (or, at least, no such interpretation exists for all possible values of $p_{E,X}(e,x)$, $p_E(e)$, and $p_X(x)$). Secondly, $f(e,x)$ depends on the values of $p_E(e)$ and $p_X(x)$. This might seem odd. One might think, for example, that it is odd that the information that a particular fitness landscape conveys about a particular organismal type depends strongly on the how common that type of fitness landscape is among environments. It seems reasonable that the information in a recipe for a particular type of pie should be independent of how often that recipe is printed. Thus it might also seem reasonable that the information that a particular fitness landscape provides about an organismal type should be independent of how common that sort of fitness landscape is within the habitat.

A third problem with using $f(e,x)$ to measure the information created by natural selection is, essentially, identical to a problem regarding Shannon Information that was mentioned in the main text. In particular, $f(e,x)$ does not indicate whether a particular association is in line with the predictions of natural selection. The fact that $f(e,x)$ shares this shortcoming with Shannon Information is, of course, not surprising, given that, as we have seen, Shannon Information is the mean of $f(e,x)$. To understand this problem more clearly, it helps to consider a situation in which $p_E(e) > p_X(x)$. In this case, if we do not change $p_E(e)$ or $p_X(x)$, then $f(e,x)$ is maximised when $p_{E,X}(e,x)$ takes on its largest-possible value, which is $p_X(x)$ (i.e., maximisation happens when $p_{E,X}(e,x) = p_X(x)$). This means that every population member with organismal-type $x$ occurs in an environment of type $e$. The problem here is that the maximum-possible value of $f(e,x)$ obtains in this case regardless of whether type-$x$ individuals have the fittest possible organismal type in environments of type $e$, or if they are the least fit organismal type in environments of type $e$. Thus, the value of $f(e,x)$ does not tell us whether the theory of natural selection provides accurate predictions about where different types of organisms will be found. A similar analysis applies when $p_E(e) \leq p_X(x)$.

A final problem with using $f(e,x)$ to measure the information created by natural selection arises due to the finite nature of real-world populations. In particular, examination of equation (S10.1) reveals that $f(e,x)$ can never be greater than $\log_2(N)$, where $N$ is the total number of individuals living in the collection of environments of all types[e]. This is problematic because, with typical organisms having genomes consisting of millions or billions of base pairs, we expect that the reproductive information associated with these genomes could potentially be many thousands or millions of bits. However, even if we had one individual for every atom in the visible universe ($\sim 10^{80}$), then the highest-possible value of $f(e, x)$ would be less than 300 bits. This suggests that calculating $f(e, x)$ from data derived from real populations is very unlikely to provide satisfactory results.

---
[e]If, for particular values of $e$ and $x$, we have $p_{E,X}(e,x) > 0$, then we must have $p_X(x) > 0$ and $p_E(e) > 0$. However, an empirical estimate of $p_X(x)$ cannot be less than $1/N$, which occurs when there is only one individual with organismal type $x$. Similarly, an empirical estimate of $p_E(e)$ also cannot be less than $1/N$.

**The correspondence between Shannon information and reproductive information when organismal-type frequencies are strictly in line with the predictions of natural selection**

Physical limitations on population size need not prevent us from investigating hypothetical worlds in which such limitations do not apply. In order to explore this further, it is helpful to recall the concept of *entropy*, which is, perhaps, the most fundamental quantity that is related to Shannon information[9,34,35]. Let us begin by focussing on the entropy associated with the distribution of organismal types for individuals that live in an environment of type $e$. Let $p_{X|E}(x|e)$ represent the probability of finding organismal type $x$ as a result of a random selection among these individuals. That is, $p_{X|E}(x|e) = p_{E,X}(e,x)/p_E(e)$. We will use $H_e(X)$ to denote the entropy in organismal type among individuals that live in an environment of type $e$. The value of $H_e(X)$ is given by

$$H_e(X) = \sum_{x=1}^{\Omega} p_{X|E}(x|e) \log_2 \left( \frac{1}{p_{X|E}(x|e)} \right). \quad (S10.3)$$

One reason for the importance of entropy is that, to a close approximation, $H_e(X)$ is equal to the minimum-possible average length for a binary *prefix-free code* that encodes the organismal types of individuals that have environmental label $e$[34,35]. A binary prefix-free code uses only two symbols, and has the property that no code word is a prefix of any other code word. Thus, for example, if we have a prefix-free code, then, if 010 is the code word for one organismal type, then 01011 cannot be a code word for any other organismal type - because is 010 a prefix of 01011. Note that the coding we have specified for the calculation of reproductive information is *not* a prefix-free code, because it uses all of the binary sequences that are shorter than a particular length, and some of these sequences will be prefixes of others (as long as $\Omega > 2$).

In the definition of reproductive information, we described a set of coding schemes for organismal types that are optimal, given the theory of natural selection. We also defined $\overline{L}_s$ to be the mean length of the code words that code for a particular organismal type, if we average over all optimal coding schemes. (In this case, 'optimality' means that the coding scheme tends to minimise average code-word length when the state of the population is in accord with the predictions of the theory of natural selection.) With this in mind, it would not be surprising to find that, at least in some circumstances, there is a close relationship between entropy (which gives the minimal average length of a prefix-free code) and reproductive information. In fact, as we shall now see, such a relationship exists.

Let us assume that $\Omega \geq 2$, so that there are at least two possible organismal types. Assume further that, among individuals living in environments of type $e$, the distribution of organismal types is in line with the predictions of natural selection. In particular, assume that, among these individuals, there are no two organismal types for which the fitter one is less common (or probable) than the less-fit one. (This second assumption, that frequencies of organismal types follow the prediction of natural selection, will be relaxed in the last section of this Supplementary Note.)

Under the circumstances just described, we find that $H_e(X)$ - the entropy of organismal type among individuals living in environments of type $e$ - must be relatively close to the mean value of $\overline{L}_s$ among individuals living in environments of type $e$ ($\overline{L}_s$ features in the definition of reproductive information, in equation (1) in the main text). In particular, let us use $\overline{\overline{L}}_{s,e}$ to denote the mean value of $\overline{L}_s$ among individuals living in environments of type $e$. With this notation, we can now express the following result (which is proved in Supplementary Note 17):

$$\overline{\overline{L}}_{s,e} - 1 \leq H_e(X) \leq \overline{\overline{L}}_{s,e} + 2\log_2(\log_2(\Omega)) + 5. \tag{S10.4}$$

Thus, for example, if we characterise organismal type in terms of genotype, and if we have a bacterium with a haploid genome of five million base pairs, then $H_e(X)$ must be between zero and ten million bits, depending on how genetically diverse the population is. Whatever the value of $H_e(X)$, the mean value of $\overline{L}_s$ among individuals for which $E = e$ (namely $\overline{\overline{L}}_{s,e}$) is guaranteed to be close to it. In particular, the absolute difference between $H_e(X)$ and $\overline{\overline{L}}_{s,e}$ cannot be larger than 52 bits. If, on the other hand, we consider the genome of a mammalian gamete (i.e., a haploid genome) with three billion base pairs, then the maximum entropy is about six billion bits, and the absolute difference between $H_e(X)$ and $\overline{\overline{L}}_{s,e}$ cannot be more than 70 bits.

Using equation S10.4 we can also show that, when organismal-type frequencies are in line with the theory of natural selection, there is a close relationship between the quantity $\log_2(\Omega) - H_e(X)$ and the average level of reproductive information among individuals for which $E = e$. Let $\overline{I_{R,e}(\overrightarrow{W}:x)}$ represent this average. That is to say, $\overline{I_{R,e}(\overrightarrow{W}:x)}$ is the mean level of reproductive information that the state of the fitness landscape provides about organismal type for individuals that live in environments of type $e$. As proven in Supplementary Note 17, the value of $\overline{I_{R,e}(\overrightarrow{W}:x)}$ must lie within the following limits

$$[\log_2(\Omega) - H_e(X)] - 4 < \overline{I_{R,e}(\overrightarrow{W}:x)} < [\log_2(\Omega) - H_e(X)] + 2\log_2(\log_2(\Omega)) + 6. \tag{S10.5}$$

Thus, for example, for a mammal with a gametic genome consisting of three billion base pairs, the absolute difference between $\overline{I_{R,e}(\overrightarrow{W}:x)}$ and $[\log_2(\Omega) - H_e(X)]$ cannot be greater than 71 bits.

Thus far, we have been discussing the relationship between reproductive information and the entropy in organismal type among individuals that all live in environments of one particular type (i.e., in environments characterised by one particular value of $e$). Let us now turn to the relationship between reproductive information and Shannon information. To do this, we shall consider the entire population, and not just those population members living in environments of one particular type. We shall also extend our assumption, so that the frequencies of organismal types are assumed to be in line with the theory of natural selection in environments of *all* possible types (i.e., for *all* possible values of $e$).

As noted above, $H_e(X)$ is the entropy in organismal type for individuals living in environments of type $e$. Let us now consider the entropy in *the entire* population, including individuals living in environments of *any* type. This is known as the *unconditional entropy*, and we shall write it as $H(X)$. In parallel with equation (S10.3) (which defines $H_e(X)$), the definition of $H(X)$ is

$$H(X) = \sum_{x=1}^{\Omega} p_X(x) \log_2\left(\frac{1}{p_X(x)}\right), \quad (S10.6)$$

where, as above, $p_X(x) = \sum_{e=1}^{\Omega!} p_{E,X}(e,x)$.

Using $H(X)$ we can show that, under our assumption that organismal-type frequencies are in line with the theory of natural selection, there is a close relationship between reproductive information and Shannon information. To show this, let $\overline{I_R(\vec{W}:x)}$ denote the average reproductive information that the environment provides about an individual's organismal type (as defined by equation (1) in the main text). In this case, the average is over all individuals living in environments of all possible types. That is

$$\overline{I_R(\vec{W}:x)} = \sum_{e=1}^{\Omega!} \sum_{x=1}^{\Omega} p_{E,X}(e,x) I_R(\vec{W}:x). \quad (S10.7)$$

Recall that $I_S(\vec{W};X)$ represents the Shannon information between organismal type and environmental type. As shown in Supplementary Note 17, we have

$$I_S(\vec{W};X) + \log_2(\Omega) - H(X) - 4 \;<\; \overline{I_R(\vec{W}:x)} \;<\; I_S(\vec{W};X) + \log_2(\Omega)$$
(S10.8)
$$-H(X) + 2\log_2\left(\log_2(\Omega)\right) + 6.$$

Thus, for a haploid genome with three billion base pairs, the absolute difference between $\overline{I_R(\vec{W}:x)}$ and $I_S(\vec{W};X) + \log_2(\Omega) - H(X)$ cannot be greater than 71 bits.

In general, the value of $H(X)$ will depend on a variety of factors, including the probability distribution of different environmental types. It is important to note that $H(X)$ can have a profound effect on Shannon information. To see this, let us define $\overline{H_e(X)}$ to be the mean value, across environmental types, of $H_e(X)$. That is

$$\overline{H_e(X)} = \sum_{e=1}^{\Omega!} p_E(e) H_e(X) \quad (S10.9)$$

where $p_E(e) = \sum_{x=1}^{\Omega} p_{E,X}(e,x)$. We call $\overline{H_e(X)}$ the *conditional entropy* of $X$. It can be shown[34] that, in addition to the formulation given by equation (S10.2), an equivalent expression for the Shannon information is

$$I_S(\vec{W};X) = H(X) - \overline{H_e(X)}. \quad (S10.10)$$

From this we can see that, for any given value of $\overline{H_e(X)}$, Shannon information is maximised when $H(X)$ takes on its maximum-possible value. This maximal value is $H(X) = \log_2(\Omega)$, and is achieved when, within the entire population (including all environmental types), every organismal type is equal in frequency (so that $p_X(x) = 1/\Omega$ for all possible values of $x$). When Shannon information is maximised in this way, inequality (S10.8) simplifies and we obtain

$$I_S(\vec{W};X) - 4 < \overline{I_R(\vec{W}:x)} < I_S(\vec{W};X) + 2\log_2\left(\log_2(\Omega)\right) + 6. \quad (S10.11)$$

Thus, in this case, for a genome with three billion base pairs, the absolute difference between the average reproductive information (i.e., $\overline{I_R(\vec{W}:x)}$) and Shannon information (i.e.; $I_S(\vec{W};X)$) cannot be greater than 71 bits. This suggests that, when the frequencies of the various organismal types are in line with the predictions of the theory of natural selection, we can think of the average value of reproductive information as a 'best guess' with regard to the maximum-possible value of Shannon information (that is, maximal given the conditional entropy).

## The relationship between Shannon information and reproductive information in the absence of restrictive assumptions

So far we have considered populations in which the frequencies of organismal types are in line with the predictions of natural selection. We have also assumed that, in every environment, every organismal type has a unique fitness value. In real populations these assumptions are unlikely to be met. For example, in any given environment, many organismal types are likely to be totally incapable of survival and/or reproduction, and so many organismal types will share a fitness value of zero. Also, in many cases, the number of organismal types ($\Omega$) will be very large, and thus it would be unsurprising if many of the viable organismal types were very close to each other in fitness. As a result, 'random genetic drift' is likely to play a large role in determining the relative frequencies of the various organismal types, and so frequency distributions that are not exactly in line with the predictions of natural selection should be common.

Let us now eliminate the assumption that the frequency of organismal types is in line with the predictions of the theory of natural selection. We will also eliminate the assumption that every organismal type has a unique fitness. Instead, let us consider a much more realistic situation. Let $I_R^*$ represent a positive real number. Furthermore, let $\Phi$ represent the fraction of the population that live in an environment in which the proportion of organisms for which $I_R(\vec{W}:x) > I_R^*$ is at least equal to $\Theta$, where $0 < \Theta \leq 1$. In other words, we will assume that a proportion of the population equal to $\Phi$ lives in an environment of a type for which the proportion of individuals with associated reproductive information ($I_R(\vec{W}:x)$) in excess of $I_R^*$ is greater than (or equal to) $\Theta$.

To make the preceding assumptions more clear, it helps to consider extreme (and therefore simple) cases. For example, if $\Phi = 1$, then, under our assumptions, this means that, in *every* environment, the proportion of individuals associated with a level of reproductive information ($I_R(\vec{W}:x)$) that is in excess of $I_R^*$ is, at minimum, equal to $\Theta$. On the other hand, if $\Theta = 1$ then $\Phi$ is the proportion of the population that is living in an environment in which *all* population members are associated with a value of reproductive information that is in excess of $I_R^*$. Finally, if $\Phi = 1$ *and* $\Theta = 1$, then *all* individuals in *every* environment are associated with a level of reproductive information ($I_R(\vec{W}:x)$) that is in excess of $I_R^*$.

Under our current assumptions, we have the following result, which is proved in Supplementary

Note 17:

$$I_S(\overrightarrow{W}; X) > H(X) - \log_2(\Omega) + \Theta\Phi I_R^* - 5. \quad (S10.12)$$

The right side of this inequality represents a lower bound on the Shannon information between the fitness landscape and organismal type. This lower bound is an increasing function of $H(X)$, the entropy in organismal type in the entire population (including all environmental types). The lower bound reaches its maximum-possible value (with respect to $H(X)$) in the symmetric situation mentioned above, where $p_X(x) = 1/\Omega$ for all organismal types. In this situation we have

$$I_S(\overrightarrow{W}; X) > \Theta\Phi I_R^* - 5. \quad (S10.13)$$

The lower bound on Shannon information represented by inequality (S10.13) is maximised with respect to $\Theta$ and $\Phi$ when these two quantities are equal to unity. When $\Theta = 1$ and $\Phi = 1$, all populations members have a value for reproductive information ($I_R(\overrightarrow{W} : x)$) that is in excess of $I_R^*$. If $\Theta = 1$, $\Phi = 1$, and, in addition, $p_X(x) = 1/\Omega$ for all organismal types, then Shannon information cannot be more than five bits less than $I_R^*$. This demonstrates, once again, that Shannon information and reproductive information are closely related quantities.

# Supplementary Note 11: Why reproductive information is more practical than Kolmogorov information

The main text says that reproductive information is practical in a way that Kolmogorov information is not. One reason for this is that Kolmogorov information is based on the length of the shortest-possible description (or encoding) of some object, and it can be proved that there is no general way to find such a description [34,35]. On the other hand, reproductive information uses a specific "effective method" to find an encoding (or a set of encodings) for an object [35]. This method is described in the main text. It is "effective" in that, unlike Kolmogorov information, the calculation of reproductive information can always, in principle, be carried out, given access to the necessary data. Note that this method depends on the assumption that, while the number of organismal types may be enormous, it is finite. Thus, this assumption (which is not generally made by treatments of Kolmogorov information) is another reason why reproductive information is practical in a sense that Kolmogorov information is not.

Restricting attention to cases where the number of organismal types is finite also serves to address another problem that is associated with Kolmogorov information. This is the fact that Kolmogorov information depends on an arbitrary choice of a method of encoding (which is sometimes called a "description language") [34,35]. To see how this works, consider the calculation of $\bar{L}_u$, which is used in the definition of reproductive information. The value of $\bar{L}_u$ represents the mean length of the code words that could be used to encode a particular organismal type, if we include all efficient coding schemes (i.e., both "optimal" and "non-optimal" coding schemes). (Here, "efficient" means that only the shortest $\Omega$ binary code words are used in the coding scheme.) The issue of choosing a description-method raises its head in this situation in that the length of the code word for a particular organismal type depends strongly on which coding scheme one uses. Specifically, the length of this code word can be anything between 1 and (approximately) $\log_2(\Omega)$ binary digits in length. However, because the number of organismal types is finite, the number of efficient coding schemes is also finite, and thus, we are able to specify that to calculate $\bar{L}_u$ one must average over *all* possible efficient coding schemes. This leads to a single value of $\bar{L}_u$, and no choice of a particular coding scheme is required. Something very similar happens in the case of $\bar{L}_s$. Thus, using a finite number of organismal types avoids the problem of having to choose an arbitrary coding scheme (or "description language").

## Supplementary Note 12: A case in which Kolmogorov information is much larger than reproductive information

In Supplementary Note 4 we consider a situation in which there is at least one organismal type for which the following inequality is satisfied in the limit as $\Omega \to \infty$

$$\varepsilon_1 \log_2(\Omega) < I_R(\overrightarrow{W}:x) < (1-\varepsilon_2) \log_2(\Omega). \tag{S12.1}$$

Here $\varepsilon_1$ and $\varepsilon_2$ are arbitrarily chosen small and positive real numbers (that is, $0 < \varepsilon_1, \varepsilon_2 \ll 1$). We show that, in this limiting case, among qualifying types (which means organismal types for which $\varepsilon_1 \log_2(\Omega) < I_R(\overrightarrow{W}:x)$), the proportion of organismal types for which Kolmogorov information is substantially different from reproductive information must be infinitesimal. In this note we will provide an example of one of these extremely rare organismal types.

Let us assume that $\Omega$ is very large, and that we have an organismal type for which reproductive information, $I_R(\overrightarrow{W}:x)$, is approximately 20% of $\log_2(\Omega)$ (i.e., $I_R(\overrightarrow{W}:x) \simeq 0.2 \log_2(\Omega)$). Assume further that, for this organismal type, the expected number of offspring (i.e., fitness) is equal to 3.7, and no other organismal type has this fitness. As explained in Supplementary Note 4, the Kolmogorov information, $K(\overrightarrow{W}:x)$, is defined by

$$K(\overrightarrow{W}:x) = K(x) - K(x|\overrightarrow{W}). \tag{S12.2}$$

where $K(x)$ is the algorithmic complexity of $x$ for an arbitrarily selected hypothetical computer that does not have access to the fitness landscape, and $K(x|\overrightarrow{W}))$ is the algorithmic complexity for the same arbitrarily selected hypothetical computer when it *does* have access to the fitness landscape. (Note that this computers is assumed to be "additively optimal" and "universal"[35].) It is a basic result of algorithmic information theory[35] that, for the overwhelming majority of integers in the range $[0, \Omega]$, the algorithmic complexity is equal to $\log_2(\Omega) + O(1)$. As $x$ is one of these integers, let us assume that $K(x) = \log_2(\Omega) + O(1)$.

The algorithmic complexity of $x$ for a computer that has access to the fitness landscape, $K(x|\vec{W})$, is defined as the length of the shortest binary program that, when provided as input to this computer, will result in the output of the value of $x$. We have assumed that the fitness of the organismal type designated by $x$ has a fitness of 3.7, and that no other type has the same fitness. This implies that a computer with access to the fitness landscape can output $x$ upon input of a program with a length that is of order one ($O(1)$). In particular, this program would search through the fitness values provided by the fitness landscape until it finds a value of $x$ that is associated with a fitness of 3.7. The program can then simply output the value of $x$ that it has found. These observations imply that, in this case, $K(x|\vec{W}) = O(1)$. We have already assumed that $K(x) = \log_2(\Omega) + O(1)$. Thus, $K(\vec{W}:x) = K(x) - K(x|\vec{W}) = \log_2(\Omega) + O(1)$.

We have assumed that $I_R(\vec{W}:x) \simeq 0.2 \log_2(\Omega)$. Thus, in the case under study here, we can now express the ratio of Kolmogorov information to reproductive information as:

$$\frac{K(\vec{W}:x)}{I_R(\vec{W}:x)} \simeq \frac{\log_2(\Omega) + O(1)}{0.2 \log_2(\Omega)}. \tag{S12.3}$$

We have assumed that $\Omega$ is very large, and thus, in comparison to $\log_2(\Omega)$, terms of $O(1)$ are negligible. We therefore have

$$\frac{K(\vec{W}:x)}{I_R(\vec{W}:x)} \simeq 5. \tag{S12.4}$$

Thus, we have now shown that the assumptions made here lead to a case in which reproductive information is very large, but Kolmogorov information is substantially larger.

# Supplementary Note 13: Calculating reproductive information using continuously distributed phenotypic traits

In this Supplementary Note we will consider how continuously distributed traits can be used to calculate reproductive information. We shall also observe that there is a technical problem that arises in the course of these calculations, and, finally, we will indicate how some recent results appear to provide a solution to this problem.

To begin, let us consider a simple example. Assume that there is a terrestrial animal for which the males make a mating call, and that the optimal intensity (or loudness) of this call is 20 decibels (20 $dB$) when measured at a distance of one kilometre. We will assume that calls more intense than 120 $dB$ are practically impossible. Accordingly, we will compare all possible intensity values that lie between zero decibels and 120 $dB$. For simplicity, let us assume that the fitness of males depends only on the intensity of their mating calls, and that selection on sound intensity is "stabilising", with fitness being determined by a function that is symmetric about the optimum, as illustrated in Fig. S13.1, panel ($a$). Next, consider a male that produces a mating call with an intensity of 19.9 $dB$. What is the reproductive information that is associated with this individual's phenotype?

To answer this question, it is important to recognise that, even though the number of possible sound-intensity levels that lie between zero and 120 $dB$) is infinite, we can calculate the length of the line that contains these values. The length of this line is, simply, 120 $dB$. Furthermore, we can calculate the length of the line that contains all organismal types that are at least as fit as a male that produces a mating call with a sound intensity of 19.9 $dB$. This second line has a length of 0.2 $dB$, as illustrated in Fig. S13.1, panel ($b$). If we consider equation (2) in the main text, then it may seem that we can use the ratio of the lengths of these two lines to produce an estimate of the reproductive information associated with a male that produces a mating call with a sound intensity of 19.9 $dB$. In particular, the ratio in question is $\left(\frac{0.2}{120}\right)$, and we can insert its reciprocal into equation (2) to obtain an estimate for reproductive information ($I_R$). This estimate is $I_R \approx \log_2\left(\frac{120}{0.2}\right) = 9.229$ bits.

However, a problem with the foregoing procedure arises if we consider the details of how sound intensity is measured in decibels[56]. There are a number of very similar definitions that may be used, but here we will take the decibel to be a logarithmic transformation of *acoustic energy*. Acoustic energy, in turn, can be defined as the energy that is being received from a sound source, as measured in watts per square meter ($W/m^2$). We can use acoustic energy to measure loudness (or sound intensity), instead of decibels (see Supplementary Note 18). However, doing this leads to a very different result in terms of reproductive information. In particular, if we use acoustic energy as a measure of loudness, then we find that a male that produces a mating call with an intensity of 19.9 $dB$ is associated with a reproductive-information value of $I_R \approx 37.660$ bits.

The reason for the increase in the estimated value of $I_R$ as we change from measuring sound intensity in decibels to measuring sound intensity in watts per square meter has to do with the process of exponentiation, which is how we transform decibels into watts per square meter[56] ($W/m^2$) (see Supplementary Note 18). In the current context, exponentiation tends to shrink the relative distance between low-intensity values, as compared to larger-intensity values. The optimum is set at 20 $dB$, which is on the low end of the set of all possible sound-intensity values. As a result, exponentiation tends to decrease the distance between points on the decibel scale that are close to the optimum, relative to the length of the line that contains all-possible sound-intensity values. This change is what leads to the increase in the estimate of reproductive information that occurs when we switch from the decibel scale to the acoustic-energy scale.

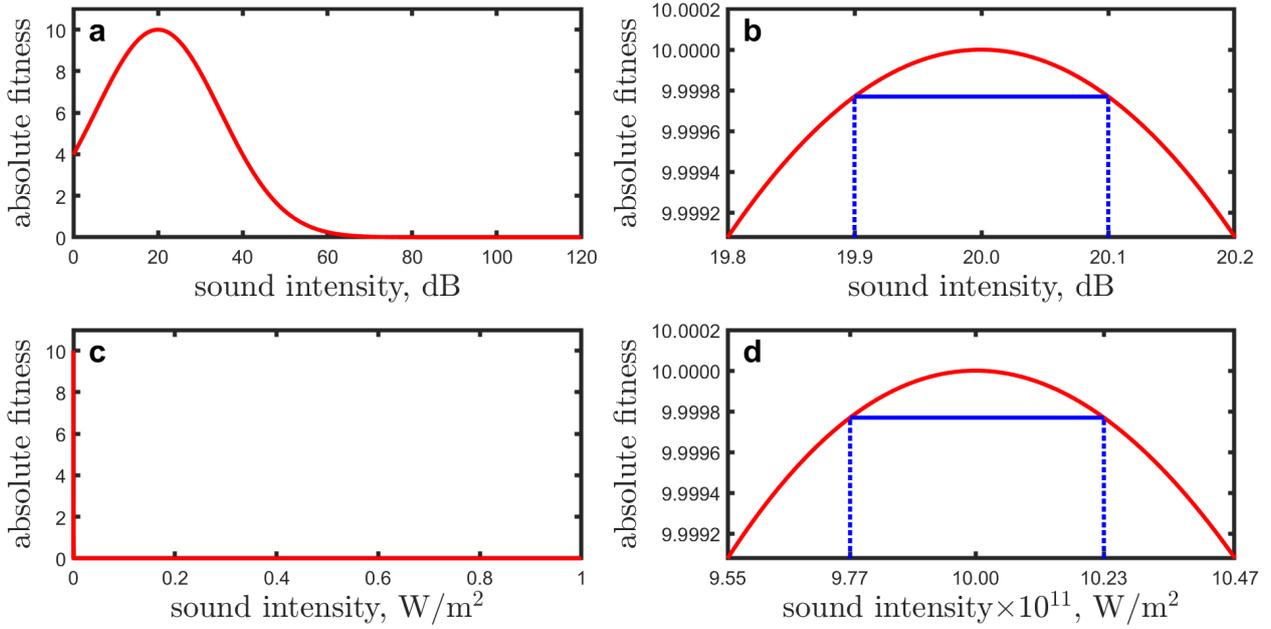

**Fig. S13.1 | Estimating reproductive information using a continuously distributed character.** Panel (a) shows a fitness landscape in which the fitness of a male is a function of the loudness (or sound intensity) of the male's mating call, as measured in decibels (dB). Panel (b) shows the same data as panel (a), except that the image is zoomed in to focus on the fitness peak. The horizontal blue line indicates the range of phenotypes that are at least as fit as an individual with a mating call that has a sound intensity of 19.9 $dB$. Panel (c) shows the same fitness landscape for the same range of sound-intensity values as that shown in Panel (a), after transforming the measure of sound intensity from decibels to acoustic energy, as measured in watts per square meter $(W/m^2)$. Note that the part of the curve that is substantially above the horizontal axis is now visually indistinguishable from the vertical axis. Panel (d) shows the same data as panel (c), except that the image is zoomed in to focus on the fitness peak. The blue horizontal blue line indicates the range of phenotypes that are at least as fit as an individual with a mating call that has a sound intensity of 19.9 $dB$. Note that 19.9 $dB$ is equivalent to $9.772 \times 10^{-11}$ $W/m^2$ (see Supplementary Note 18 for details about this transformation).

To see this graphically, consider that, if we use standard definitions, then a sound-intensity of zero $dB$ is equivalent to $10^{-12}$ $W/m^2$, when loudness is measured using acoustic energy[56]. Furthermore, an intensity of 120 $dB$ is equivalent to 1.0 $W/m^2$. Thus, when measuring in terms of acoustic energy, the line that, under our current assumptions, includes all possible sound-intensity values has a length of $(1 - 10^{-12})$ $W/m^2$. However, when we plot fitness as a function of acoustic energy over the range of possible acoustic-energy values, we find that the part of the curve that shows non-negligible fitness values is so close to the vertical axis that its separation from the vertical axis cannot be visually distinguished at this scale (see Fig. $S$13.1, panel (c)). It is only when we magnify a part of the graph that is close to the vertical axis that we can see what is going on. This is shown in Fig. $S$13.1, panel (d).

To gain a numerical understanding of the effects of the transformation from decibels to acoustic energy, consider that, if we express 19.9 $dB$ in terms of acoustic energy, then we get $9.772 \times 10^{-11}$ $W/m^2$. The line containing all acoustic-energy values that are at associated with a fitness value that is at least as high as the fitness value that is associated with an acoustic energy of $9.772 \times 10^{-11}$ $W/m^2$ has a length equal to $4.606 \times 10^{-12}$ $W/m^2$ (see Fig. S13.1, panel (d)). Our estimate of reproductive information is thus $I_R \approx \log_2 \left( \frac{1 - 10^{-12}}{4.606 \times 10^{-12}} \right) = 37.660$ bits.

The mathematical transformation that changes sound intensity (as measured in decibels) into a measure that is in terms of watts-per-square-meter is just one of an infinite number of possible mathematical transformations. The reproductive information measured will typically depend strongly on which transformation we choose. This is a problem, because it is not obvious that there is any transformation that is uniquely better and more appropriate than all the others. In the absence of such a unique transformation, it seems that there is a large degree of arbitrariness regarding the level of reproductive information that is associated with a phenotype that is defined by a continuously distributed phenotypic variable.

Fortunately, some recent research suggests a solution to this conundrum. In particular, this work shows that, for almost any one-dimensional continuously distributed phenotypic trait, there exists a special transformation that is much more convenient and natural than other transformations[57]. Thus, problems like the one just outlined can be resolved by simply agreeing to use a special transformation of this sort.

In order to explain this in more detail, imagine a continuous phenotypic trait that is controlled by some set of causal variables. These causal variables may be genetic or environmental in nature. For example, the loudness of a male's mating call might be determined primarily by that male's genotype, and by the availability of food in the male's immediate vicinity. The new results include a theorem that shows that, for almost any such causal relationship, there exists a special transformation such that, given *any* values of the causal variables, the entropy of the continuous phenotypic trait will *always* have the same value. This means that the decrease in "uncertainty" about the value of the phenotypic trait that occurs when we learn the values of the causal variables is always the same, regardless of which values of the causal variable we observe[9,34]. Thus, for example, after application of the special transformation, we might find that poorly fed males with a particular genotype produce mating calls that, on average, are quieter than the calls produced by better-fed males with a different particular genotype. However, the *entropy* that characterises the distribution of loudness values for one of these groups of males will be *identical* to the entropy associated with the distribution of loudness values for the other group of males. Furthermore, the value of the "universal entropy" in the phenotypic trait that arises as a result of the special transformation has significance. In particular, the magnitude of the universal entropy is equal to the *channel capacity* that is associated with the causal relationship. The channel capacity, in turn, is equal to the maximum rate at which information could be sent, if one used the causal relationship as a communications system.

The new results open the possibility of further developments that could enhance their scope and practicality. For example, at present, the new results have only been proved in the context of one-dimensional continuous phenotypic variables. It is plausible that multi-dimensional versions of the results will be possible, but this has not yet been demonstrated. Nevertheless, the new results do suggest that, eventually, practical and unambiguous measurements of reproductive information will be possible for organisms that are described entirely in terms of their phenotypic characteristics.

## Supplementary Note 14: Previous studies

In this note we present a few comments regarding three prominent previous studies that have proposed methods to measure the information that is created by natural selection. These studies are, in chronological order, a publication by Kimura[15] in 1961, a publication by Frank[26] in 2012, and, finally, a study by Hledik, Barton, and Tkacik[19], which was published in 2022. This last study builds on previous work by Barton[18].

### Kimura and "genetic information"

An early information-theoretic study of evolution by natural selection was published by M. Kimura only 13 years after Claude Shannon's seminal paper of 1948 initiated the modern quantitative study of information[9,15]. Kimura tells his readers that he wishes to measure "*genetic information.*" While Kimura does not explicitly define genetic information, he does make various comments such as: "New genetic information was accumulated in the process of adaptive evolution, determined by natural selection acting on random mutations." This suggests that Kimura's concept of genetic information is similar to the concept of reproductive information, which we have presented in the current work.

Kimura chooses to measure genetic information by considering how gene frequencies change over time as a result of selection. Thus, his measure takes, as input, the state of the population at two different points in time. We call this sort of methodology the *historical method for measuring the information created by natural selection* (or the *historical method* for short).

Unfortunately, the historical method is simply not in line with modern ideas about the nature of information. Modern information theory is about the relationship between two phenomena. It tells us how much knowing the state of one phenomenon tells us about the state of another phenomenon[9,12,34,35]. Thus, from the perspective of modern information theory, the information about the selective environment that is embodied by the genomes of organisms in natural populations cannot be calculated by simply considering how the frequencies of various types of genomes have changed in the past.

To better understand the problems with Kimura's approach, let us consider the details of how Kimura calculates genetic information. The simplest case is when a selectively advantageous allele at a genetic locus in a haploid population has an initial frequency of $p$. Kimura tells us that, if this organismal type becomes fixed, then the amount of genetic information that will be created as a result is equal to $\log_2(1/p)$ bits. This assertion was recently reiterated by NH Barton, who said that $\log_2(1/p)$ measures the "ultimate gain in information" that is associated with the fixation of an allele [18]. However, this cannot be considered to be a reasonable measure, as it allows the amount of information gained to take *any* positive value, no matter how large, so long as $p$ is sufficiently small. This result is contrary to the most basic information-theoretic ideas. For example, let us characterise individuals in terms of their genotypes, and let us assume that only two genotypes are possible ($\Omega = 2$). Now, consider the case where $p < 1/2$. In this case, Kimura's formulation leads to the conclusion that fixation involves the creation of more than one bit of information. But this is clearly impossible, at least by the lights of modern information theory. If $\Omega = 2$ then organisms can only be in one of two possible states. A system that can take two states can embody, at most, one bit of information [9,34]. This is because one bit of information is, in essence, the answer to a yes-or-no question, and a system that can only be in one of two states can answer, at most, only one yes-or-no question. If, for example, we are in a world where the fittest genotype is the only one that we ever find in any environment, then knowledge of a randomly selected individual's genotype can answer a question as to which genotype is fittest in the local environment. However, if $\Omega = 2$ then there are only two possible answers to this question, and so the number of bits created by selection can never be greater than one. One bit is the absolute limit in this case because the answer to one yes-or-no question is sufficient to indicate which of the two possible genotypes is most fit.

## S. Frank and "the information accumulated by natural selection"

A more recent example of the historical method for measuring the information created by natural selection is a study [26] published by Frank in 2012. To measure the information created by natural selection, Frank uses a quantity known as the *Jeffreys divergence*. The Jeffreys divergence takes, as input, two frequency distributions of some variable. Frank's method involves using the distribution of organismal types before an episode of selection as one of these distributions, and it uses the distribution of organismal types after an episode of selection as the other distribution. Frank says that, once this is done, the Jeffreys divergence will measure "the information accumulated by natural selection" during the episode of selection.

Frank's methodology suffers from the same core problem as does the information-theoretic formulation of Kimura. In particular, it attempts to calculate a quantity of information by considering how a population changes over time. However, as pointed out above, the change in a population is not, in general, a good measure of the information that members of the population embody about their selective environment. (Or, at least, this is so if one defines information the traditional manner that was established by Shannon [9], and by the researchers who first described Kolmogorov information [10–12]).

The details of Frank's calculations are different from those of Kimura. For example, Frank's method does not allow calculation of the information gained when an organismal type becomes fixed in a population. This is because, with Frank's formulation, such a calculation requires division by zero. Nevertheless, like Kimura's method, Frank's method can produce apparently nonsensical results. For example, assume that there are only two possible organismal types, and the one that is more fit has an initial frequency of $p$, where $0 < p < 1/2$. Assume further that, over time, the frequency of the fitter type increases from $p$ to $1 - p$. Using Frank's formulation, this leads to the following expression for "information accumulated by natural selection":

$$2 \log_2 \left( \frac{1-p}{p} \right) + 4p \log_2 \left( \frac{p}{1-p} \right). \tag{S14.1}$$

As with Kimura's formulation, this means that the "information accumulated by natural selection" can take on *any* positive value, no matter how large, so long as $p$ is sufficiently small. Furthermore, if $p \leq 0.297$ then Frank's formulation leads to the conclusion that the "information accumulated by natural selection" is greater than one bit. Once again, if we use the standard concepts of modern information theory, this is absolutely impossible given that there are only two possible organismal types[34]. With two organismal types, the identity of an individual's type can only convey, at most, one bit of information about the environment[9]. Similarly, when there are only two possible organismal types, data about the environment cannot provide more than one bit of information about an individual's type.

## M. Hledik, N. Barton, G. Tkacik, and "genotype-level information"

An even more recent attempt to quantify the information that is created by natural selection is the work of M. Hledik, N. Barton, and G. Tkacik[18,19], (whom we shall refer to as HBT). In this case, the authors study finite populations in which stochastic processes may be important. They consider multiple realisations of an evolutionary process in which a population evolves under selection. In most of their examples, all realisations begin with the population in one particular configuration (i.e., with a particular initial distribution of organismal types that is the same in all realisations of the process).

HBT study three similar and closely related types of "information." For our purposes, perhaps the most relevant and relatable of these is what they call "*genotype-level information*", which they represent with the symbol $D(G)$. HBT use the symbol $\psi^G(g)$ to represent the frequency of genotype $g$ at some point in time after the initiation of the evolutionary process, if one averages over all possible realisations of the evolutionary process. Similarly, they use the symbol $\phi^G(g)$ to represent the frequency of genotype $g$ at some point in time after the initiation of the evolutionary process, if we assume that there are no differences between genotypes in fitness, and if, once again, we average over all possible realisations of the evolutionary process. Thus, $\psi^G$ represents the averaged distribution of genotypes in the presence of selection, and $\varphi^G$ represents the averaged distribution of genotypes under selective neutrality (assuming that both distributions are calculated at some particular point in time after the initiation of the evolutionary process). To measure "genotype-level information", HBT choose to measure the *Kullback-Leibler* divergence[34,58] from the averaged neutral distribution of genotypes ($\phi^G$) to the averaged distribution of genotypes under selection ($\psi^G$). HBT use $D(G)$ to represent genotype-level information. The following equation, therefore, provides their measure of genotype-level information:

$$D(G) = \sum_{g=1}^{\Omega} \psi^G(g) \log_2 \left( \frac{\psi^G(g)}{\varphi^G(g)} \right). \tag{S14.2}$$

Here, we are using $\Omega$ to represent the total number of possible genotypes, and we assume that these genotypes are numbered as 1, 2, 3, ..., $\Omega$.

HBT do not explain how investigators can study multiple realisations of evolutionary processes in natural populations. Aside from this practical issue, HBT also offer no reason for their use of the Kullback-Leibler divergence, other than to say that it is a "central quantity in information theory." While this is true, the Kullback-Leibler divergence does not, in general, measure information in the same sense that information is measured by the two main contemporary quantitative definitions of information[9,12,34,35,58] (e.g., Shannon information and Kolmogorov information). An exception to this is that the Kullback-Leibler divergence can be used to calculate Shannon information, but in this case the two distributions used are very different from those used by HBT. Another problem is that the Kullback-Leibler divergence is not symmetric in the sense that the Kullback-Leibler divergence from $\varphi^G$ to $\psi^G$ is not, in general, equal to the Kullback-Leibler divergence from $\psi^G$ to $\varphi^G$. Equation (S14.2) shows that HBT chose to equate genotype-level information with the Kullback-Leibler divergence from $\varphi^G$ to $\psi^G$, instead of the other way around. However, their reasons for this choice are unclear.

From equation (S14.2) we can see that genotype-level information is a property that describes an entire population. It does not apply to any particular individual. As such, it appears that genotype-level information cannot be used to assess the amount of information about the selective environment that is embodied by the genome of a particular individual, and vice-versa. This is unfortunate, as it greatly limits the range of scientific questions that can be addressed using the concept of genotype-level information.

Note that the same critique applies both to Kimura's concept of genetic information, and to the "information accumulated by natural selection" that Frank claims to measure by use of Jeffery's Divergence[15,26].

Equation (S14.2) shows that $D(G)$ depends very strongly on the distribution of genotypes under neutrality, $\varphi^G$. This distribution can take virtually any form, depending, in part, on the details of the mutation process. With this in mind, allowing $\varphi^G$ to have such an important role in determining genotype-level information seems odd. The only type of mutation that HBT mention explicitly are point mutations, and per-nucleotide-point-mutation rates in DNA-based organisms are typically very small[59] - on the order of $10^{-8}$ mutations per nucleotide, per generation. As such, it is not difficult to imagine situations in which mutation will have virtually no effect on evolution at nucleotide sites that are subject to selection[39,41], even if selection coefficients are as small as $10^{-6}$. Nevertheless, equation (S14.2) tells us that, even in these scenarios, $\varphi^G$ (and thus the relative values of the various mutation rates) has a large role in determining the value of $D(G)$. Indeed, if a change in radiation or some other phenomenon causes a change in mutation rates, then, even if mutation rates are always too low to ever have any substantial effect on evolutionary dynamics, this change can have a very large effect on genotype-level information by causing a change to $\varphi^G$. It is hard to understand why mutation should decide the quantity of genotype-level information. This problem is particularly apparent in cases in which mutation has virtually no effect on the dynamics of evolution.

In the examples they present, HBT explicitly assume that mutation is symmetric, so that all mutation rates from one allele to another are the same. They claim that this ensures that, for every possible value of $g$, $\varphi^G(g) = 1/\Omega$. However, their formulation is much more general than this, and it allows for the more biologically realistic possibility of non-symmetric mutation. Furthermore, HBT are incorrect in their assertion that symmetry in mutation rates inevitably leads to a uniform neutral genotypic distribution. For example, a single biallelic locus in a very large diploid population with symmetric mutation can be expected, in the absence of selection, to be found in a state where each of the homozygotes have a frequency of about $\frac{1}{4}$, while the heterozygote has a frequency of about $\frac{1}{2}$. This is true if the population is asexual, or if it is a random-mating sexual population[39,41].

If we consider the more general case of a non-uniform neutral distribution ($\varphi^G$), then, inevitably, for some values of $g$, we will have $\varphi^G(g) < 1/\Omega$. This observation leads to the revelation of a profound problem with HBT's analysis. To see this, consider a situation in which a genotype for which $\varphi^G(g) < 1/\Omega$ inevitably becomes fixed (or nearly fixed) as a result of selection. According to HBT's formulation (i.e., equation (S14.2)) this leads to the creation of more than $\log(\Omega)$ bits of genotype-level information. Indeed, fixation can create *any* finite amount of genotype-level information, no matter how large, so long as the mutation rates ensure that $\varphi^G(g)$ is sufficiently small for the genotype that becomes fixed. This should be impossible, as a choice among $\Omega$ alternatives cannot possibly convey more than $\log(\Omega)$ bits of information[9,34]. Thus, for example, as mentioned above, if $\Omega = 2$, then it should be impossible for evolution to create more than one bit of information.

Another problem with HBT's formulation has to do with the effects of epistasis. These effects are discussed in the main text, and in Supplementary Note 2. As we have demonstrated, when there are epistatic interactions between loci in the determination of fitness, it is entirely possible for a hypothetical population to evolve until the vast majority of individuals have a genotype that is a local optimum, so that any single mutation of this genotype causes a decrease in fitness. Furthermore, because epistasis can lead to the existence of a very large number of such local optima, the genomes of members of a population that has becomes nearly fixed on a local optimum will not necessarily provide more than a few bits of information about the fitness landscape in which they evolved (see the main text, and Supplementary Note 2).

Nevertheless, if we use HBT's assumption that all $\varphi^G(g)$ values are identical, then HBT's formulation suggests that genotype-level information is maximised whenever the population inevitably becomes fixed on a particular local optimum. This is true regardless of whether the local optimum in question provides the highest fitness possible, or provides the lowest fitness associated with any of a very large number of local optima (see equation (S14.2)). Indeed, as we have seen, in a highly epistatic fitness landscape, a hypothetical population with a genome consisting of millions of loci that is initiated at a random location in "genotype space" can be expected to become fixed (or nearly fixed) on a local optimum after only a very small number of loci have undergone changes from their initial random configurations. It seems implausible that a few substitutions into an initially random genome can lead inevitably to the accumulation of millions or billions of bits of information, but this is exactly what is implied by HBT's formulation.

Of course, a population initiated at a given location in "genotype space" might not inevitably evolve until it becomes fixed (or nearly fixed) on one particular local optimum. Instead, there may be several local optima that are commonly reached from a given location, depending on the order in which mutations arise and become common. However, in a highly epistatic fitness landscape, the number of optima typically reached from any given initial position is likely to be extremely small in comparison to the total number of local optima[42–44]. From consideration of equation (S14.2) we can see that this implies that our conclusion should remain unchanged. Even if there are a few different local optima that may be achieved starting from some particular point in genotype space, it is still the case that, using HBT's methodology, a population that rarely requires more than a few beneficial mutations to reach a local optimum from a random initial position in genotype space can, nevertheless, appear to have acquired a very large amount of genotype-level information in so doing.

In a worked example, HBT calculate genotype-level information at various points in time after an evolutionary process begins. One might try to salvage the case for the biological relevance of HBT's analysis by insisting on considering only extremely long time scales, so that genotypes with multiple new mutations have a chance to arise, and thus move the population away from local optima during an adaptive episode. For a similar reason one might also try very large (and possibly very unrealistic) population sizes, so that all possible genotypes arise regularly via mutation. With this in mind it is worth noting that, if reproduction is sexual, then even if a population is infinite in size, the population can remain nearly fixed on *any* local optimum, even if one waits an infinite period of time (see Supplementary Note 15). Thus, by starting a sexual population sufficiently close to a particular local optimum, one might ensure that, in every realisation of the evolutionary process, the population will forever remain nearly fixed upon that local optimum. This means that, for sexual populations, even infinite time and infinite population sizes cannot resolve the paradoxical situation we have observed. In particular, HBT's methods can allow one to conclude that a population is associated with a very large amount of genotype-level information, even when, in fact, genomes can provide very little information about the fitness landscape in which they evolved. With these considerations in mind, it seems fair to conclude that HBT's analysis has little utility if we measure information in a way that is in line with the best known modern methods[34,35], and if our objective is to calculate the amount of information created by natural selection.

## Supplementary Note 15: Stability of local optima

In this section of the Supplementary Information we establish the conditions under which a locally optimal genotype, in a sexual population, is stable to invasion by all other genotypes.

**Assumptions**

- The population is effectively infinite, thus the dynamics is deterministic.

- Generations are discrete.

- Adults are haploid, with $n$ biallelic loci, with $n \geq 2$.

- The lifecycle, starting with adults, is: (i) adults produce haploid offspring (involving free recombination but neglecting mutation) by random mating, where each adult has the same expected number of offspring; (ii) viability selection acts on the offspring, leading to the adults of the next generation.

**Analysis**

We begin by considering a genotype which is locally optimal. To illustrate matters, we first consider the case of $n = 3$ loci. We will then extend the pattern we observe to $n$ loci.

We write the eight possible haploid genotypes at the three loci as $abc$, $abC$, $aBc$, $Abc$, $aBC$, $AbC$, $ABc$ and $ABC$.

We assume $abc$ is locally optimal, and shall investigate its stability, by considering $abc$ at high frequency and all other genotypes at very low frequencies. In particular, we take $abc$ to have an initial frequency of $1 - \delta$ while genotypes $abC$, $aBc$, $Abc$, $aBC$, $AbC$, $ABc$, and $ABC$ have initial frequencies of $\Delta_{abC}$, $\Delta_{aBc}$, $\Delta_{Abc}$, $\Delta_{aBC}$, $\Delta_{AbC}$, $\Delta_{ABc}$ and $\Delta_{ABC}$, respectively. We have

$$\delta = \Delta_{abC} + \Delta_{aBc} + \Delta_{Abc} + \Delta_{aBC} + \Delta_{AbC} + \Delta_{ABc} + \Delta_{ABC} \tag{S15.1}$$

so that the frequencies of all genotypes sum to unity.

We shall work to linear order in the $\Delta$'s, and neglect higher order terms[f].

We now consider different matings, and their contribution to the adults of the next generation.

---
[f]We interpret $\delta$ as being of linear order in the $\Delta$'s.

We use fitnesses, relative to the fitness of *abc*. Thus the relative fitness of *abc* is $w_{abc} = 1$, and we write the relative fitness of all other genotypes as $w_{abC}$, $w_{aBc}$, ... etc.

1. The mating of *abc* with *ABC* leads to a contribution to the frequency of[g]

   *abC* in adults of the next generation of $w_{abC}\Delta_{ABC}/2^2$,

   *aBc* in adults of the next generation of $w_{aBc}\Delta_{ABC}/2^2$,

   *Abc* in adults of the next generation of $w_{Abc}\Delta_{ABC}/2^2$,

   *aBC* in adults of the next generation of $w_{aBC}\Delta_{ABC}/2^2$,

   *AbC* in adults of the next generation of $w_{AbC}\Delta_{ABC}/2^2$,

   *ABc* in adults of the next generation of $w_{ABc}\Delta_{ABC}/2^2$,

   *ABC* in adults of the next generation of $w_{ABC}\Delta_{ABC}/2^2$.

2. The mating of *abc* with *ABc* leads to a contribution to the frequency of

   *aBc* in adults of the next generation of $w_{aBc}\Delta_{ABc}/2$,

   *Abc* in adults of the next generation of $w_{Abc}\Delta_{ABc}/2$,

   *ABc* in adults of the next generation of $w_{ABc}\Delta_{ABc}/2$,

3. The mating of *abc* with *AbC* leads to a contribution to the frequency of

   *abC* in adults of the next generation of $w_{abC}\Delta_{AbC}/2$,

   *Abc* in adults of the next generation of $w_{Abc}\Delta_{AbC}/2$,

   *AbC* in adults of the next generation of $w_{AbC}\Delta_{AbC}/2$,

4. The mating of *abc* with *aBC* leads to a contribution to the frequency of

   *abC* in adults of the next generation of $w_{abC}\Delta_{aBC}/2$,

   *aBc* in adults of the next generation of $w_{aBc}\Delta_{aBC}/2$,

   *aBC* in adults of the next generation of $w_{aBC}\Delta_{aBC}/2$,

5. The mating of *abc* with *Abc* leads to a contribution to the frequency of

   *Abc* in adults of the next generation of $w_{Abc}\Delta_{Abc}$.

---

[g]Under random mating, the frequency that *abc* mates with *ABC* is $2 \times (1-\delta) \times \Delta_{ABC} \simeq 2\Delta_{ABC}$. Given such a mating, free recombination results in all $2^3$ possible genotypes being present in the offspring, each in the proportion $2^{-3}$. The frequency of any genotype, such as *abC*, that is contributed to the adults in the next generation is given by $2\Delta_{ABC}$ when: (i) multiplied by the fraction of offspring of this type, namely $2^{-3}$, (ii) multiplied by the relative fitness of the genotype, namely $w_{abC}$, (iii) divided by the mean relative fitness of the population, which primarily arises from *abc* with *abc* matings, and has the value $1 + O(\Delta)$. Thus to linear order in the $\Delta$'s, the mating of *abc* with *ABC* results in a contribution to the frequency of *abC* in adults of $2w_{abC}\Delta_{abC}/2^3 = w_{abC}\Delta_{abC}/2^2$.

6. The mating of *abc* with *aBc* leads to a contribution to the frequency of *aBc* in adults of the next generation of $w_{aBc}\Delta_{aBc}$.

7. The mating of *abc* with *abC* leads to a contribution to the frequency of *abC* in adults of the next generation of $w_{abC}\Delta_{abC}$.

Thus the frequency of the different genotypes in the next generation (denoted by a prime), are given by

$$\Delta'_{abC} = w_{abC}\left(\Delta_{abC} + \frac{\Delta_{aBC}}{2} + \frac{\Delta_{AbC}}{2} + \frac{\Delta_{ABC}}{2^2}\right)$$
$$\Delta'_{aBc} = w_{aBc}\left(\Delta_{aBc} + \frac{\Delta_{aBC}}{2} + \frac{\Delta_{ABc}}{2} + \frac{\Delta_{ABC}}{2^2}\right) \quad \text{(S15.2)}$$
$$\Delta'_{Abc} = w_{Abc}\left(\Delta_{Abc} + \frac{\Delta_{AbC}}{2} + \frac{\Delta_{ABc}}{2} + \frac{\Delta_{ABC}}{2^2}\right)$$

and

$$\Delta'_{aBC} = w_{aBC}\left(\frac{\Delta_{aBC}}{2} + \frac{\Delta_{ABC}}{2^2}\right)$$
$$\Delta'_{AbC} = w_{AbC}\left(\frac{\Delta_{AbC}}{2} + \frac{\Delta_{ABC}}{2^2}\right) \quad \text{(S15.3)}$$
$$\Delta'_{ABc} = w_{ABc}\left(\frac{\Delta_{ABc}}{2} + \frac{\Delta_{ABC}}{2^2}\right)$$

and

$$\Delta'_{ABC} = w_{ABC}\frac{\Delta_{ABC}}{2^2}. \quad \text{(S15.4)}$$

Let $w_1$ be the largest of $w_{abC}$, $w_{aBc}$, and $w_{Abc}$:

$$w_1 = \max(w_{abC}, w_{aBc}, w_{Abc}). \quad \text{(S15.5)}$$

Let $w_2$ be the largest of $w_{aBC}$, $w_{AbC}$ and $w_{ABc}$

$$w_2 = \max(w_{aBC}, w_{AbC}, w_{ABc}) \quad \text{(S15.6)}$$

and write

$$w_3 = w_{ABC}. \quad \text{(S15.7)}$$

Then with

$$\Delta_1 = \Delta_{abC} + \Delta_{aBc} + \Delta_{Abc} \quad \text{(S15.8)}$$

$$\Delta_2 = \Delta_{aBC} + \Delta_{AbC} + \Delta_{ABc} \quad \text{(S15.9)}$$

$$\Delta_3 = \Delta_{ABC} \quad \text{(S15.10)}$$

we have, by separately summing equations (S15.2), (S15.3), and (S15.4), and using equations (S15.5)-(S15.7), that

$$\Delta'_1 \leq w_1 \left( \Delta_1 + \Delta_2 + \frac{3}{4}\Delta_3 \right)$$

$$\Delta'_2 \leq \frac{w_2}{2}\Delta_2 + \frac{3w_2}{4}\Delta_3 \tag{S15.11}$$

$$\Delta'_3 = \frac{w_3}{4}\Delta_3.$$

As long as $w_3/4 < 1$ it follows that as time progresses $\Delta_3 \longrightarrow 0$. This behaviour of $\Delta_3$ affects $\Delta_2$: as long as $w_2/2 < 1$ it results in $\Delta_2 \longrightarrow 0$. Lastly, by virtue of *abc* being a local optimum we have $w_1 < 1$ and this, combined with the results for $\Delta_2$ and $\Delta_3$ leads to $\Delta_1 \to 0$.

More formally, we can consider the case of equality of equation(S15.11 ), and write the corresponding set of equations as $\boldsymbol{\Delta}' = \mathbf{M}\boldsymbol{\Delta}$ where $\boldsymbol{\Delta}$ is a three component column vector and $\mathbf{M}$ is a $3 \times 3$ matrix. The eigenvalues of $\mathbf{M}$ are $w_3/4$, $w_2/2$ and $w_1$. Hence stability of *abc* is guaranteed under the conditions stated.

**General case**

Let us now consider the general case of $n$ loci, with $n \geq 2$, using similar notation to that used above, for genotype frequencies. The locally optimal genotype that we now focus upon - the focal genotype - is schematically written as *abc...z*. We again use fitnesses, relative to the fitness of the focal genotype, hence the relative fitness of *abc...z* is $w_{abc...z} = 1$.

The single genotype that has a different allele at all $n$ loci, compared with the focal genotype, is $ABC...Z$ and we write its relative fitness as $w_n$. We argue that only *abc...z* matings with $ABC...Z$ produce the $ABC...Z$ genotype. Furthermore, writing the frequency of the $ABC...Z$ genotype as $\Delta_n \equiv \Delta_{ABC...Z}$, the frequency in the next generation it is $\Delta'_n = 2\frac{w_n}{2^n}\Delta_n$. Thus, this genotype will decrease in frequency over time if

$$\frac{w_n}{2^{n-1}} < 1. \tag{S15.12}$$

Consider next the $n$ different genotypes where one locus has the same allele as the focal genotype, and $n-1$ loci have different alleles, e.g., $aBC...Z$, $AbC...Z$, etc. These genotypes are produced in two different ways: (i) from the mating of *abc...z* with $ABC...Z$, and (ii) from the mating of *abc...z* with these $n$ genotypes. With $\Delta_{n-1}$ denoting the sum of the frequencies of these $n$ genotypes (i.e., $\Delta_{n-1} = \Delta_{aBC...Z} + \Delta_{AbC...Z} + ...$), and with $w_{n-1}$ the largest relative fitness of the $n$ genotypes described above, we arrive at $\Delta'_{n-1} \leq w_{n-1} 2 \left( \frac{\Delta_{n-1}}{2^{n-1}} + \frac{n\Delta_n}{2^n} \right)$ or

$$\Delta'_{n-1} \leq \left( \frac{w_{n-1}}{2^{n-2}}\Delta_{n-1} + \frac{n}{2^{n-1}}w_{n-1}\Delta_n \right). \tag{S15.13}$$

In the case where the equality of the above result holds, we can show that with equation (S15.12) applying, we have that $\Delta_{n-1} \to 0$ when $\frac{w_{n-1}}{2^{n-2}} < 1$. Continuing in this way, with $w_m$ the maximum relative fitness of the set of all genotypes that differ from the focal genotype at $m$ loci, we arrive at the following condition for stability of the focal genotype

$$\frac{w_m}{2^{m-1}} < 1 \tag{S15.14}$$

which holds for $m = 1, 2, \ldots n$.

For $m = 1$, equation (S15.14) applies because the focal genotype is a local optimum, thus all genotypes associated with a single difference to the focal genotype have a relative fitness which is less than the relative fitness of the focal genotype, thus $w_1 < 1$. For any other genotype, equation (S15.14) represents non-trivial conditions on the maximum fitness of a set of genotypes, such that when these conditions apply, the focal genotype will persist in the population, i.e., is stable, in the presence of all other genotypes, when they start at low frequencies.

## Supplementary Note 16: A minor caveat

In Supplementary Note 2 we consider fitness landscapes that have two local optima, and we say that flawless communication using such landscapes "requires that each possible message in the codebook is associated with a minimum of two genotypes." This is true. However, it is not sufficient. In fact, for flawless communication we must also ensure that, for any message we might send, the two genotypes that are associated with the intended messages can both simultaneously be local optima. In order to satisfy this requirement, it is necessary that, for each possible message, the two genotypes that are associated with that message differ from each other at two or more loci (or nucleotide sites).

## Supplementary Note 17: Some proofs regarding entropy and Shannon information

This note provides some mathematical details to explain and prove various results that are presented in Supplementary Note 10. These results have to do with the relationship between reproductive information and Shannon information.

We begin by proving the result represented by equation (S10.4) in Supplementary Note 10. This result was derived under the assumption that, for individuals in environments of type $e$, every possible organismal type has a unique fitness value, which is shared with no other organismal type. Furthermore, the result is also derived under the assumption that, within an environment of type $e$, there are no two organismal types for which the fitter one is less common than the less-fit one. Equation (S10.4) in Supplementary Note 10 claims that, under these assumptions, we have

$$\overline{\overline{L}}_{s,e} - 1 \leq H_e(X) \leq \overline{\overline{L}}_{s,e} + 2\log_2\left(\log_2\left(\Omega\right)\right) + 5. \tag{S17.1}$$

Here, $\overline{\overline{L}}_{s,e}$ represents the mean value of $\overline{L}_s$ among individuals that live in environments of type $e$, and $H_e(X)$ represents the entropy in organismal type among individuals that live in environments of type $e$.

Let us begin by establishing that, as stated by inequality (S17.1), $\overline{\overline{L}}_{s,e} - 1 \leq H_e(X)$. To do this, we begin by defining a symbol to represent the shortest possible average length of a prefix-free code for the organismal types of individuals that live in environments of type $e$. We will use $\overline{\ell}_{PF,e}$ for this purpose. Note that this code, along with all other codes mentioned in this Supplementary Note, are binary. Prefix-free codes are defined in Supplementary Note 10.

According to Shannon's *noiseless-coding theorem*[35], the value of $\overline{\ell}_{PF,e}$ must be close to $H_e(X)$. In particular

$$H_e(X) \leq \overline{\ell}_{PF,e} \leq H_e(X) + 1 \tag{S17.2}$$

For the purposes of the first sections of Supplementary Note 10 (and thus for the current analysis as well) we have assumed that each organismal type has as unique fitness value. As such, there is only one optimal coding scheme (as defined in Supplementary Note 5). Furthermore, as we shall see, under these assumptions, it is straightforward to show that $\overline{\overline{L}}_{s,e}$ must be equal to the smallest possible average code length for a prefix-allowing coding scheme that encodes the organismal types of individuals that live in environments of type $e$. Here, a *prefix-allowing coding scheme* is one in which there is no prohibition against one binary codeword constituting the first part of another code word. Thus, if our coding scheme included the binary codeword 01, then there would be no prohibition against the coding scheme also using the codeword 010. Both the optimal and the non-optimal coding schemes described in the main text (and in Supplementary Note 5) are prefix-allowing codes.

The optimal coding schemes described in Supplementary Note 5 use the shortest binary sequences to encode the organismal types, and they assign the shortest codes to the fittest types. Under the assumption of the present Supplementary Note (i.e., that fitter organismal types are the most common organismal types), this means that they also assign the shortest codes to the most frequent types. As such, under the assumptions of the present Supplementary Note, $\overline{\overline{L}}_{s,e}$ represents the minimum possible average code-word length for a coding scheme that encodes the $\Omega$ organismal types using binary sequences. (Here, we are averaging over all of the individuals that live in environments of type $e$.) As a result of these considerations, we know that $\overline{\overline{L}}_{s,e} \leq \overline{\ell}_{PF,e}$. Furthermore, from the noiseless coding theorem (equation (S17.2)) we know that $\overline{\ell}_{PF,e} \leq H_e(X) + 1$. Combining these last two observations, we see that $\overline{\overline{L}}_{s,e} \leq H_e(X) + 1$, and thus $\overline{\overline{L}}_{s,e} - 1 \leq H_e(X)$, as stated by the inequality (S17.1).

It remains to show that $H_e(X) \leq \overline{\overline{L}}_{s,e} + 2\log_2(\log_2(\Omega)) + 5$. To do this, we will describe a prefix-free coding scheme, and we will show that, for every organismal type, the length of the code word assigned to the organismal type by this coding scheme is less than $2\log_2(\log_2(\Omega)) + 5$ binary digits longer than the code word for this organismal type under the optimal prefix-allowing coding scheme for individuals that live in environments of type $e$. As $\overline{\overline{L}}_{s,e}$ is average value of these optimal prefix-allowing code words among individuals living in environments of type $e$, this implies that, for these individuals, the average code-word length for the most efficient prefix-free coding scheme (i.e., $\bar{\ell}_{PF,e}$) must be less than $\bar{\bar{L}}_{s,e} + 2\log_2(\log_2(\Omega)) + 5$. This implication holds because the prefix-free coding scheme that we will describe will not typically be maximally efficient, and thus the average length for the code words it produces can cannot be less than $\bar{\ell}_{PF,e}$, which represents the smallest-possible mean code-word length for a prefix-free code. We know, from equation (S17.2), that $H_e(X) \leq \bar{\ell}_{PF,e}$. Thus, if we can show that $\bar{\ell}_{PF,e} < \bar{\bar{L}}_{s,e} + 2\log_2(\log_2(\Omega)) + 5$, then we know that $H_e(X) < \bar{\bar{L}}_{s,e} + 2\log_2(\log_2(\Omega)) + 5$, and so this would complete proof of the inequality (S17.1).

Let us now specify the prefix-free coding scheme that we require to prove inequality (S17.1). Our method is inspired by procedures described by Vitányi and Li[35]. To begin, let us encode the positive integers $(1, 2, 3, ...)$ using all possible binary sequences, as arranged in length-lexicographic order. Thus, we encode $1, 2, 3, 4, 5, 6, 7, 8...$ with binary sequences $0, 1, 00, 01, 10, 11, 000, 001, ...$, respectively. Using this coding scheme, let $B(y)$ represent the binary code for $y$, where $y$ is a positive integer. Note that because this coding scheme uses *all* possible binary sequences, it is a prefix-allowing code for the positive integers. We will call this coding scheme the *standard prefix-allowing code for the integers*.

Let $R_e(x)$ represent the *fitness rank* for organismal-type $x$ when it occurs in environments of type $e$. The value of $R_e(x)$ is equal to 1 for the fittest organismal type in environments of type $e$, and it is equal to 2 for the second-fittest type, it is equal to 3 for the third-fittest type, and so on. Under our current assumptions it is straightforward to show that the optimal coding scheme (as defined in Supplementary Note 5) is one in which the binary code for organismal type $x$ (when it occurs in environments of type $e$) is $B(R_e(x))$.

To specify our prefix-free code for organismal type $x$ (when it occurs in environments of type $e$), let us define $\ell(y)$ to be the number of binary digits in the standard prefix-allowing code for the integer $y$. Furthermore, let $\ell^* = \ell(R_e(x))$. Thus, $\ell^*$ is the length, in binary digits, of the optimal (and prefix-allowing) code for organismal-type $x$, when optimised for environments of type $e$. Let us also define $\ell^{**}$ to be the length, in binary digits, of the standard prefix-allowing code for $\ell^*$. Thus $\ell^{**} = \ell(\ell^*)$.

Let us write our prefix-free code from left to right. Our prefix-free code for organismal-type $x$ (when it occurs in environments of type $e$) is illustrated in Figure $S17.1$. The code begins, on the left side, with $\ell^{**}$ ones. Next comes a zero. Next is the standard prefix-allowing code for $\ell^*$ (that is, $B(\ell^*)$). Finally, at the far right we have the standard prefix-allowing code for $R_e(x)$ (which is the fitness rank of organismal type $x$ in environments of type $e$).

$$\underbrace{111\cdots 1111}_{\ell^{**}\text{ones}}\ 0\ \underbrace{10100\cdots 1101}_{B(\ell^*)}\ \underbrace{10101100010\cdots 1010}_{B[R_e(x)]}$$

**Fig. S17.1 | A prefix-free code word for organismal type $x$, when it occurs in environments of type $e$.** The prefix-free code illustrated here is described in the text.

For a prefix-free coding scheme, one consequence of the lack of prefixes is that we can read a binary code word from the beginning, and we will always know when we have reached the last digit. With this in mind, let us now consider how we can use the code word illustrated in Fig. $S17.1$ to find the value of $x$ in a prefix-free manner (i.e., so that, when we read the code word from left to right, we know when we have reached the end of the code word). To do this, we start at the left, and then move right, digit by digit, counting the number of ones that we find before we reach the zero. The resulting integer gives us the number of binary digits in $B(\ell^*)$, which is the standard prefix-allowing code for $\ell^*$. The zero is followed by $B(\ell^*)$, which we can now read unambiguously, and thus, we can unambiguously know the value of $\ell^*$. The reason for this lack of ambiguity is that we know the length of $B(\ell^*)$, so we know exactly where it begins (just after the zero), and we know where it ends (because we know how long it is).

The remaining binary digits in the code word shown is Fig. $S17.1$ constitute the standard prefix-allowing code for $R_e(x)$. We can use these digits to find $R_e(x)$ unambiguously, even if the code word illustrated in Fig. $S17.1$ is immediately followed by more binary digits. The reason for this lack of ambiguity is that the previous part of the code word allows us to know the value of $\ell^*$, which is the length of the standard prefix-allowing binary code for $R_e(x)$. Thus, we can find the value of $R_e(x)$, even if the code word shown in Fig $S17.1$ is immediately followed by additional binary digits.

Recall that, under our current assumptions, each organismal type has a unique fitness rank ($R_e(x)$), and this fitness rank is used to encode $x$ in the optimal (prefix-allowing) coding scheme for environments of type $e$. Thus, so long as we have access this optimal coding scheme, the value of $R_e(x)$ allows us to find the value of $x$. Equivalently, using the codeword illustrated in Fig. $S17.1$, we can also find the value of $x$ if we just have access to the fitness landscape for environments of type $e$. This is because access to the fitness landscape allows us to find the organismal type that has fitness-rank $R_e(x)$, and this is $x$.

Next, we shall calculate an upper bound for the length of the code word illustrated in Fig. $S17.1$. Starting on the right, we have the binary sequence $B(R_e(x))$, and this has a length of $\ell^*$, as stated above. Next, as we move to the left, we have a binary sequence which constitutes the standard binary prefix-allowing code for $\ell^*$. As stated above, the length of this sequence (in binary digits) is denoted as $\ell^{**}$. Next, to the left, we have a zero, and finally, to the left of the zero, we have $\ell^{**}$ ones. Thus, the length of the prefix-free-binary code word is $\ell^* + 1 + 2\ell^{**}$.

We know that, for any positive integer $y$, $\ell(y) \leq \log_2(y) + 1$ (see equation S8.1). Furthermore, clearly, $\ell(R_e(x)) \leq \ell(\Omega)$. This last inequality holds because $R_e(x)$ is a fitness rank among types, and thus the maximum-possible value of $R_e(x)$ is $\Omega$. Recall that $\ell^* = \ell(R_e(x))$. Thus, we know that $\ell^* \leq \ell(\Omega)$.

Since $\Omega$ is a positive integer, we know that $\ell^* \leq \log_2(\Omega) + 1$. Recall that $\ell^{**} = \ell(\ell^*)$. Thus, we have $\ell^{**} \leq \log_2(\log_2(\Omega) + 1) + 1$. It is straightforward to show that, for any positive integer $y$ such that $y \geq 2$, we have $\log_2\left[\log_2(y) + 1\right] \leq \log_2\left[\log_2(y)\right] + 1$. Thus, $\log_2(\log_2(\Omega) + 1) + 1 \leq \log_2(\log_2(\Omega)) + 2$. This implies that $\ell^{**} \leq \log_2(\log_2(\Omega)) + 2$.

We can now calculate a maximum-possible value for the mean length of the code word displayed in Fig. $S$17.1. As noted above, this code word has a length of $\ell^* + 1 + 2\ell^{**}$. Recall that $\overline{\overline{L}}_{s,e}$ represents the mean value of $\overline{L}_s$ among individuals that live in environments of type $e$. The quantity $\overline{L}_s$ is the mean value of the length of all of the code words that could be used for a particular organismal type, when averaging among all of the coding schemes that are optimal, given the local fitness landscape. However, as we have assumed, for the purposes of the current Supplementary Note, that each type has a unique fitness value in environment $e$, we know that, for environment $e$, there is only one optimal code word for organismal type $x$, and its length is equal to $\ell^*$, where $\ell^* = \ell(R_e(x))$. Thus, under our current assumptions we have

$$\overline{\overline{L}}_{s,e} = \sum_{x=1}^{\Omega} p_{X|E}(x|e)\,\ell(R_e(x)), \qquad (S17.3)$$

where $p_{X|E}(x|e)$ is the frequency of type $x$ in environments of type $e$, as defined in Supplementary Note 10.

We have noted that $\ell^{**} \leq \log_2(\log_2(\Omega)) + 2$. As our prefix-free code word has a length of $\ell^* + 1 + 2\ell^{**}$, this implies that, for every organismal type in environment $e$, the code word must have a length that is no greater than $\ell^* + 2\log_2(\log_2(\Omega)) + 5$. We know that $\ell^* = \ell(R_e(x))$, and thus, in light of equation (S17.3), we know that the mean value of our prefix-free codeword (when we average over individuals that live in environments of type $e$) is no greater than $\overline{\overline{L}}_{s,e} + 2\log_2(\log_2(\Omega)) + 5$. The value of $\overline{\ell}_{PF,e}$ represents the shortest-possible mean value of prefix-free code words that encode the organismal types of individuals living in environments of type $e$. As $\overline{\ell}_{PF,e}$ is the shortest-possible mean value, we know that it must be less than or equal to the mean value of code words generated as in Fig. $S$17.1. As such, it must be the case that $\overline{\ell}_{PF,e} \leq \overline{\overline{L}}_{s,e} + 2\log_2(\log_2(\Omega)) + 5$. We know, from Shannon's Noiseless Coding Theorem (equation (S17.2)) that $H_e(X) \leq \overline{\ell}_{PF,e}$. As such, we now also know that $H_e(X) \leq \overline{\overline{L}}_{s,e} + 2\log_2(\log_2(\Omega)) + 5$. Above, we have already proved that $\overline{\overline{L}}_{s,e} - 1 \leq H_e(X)$, and thus our proof of inequality (S17.1) is now complete.

Let us next proceed to the derivation of the inequality (S10.5) in Supplementary Note 10. From Supplementary Note 19 we know that, for any integer $y$ such that $y \geq 2$, we have $\log_2(y) - 3 < \mu(y) < \log_2(y) + 1$, where $\mu(y)$ is the mean of the first $y$ binary codewords (i.e., the first in length-lexicographical order). Furthermore, from Supplementary Note 8, we note that $\overline{L}_u = \mu(\Omega)$. Thus

$$\overline{L}_u < \log_2(\Omega) + 1 \tag{S17.4}$$

and

$$\overline{L}_u > \log_2(\Omega) - 3. \tag{S17.5}$$

From inequality (S17.1) we know that $\overline{\overline{L}}_{s,e} - 1 \leq H_e(X)$. Let us now add $\log_2(\Omega) - 3$ to the left side of this inequality, and $\overline{L}_u$ to the right side. After rearranging, this yields (in light of inequality (S17.5)) $\log_2(\Omega) - H_e(X) - 4 < \overline{L}_u - \overline{\overline{L}}_{s,e}$. Also, from inequality (S17.1), we know that $H_e(X) \leq \overline{\overline{L}}_{s,e} + 2\log_2(\log_2(\Omega)) + 5$. Let us add $\overline{L}_u$ to the left side of this inequality, and let us add $\log_2(\Omega) + 1$ to the right side. After rearranging, and in light of inequality (S17.4), this yields $\overline{L}_u - \overline{\overline{L}}_{s,e} < \log_2(\Omega) - H_e(X) + 2\log_2(\log_2(\Omega)) + 6$. Note, from equation (1) in the main text that the reproductive information provided by the fitness landscape in a particular area that is about the organismal type (i.e., $I_R(\overrightarrow{W}:x)$) is defined by $I_R(\overrightarrow{W}:x) = \overline{L}_u - \overline{L}_s$. As noted above, the value of $\overline{L}_u$ is given by $\overline{L}_u = \mu(\Omega)$, and thus it is the same for all population members. As $\overline{\overline{L}}_{s,e}$ is the mean value of $\overline{L}_s$ in environments of type $e$, this imples that $\overline{L}_u - \overline{\overline{L}}_{s,e}$ is the mean value of reproductive information in environments of type $e$. In Supplementary Note 10 this mean value is denoted by $\overline{I_{R,e}(\overrightarrow{W}:x)}$, and thus we have $\overline{I_{R,e}(\overrightarrow{W}:x)} = \overline{L}_u - \overline{\overline{L}}_{s,e}$. Collectively, these considerations lead to

$$[\log_2(\Omega) - H_e(X)] - 4 < \overline{I_{R,e}(\overrightarrow{W}:x)} < [\log_2(\Omega) - H_e(X)] + 2\log_2(\log_2(\Omega)) + 6. \tag{S17.6}$$

This is identical to inequality (S10.5) in Supplementary Note 10, and so the verification of these inequalities is now complete.

Let us now proceed to prove inequality (S10.8) from Supplementary Note 10. First, consider the inequality on the left in inequality (S17.6) (that is, $[\log_2(\Omega) - H_e(X)] - 4 < \overline{I_{R,e}(\overrightarrow{W}:x)}$). Given that this inequality is true, it must also be true that

$$\sum_{e=1}^{\Omega!} p_E(e)\left([\log_2(\Omega) - H_e(X)] - 4\right) < \sum_{e=1}^{\Omega!} p_E(e)\overline{I_{R,e}(\overrightarrow{W}:x)}. \tag{S17.7}$$

In Supplementary Note 10 the conditional entropy of $x$ (i.e., $\overline{H_e(x)}$) was defined as $\overline{H_e(X)} = \sum_{e=1}^{\Omega!} p_E(e)H_e(X)$. Furthermore, it is straightforward to verify that $\sum_{e=1}^{\Omega!} p_E(e)\overline{I_{R,e}(\overrightarrow{W}:x)} = \sum_{e=1}^{\Omega!}\sum_{x=1}^{\Omega} p_{E,X}(e,x)I_R(\overrightarrow{W}:x)$, and thus, from equation (S10.7) in Supplementary Note 10, we know that the right side of inequality (S17.7) is equal to $\overline{I_R(\overrightarrow{W}:x)}$, the mean value of reproductive information, if we average over all of the individuals living in all possible environments. We know that $\log_2(\Omega)$ and the number 4 are constants, and so we can simplify inequality (S17.7) so that it reads

$$\log_2(\Omega) - \overline{H_e(X)} - 4 < \overline{I_R(\overrightarrow{W}:x)}. \tag{S17.8}$$

The unconditional entropy ($H(X)$) is defined by equation (S10.6) in Supplementary Note 10. Let us both add and subtract $H(X)$ to the left side of equation (S17.8) to obtain

$$\log_2(\Omega) + \left[H(X) - \overline{H_e(X)}\right] - H(X) - 4 < \overline{I_R(\overrightarrow{W}:X)}. \tag{S17.9}$$

We can now use equation (S10.10) from Supplementary Note 10 to restate this inequality as

$$\log_2(\Omega) + I_S(\overrightarrow{W}; X) - H(X) - 4 < \overline{I_R(\overrightarrow{W} : X)}. \tag{S17.10}$$

A nearly identical procedure allows us to rewrite the inequality on the right in inequality (S17.6) as

$$\overline{I_R(\overrightarrow{W} : X)} < \log_2(\Omega) + I_S(\overrightarrow{W}; X) - H(X) + 2\log_2(\log_2(\Omega)) + 6. \tag{S17.11}$$

Comparison of inequalities (S17.10) and (S17.11) to inequality (S10.8) in Supplementary Note 10 shows that, collectively, these two inequalities complete our confirmation of inequality (S10.8) in Supplementary Note 10.

Finally, let us prove inequality (S10.12) in Supplementary Note 10. Consider an organismal type for which $I_R(\overrightarrow{W} : x) > I_R^*$, where $I_R^*$ is a positive real number, as defined in Supplementary Note 10. Call this the *focal type*. Let $\alpha^* = I_R^* - 4$. We can show, from inequality (2) in the main text, that, for the fitness landscape that describes the focal-type's environment, the number of organismal types that have a fitness that is *at least* the fitness of the focal type must be no greater than $\Omega/2^{\alpha^*}$.

Next, consider an environmental type (i.e., a value of $e$) for which the proportion of individuals for which $I_R(\overrightarrow{W} : x) > I_R^*$ is at least equal to $\Theta$. We consider the organisms who live in this type of environment to fall into two groups, with regard to their organismal types. One group has organismal types for which $I_R(\overrightarrow{W} : x) > I_R^*$. Let us call this the *high group*. The other group has organismal types for which $I_R(\overrightarrow{W} : x) \leq I_R^*$. Call this the *low group*. Note that no organismal types that appear among individuals in the high group can possibly occur among individuals in the low group, and vice versa. In this sense, the two groups are *disjoint*. It can be shown that, if a population is made up of two disjoint groups with entropies on some characteristic of $H_a$ and $H_b$, and if the proportion of individuals in these two groups is $q$ and $1 - q$, respectively, then the entropy for the *entire* population is given by $q \log_2(H_a) + (1 - q) \log_2(H_b) - q \log_2(q) - (1 - q) \log_2(1 - q)$.

Within the high group no more than $\Omega/2^{\alpha^*}$ organismal types can possibly be represented (as indicated above). The maximum number of organismal types that can possibly be represented within the low group is clearly less than $\Omega$. Furthermore, since $0 \leq q \leq 1$, we know that $-q \log_2(q) - (1 - q) \log_2(1 - q) \leq 1$. Finally, note that $\log_2(\Omega/2^{\alpha_*}) = \log_2(\Omega) - \alpha^*$. As entropy cannot be greater than the logarithm of the maximum number of different types that can appear in a population, and as the proportion of individuals who live in the environments of the type under consideration here is at least $\Theta$, the preceding considerations imply that, for environments of the type under consideration, the maximum-possible entropy in the value of $x$ must be less than

$$\Theta\left(\log_2(\Omega) - \alpha^*\right) + (1 - \Theta)\log_2(\Omega) + 1. \tag{S17.12}$$

This expression simplifies to $1 + \log_2(\Omega) - \Theta \alpha^*$.

As stated in Supplementary Note 10, the conditional entropy is defined by

$$\overline{H_e(X)} = \sum_{e=1}^{\Omega!} p_E(e) H_e(X). \tag{S17.13}$$

We have assumed that a proportion $\Phi$ of individuals live in environments of the type just described. In particular, they live in environments for which a proportion of individuals that is at least equal to $\Theta$ have a value of reproductive information that satisfies $I_R(\overrightarrow{W}:x) > I_R^*$. The total number of possible organismal types is $\Omega$, and thus, within all *other* environments, the entropy cannot be greater than $\log_2(\Omega)$. Collectively, these observations lead to the conclusion that conditional entropy is bounded above by

$$\overline{H_e(X)} < \Phi\left(1 + \log_2(\Omega) - \Theta\alpha^*\right) + (1-\Phi)\log_2(\Omega) \tag{S17.14}$$

After simplification, this allows us to state

$$\overline{H_e(X)} < \Phi + \log_2(\Omega) - \Phi\Theta\alpha^*. \tag{S17.15}$$

Equation (S10.10) in Supplementary Note 10 tells us that $I_S(\overrightarrow{W};X) = H(X) - \overline{H_e(X)}$. Using this, along with inequality (S17.15), we obtain

$$I_S(\overrightarrow{W};X) > H(X) - \log_2(\Omega) + \Phi\Theta\alpha^* - \Phi. \tag{S17.16}$$

We have defined $\alpha^*$ as $\alpha^* = I_R^* - 4$. In addition, we note that the maximum-possible value of $\Phi\Theta$ is unity and that the maximum-possible value of $\Phi$ is also unity. Thus we have

$$I_S(\overrightarrow{W};X) > H(X) - \log_2(\Omega) + \Theta\Phi I_R^* - 5. \tag{S17.17}$$

As this is identical to equation (S10.12) in Supplementary Note 10, the proof is now complete.

## Supplementary Note 18: Calculation details for reproductive information using continuously distributed phenotypic traits

In this note, we will provide some details on the calculations used to produce Supplementary Note 13.

In order to calculate sound intensity in decibels, we start with a measure of acoustic energy, which is the energy being received from a sound source. Acoustic energy is typically measured in watts per square meter[56] ($W/m^2$). To calculate sound intensity in decibels, we must choose a small but non-zero "reference" sound energy. The value that is typically used (and which we use in our calculations) is $10^{-12}$ $W/m^2$. This value has been shown experimentally to be close to the lower threshold of human hearing. Let us use $S_W$ to represent sound intensity as measured in watts per square meter, and let us use $S_{dB}$ to represent sound intensity as measured in decibels. Using standard definitions[56], the relationship between $S_W$ and $S_{dB}$ is as follows:

$$S_{dB} = 10 \log_{10}\left(\frac{S_W}{10^{-12}}\right). \tag{S18.1}$$

Equivalently, the relationship can be expressed as:

$$S_W = 10^{-12}\left(10^{S_{dB}/10}\right). \tag{S18.2}$$

Equations S18.1 and S18.2 were used to produce Fig. $S$13.1 in Supplementary Note 13, and they were also used to calculate numerical values that are mentioned in Supplementary Note 13.

# Supplementary Note 19: A result regarding the sum of the length of binary code words

In this note we prove an additional inequality related to the length of binary code words. The proof draws on some results derived in Note 8 of the Supplementary Information.

With $b = 2, 3, ...$ the required inequality takes the form

$$\log_2(b) - 3 < \mu(b) < \log_2(b) + 1 \tag{S19.1}$$

where $\mu(b)$ is the mean length of first $b$ binary code words.

**Proof**

To prove inequality (S19.1) we begin with the length of binary code word $b$, which we write as $\lambda(b)$. This is given by

$$\lambda(b) = \lfloor \log_2(b+1) \rfloor \tag{S19.2}$$

where $\log_2(b)$ denotes the logarithm of $b$ to base 2 and $\lfloor b \rfloor$ denotes the largest integer $\leq b$.

For $b > 1$ we have

(i) $\lfloor \log_2(b+1) \rfloor \leq \log_2(b+1) < \log_2(b) + 1$,

(ii) $\lfloor \log_2(b+1) \rfloor > \log_2(b+1) - 1 > \log_2(b) - 1$.

Thus $\lambda(b)$ satisfies the inequality

$$\log_2(b) - 1 < \lambda(b) < \log_2(b) + 1. \tag{S19.3}$$

Next, we note that the relation between $\mu(b)$ and $\lambda(b)$ is given in Eq. (S8.4):

$$\mu(b) = \lambda(b) - \Delta(b) \tag{S19.4}$$

and for the purposes of this note all we need to know about $\Delta(b)$ is inequality (S8.8), namely

$$0 \leq \Delta(b) < 2. \tag{S19.5}$$

We proceed by combining inequality (S19.3) and equation (S19.4) yielding

$$\log_2(b) - 1 < \mu(b) + \Delta(b) \tag{S19.6}$$

and

$$\mu(b) + \Delta(b) < \log_2(b) + 1. \tag{S19.7}$$

Using inequality (S19.5) allows us to write

$$\log_2(b) - 1 < \mu(b) + \Delta(b) < \mu(b) + 2 \tag{S19.8}$$

and

$$\mu(b) \leq \mu(b) + \Delta(b) < \log_2(b) + 1. \tag{S19.9}$$

From these last two inequalities we obtain $\log_2(b) - 3 < \mu(b)$ and $\mu(b) < \log_2(b) + 1$, thereby establishing inequality (S19.1).

# References used in the Supplementary Material